\documentclass[preprint,aps,eqsecnum]{revtex4} 

\usepackage{bm,amsfonts,amsmath}
\usepackage[mathscr]{eucal}

\def\eqref#1{(\ref{#1})}
\def\eqrefs#1#2{(\ref{#1}) and~(\ref{#2})}
\def\eqsref#1#2{(\ref{#1})--(\ref{#2})}

\def\secref#1{Sec.~\ref{#1}}

\def\Ref#1{Ref.\cite{#1}}

\def\tableref#1{Table~\ref{#1}}

\def\EQ{\begin{equation}}
\def\EQs{\begin{eqnarray}}
\def\endEQ{\end{equation}}
\def\endEQs{\end{eqnarray}}

\def\Items{\begin{enumerate}}
\def\endItems{\end{enumerate}}

\def\AR{\begin{array}}
\def\endAR{\end{array}}
\def\TAB{\begin{tabular}}
\def\endTAB{\end{tabular}}

\def\eqtext#1{\hbox{\rm{#1}}}

\def\mstrut{\mathstrut}
\def\hp#1{\hphantom{#1}}

\def\up#1{{\mstrut}_{\mstrut}^{\mstrut #1}}
\def\down#1{{\mstrut}^{\mstrut}_{\mstrut #1}}
\def\mixed#1#2{{\mstrut}^{\mstrut #1}_{\mstrut #2}}
\def\downup#1#2{{\mstrut}_{\mstrut #1}^{\hp{#1}\mstrut #2}}
\def\updown#1#2{{\mstrut}^{\mstrut #1}_{\hp{#1}\mstrut #2}}

\def\Rnum{\mathbb{R}}
\def\Cnum{\mathbb{C}}
\def\frac#1#2{{\textstyle {#1\over#2}}}

\def\der#1{\partial\down{#1}}
\def\coder#1{\partial\up{#1}}
\def\D#1{D\down{#1}}
\def\coD#1{D\up{#1}}
\def\totder{D}
\def\extder{d}

\def\flat#1{\eta\down{#1}}
\def\invflat#1{\eta\up{#1}}
\def\vol#1#2{\epsilon\downup{#1}{#2}}
\def\invvol#1#2{\epsilon\updown{#1}{#2}}
\def\id#1#2{\delta\updown{#1}{#2}}

\def\x#1#2{x\mixed{#1}{#2}}

\def\curr#1{\Psi\up{#1}}
\def\curl#1#2{\Theta\mixed{#1}{#2}}
\def\triv#1#2{\D{#1}\Theta\up{#2}}

\def\kv#1#2{\xi\mixed{#1}{#2}}
\def\kvp#1#2{\xi'\mixed{#1}{\! #2}}
\def\ky#1#2{Y\down{#1}\up{#2}}
\def\duky#1#2{{*Y}\down{#1}\up{#2}}

\def\Lie#1{{\cal L}_{#1}}
\def\wLie#1{\hat{{\cal L}}_{#1}}

\def\dilop{{\mathscr D}}

\def\div{{\rm div\,}}
\def\trfr{{\rm trfr}}
\def\mod#1{\;\;\mbox{mod}\;\;{#1}} 
\def\modcurl{\eqtext{ mod curls}}
\def\const{{\rm const}}
\def\i{{\rm i}}

\def\Int{{\displaystyle\int}}

\newtheorem{thm}{Theorem}[section]
\newtheorem{lem}[thm]{Lemma}
\newtheorem{cor}[thm]{Corollary} 
\newtheorem{prop}[thm]{Proposition} 
\newtheorem{defn}[thm]{Definition}

\def\STR{1.0}
\def\mystretch#1{\renewcommand{\arraystretch}{#1}}

\def\Proof#1{{\bf Proof{#1}:}}
\def\endProof{\hfill$\Box$}

\def\l{\left}
\def\r{\right}

\def\Rsp#1{R\up{#1}}
\def\Jsp#1{J\up{#1}}
\def\divfrRsp#1{\tilde R\up{#1}}

\def\eval#1#2{ \left. {#1} \right|_{#2}}

\def\F#1#2{F\updown{#1}{#2}}
\def\duF#1#2{{*F}\updown{#1}{#2}}
\def\Fp#1#2{F'\updown{#1}{#2}}

\def\A#1#2{A\downup{#2}{#1}}
\def\Ap#1#2{A'\down{\! #2}\up{#1}}
\def\tA#1#2{{\tilde A}\downup{#2}{#1}}
\def\tAp#1#2{{\tilde A'}\down{\! #2}\up{#1}}

\def\divfrA{\tilde A}

\def\Je#1{J_{\rm e}^{#1}}
\def\Jm#1{J_{\rm m}^{#1}}

\def\SO{\mathfrak{so}}
\def\U{\mathfrak{u}}

\def\X{{\bf X}}
\def\prX{{\rm pr} {\bf X}}

\def\sX{\X_{\it scal}}
\def\dX{\X_{\it dual}}
\def\hX#1{\X_{\xi^H_{#1}}}
\def\cX#1{\X_{\xi_{#1}}}
\def\rbX#1{\X_{\gamma_{#1}}}
\def\durbX#1{\X_{*\gamma_{#1}}}

\def\scurr#1{\Psi^{#1}_{\it scal}}
\def\dcurr#1{\Psi^{#1}_{\it dual}}
\def\ccurr#1{\Psi^{#1}_{\xi}}
\def\cducurr#1{\Psi'{}^{#1}_{\xi}}
\def\rbcurr#1{\Psi^{#1}_{\gamma}}
\def\rbducurr#1{\Psi^{#1}_{*\gamma}}

\def\tX{{\bf Z}}
\def\ctX#1{\tX_{\xi_{#1}}}
\def\rbtX#1{\tX_{\gamma_{#1}}}
\def\kyX{\tX_{Y}}

\def\tcurr{\Phi}
\def\dtcurr#1{\tcurr^{#1}_{\it dual}}
\def\rbtcurr#1{\tcurr^{#1}_{\gamma}}
\def\ctcurr#1{\tcurr^{#1}_{\xi}}
\def\ductcurr#1{\tcurr'{}^{#1}_{\xi}}
\def\wavecurr#1{\Upsilon^{#1}}

\def\kycurr#1{\tcurr^{#1}_{Y}}
\def\dukycurr#1{\tcurr'{}^{#1}_{Y}}

\def\consT#1#2{T\updown{#1}{#2}}

\def\Parder#1{ {\partial \over {\partial{#1}}} }
\def\parder#1{\partial/\partial{#1}}

\def\Q#1#2{Q\mixed{#1}{#2}}
\def\Qp#1#2{Q'\mixed{#1}{\!#2}}

\def\t#1{\widetilde{#1}}
\def\P#1#2{P\mixed{#1}{#2}}
\def\duP#1#2{{*P}\updown{#1}{#2}}
\def\*P{\duP{}{}}

\def\hook{\lrcorner}

\def\llra{\longleftrightarrow}

\def\lra{\longrightarrow}
\def\Lra{\Longrightarrow}
\def\ra{\rightarrow}

\def\c#1{c\down{#1}}
\def\k#1#2#3{k_{#1}\updown{#2}{#3}}
\def\kp#1#2#3{k'_{#1}\updown{#2}{#3}}
\def\kb#1#2#3{\bar k_{#1}\updown{#2}{#3}}

\def\gam#1#2{\gamma\downup{#1}{#2}}
\def\dugam#1#2{{*\gamma}\downup{#1}{#2}}

\def\z#1#2{\zeta\downup{#1}{#2}}
\def\zp#1#2{\zeta'\down{\! #1}\up{#2}}
\def\duz#1#2{{*\zeta}\downup{#1}{#2}}
\def\duzp#1#2{{*\zeta}'\down{\! #1}\up{#2}}

\def\T#1#2#3{\Theta_{#1}\updown{#2}{#3}}
\def\Tp#1#2{\Theta'\updown{#1}{#2}}
\def\bb#1#2#3{b_{#1}\updown{#2}{#3}}

\def\aa#1{a\down{#1}}
\def\aap#1{a'\down{#1}}

\def\a#1#2{a\downup{#1}{#2}}

\def\ap#1#2{a'\down{\!#1}\up{#2}}

\def\ba#1#2{\bar{a}\down{\!#1}\up{#2}}
\def\bap#1#2{\bar{a}'\downup{#1}{#2}}
\def\ta#1#2{\tilde{a}\downup{#1}{#2}}
\def\tap#1#2{\tilde{a}'\down{\!#1}\up{#2}}
\def\ah#1#2{\hat a\downup{#1}{#2}}
\def\aph#1#2{{\hat a}'\down{\!#1}\up{#2}}

\def\eq/{equation}
\def\sym/{symmetry}
\def\syms/{symmetries}
\def\adjsym/{adjoint-\sym/}
\def\adjsyms/{adjoint-\syms/}
\def\Sym/{Symmetry}
\def\Syms/{Symmetries}
\def\Adjsym/{Adjoint-\sym/}

\def\rbs/{rotations/boosts}
\def\rb/{rotation/boost}
\def\rbvec/{internal \rb/}
\def\rbsvec/{internal \rbs/}
\def\cvec/{internal conformal}
\def\cdualvec/{internal dual-conformal}

\def\solsp/{solution space}
\def\potsys/{potential system}
\def\Potsys/{Potential system}
\def\Meq/{Maxwell's equations}
\def\Minksp/{Minkowski space}
\def\conslaw/{conservation law}
\def\conscurr/{conserved current}
\def\Kvec/{Killing vector}
\def\CKvec/{conformal \Kvec/}
\def\HKvec/{homothetic \Kvec/}
\def\KV/{Killing-vector}
\def\CKV/{conformal \KV/}
\def\HKV/{homothetic \KV/}
\def\KY/{Killing-Yano}

\def\loccohom#1{local #1-form cohomology}
\def\horiz/{differential}
\def\geom/{geometric}
\def\jps/{joint potential system}
\def\mps/{magnetic potential system}
\def\eps/{electric potential system}
\def\dutr/{duality transformation}

\def\ie/{i.e.}
\def\eg/{e.g.}

\begin{document}

\title{ 
Symmetries, conservation laws, and cohomology of 
Maxwell's equations using potentials 
}

\author{Stephen C. Anco}
\email{sanco@brocku.ca}
\affiliation{
Department of Mathematics, \\
Brock University,  \\
St.Catharines, ON Canada L2S 3A1 }

\author{Dennis The}
\email{dthe@math.mcgill.ca}
\affiliation{
Department of Mathematics, \\
McGill University, \\
Montreal, QC Canada H3A 2K6}

\begin{abstract}
New nonlocal symmetries and conservation laws are derived for
Maxwell's equations in 3+1 dimensional Minkowski space 
using a covariant system of joint vector potentials 
for the electromagnetic tensor field and its dual. 
A key property of this system, 
as well as of this class of new symmetries and conservation laws,
is their invariance under the duality transformation 
that exchanges the electromagnetic field with its dual. 
(In contrast the standard potential system 
using a single vector potential is not duality-invariant.)
The nonlocal symmetries of Maxwell's equations
come from an explicit classification of
all symmetries of a certain natural geometric form admitted by 
the joint potential system in Lorentz gauge. 
In addition to scaling and duality-rotation symmetries,
and the well-known Poincar\'e and dilation symmetries
which involve homothetic Killing vectors, 
the classification yields new geometric symmetries 
involving Killing-Yano tensors 
related to rotations/boosts and inversions. 
The nonlocal conservation laws of Maxwell's equations 
are constructed from these geometric symmetries by applying 
a conserved current formula that uses the joint potentials 
and directly generates conservation laws 
from any (local or nonlocal) symmetries of Maxwell's equations. 
This formula is shown to arise through a series of mappings that relate, 
respectively, 
symmetries/adjoint-symmetries of the joint potential system
and adjoint-symmetries/symmetries of Maxwell's equations.
The mappings are derived as by-products of the study of 
cohomology of closed 1-forms and 2-forms locally constructed 
from the electromagnetic field and its derivatives to any finite order
for all solutions of Maxwell's equations. 
In particular it is shown that the only nontrivial cohomology 
consists of the electromagnetic field (2-form) itself
as well as its dual (2-form),
and that this 2-form cohomology is killed by the introduction of 
corresponding potentials. 
\end{abstract}


\maketitle

\section{Introduction}
\label{sec:intro}

Two basic aspects in the study of field \eq/s in mathematical physics
are their \sym/ structure and their \conslaw/ structure.
Geometrically speaking,
symmetries are infinitesimal transformations of the fields
under which all solutions are mapped into solutions.
Symmetries of local form in the fields and partial derivatives
of the fields to a finite order (generalizing classical point symmetries)
are the basis for construction of exact (invariant) solutions
and provide an important connection with separation of variables
in certain cases.
In the situation where a system of field \eq/s has a Lagrangian,
every local \sym/ that leaves the Lagrangian invariant to within a divergence
yields a \conslaw/, 
namely a current (density and flux) 
whose space-time divergence 
(\ie/ time derivative of density plus spatial divergence of flux)
vanishes on all solutions of the field equations, 
through Noether's theorem. 
Conservation laws given by currents of local form
determine physically important conserved quantities such as
energy, momentum, angular momentum, mass, charge, etc.
which are constants of motion 
central to an analysis of the time evolution of the fields.
An infinite hierarchy of local \syms/ and \conslaw/s
of increasing higher order is a hallmark of
complete integrability of field equations.
Nonlocal \syms/ and nonlocal \conslaw/s,
involving other than a local form,
such as essential dependence on potentials or integrals of the fields,
have been less studied
but are also important and useful in the study of field equations ---
for instance, they yield exact solutions and conserved quantities
that are not obtainable from local \syms/ or local \conslaw/s
\cite{AncBlu:1996JMP}. 

For the source-free \Meq/ in classical electromagnetic field theory
in 3+1 dimensional Minkowski space,
a complete explicit classification of all local \syms/
and local \conslaw/s in a unified coordinate-invariant form
has been recently carried out by Anco \& Pohjanpelto 
\cite{AncPoh:2004,AncPoh:2001}.
The symmetry classification was obtained by
solving the \sym/ determining \eq/s
through the use of spinor techniques 
(and properties of Killing spinors).
Since \Meq/ are a non-Lagrangian system,
Noether's theorem cannot be used to find \conslaw/s from \syms/.
A scaling formula that generates all nontrivial conserved currents
from \adjsyms/ was instead used to classify \conslaw/s
through solving the similar \adjsym/ determining equations.
This formula is a variant of a general \conslaw/ formula
that applies to any
PDE system admitting a scaling \sym/ and produces
all local conserved currents with nonzero scaling weight 
\cite{Anc:2003JPhysA}.

The local \conslaw/s of \Meq/ comprise
the well-known stress-energy currents and Lipkin's zilch currents
\cite{BesHag,Lip,Kib,Fai,Mor}, 
and new chiral currents which contain first derivatives of
the electromagnetic field and hence are of one order higher in derivatives
compared to the stress-energy currents.
Associated with these \conslaw/s are corresponding conserved tensors.
The local \syms/ of \Meq/ consist of 
infinitesimal scaling and duality-rotation transformations, 
infinitesimal Poincar\'e, dilation and conformal transformations
\cite{Bat:1909,Cun:1909,Ibr:1968},
and in addition 
infinitesimal chiral transformations
which are, again, of one order higher in derivatives.
Second order \syms/ and conserved currents of chiral type were
first discovered by Fushchich \& Nikitin 
\cite{FusNik:1983,FusNik:1987book,FusNik:1992}
several years ago
and possess the striking feature of odd parity
under exchange of electric and magnetic fields,
in contrast to the even parity of the stress-energy and zilch currents
as well as that of the Poincar\'e \syms/ and the dilation/conformal \syms/.
Due to the covariant linear nature of \Meq/,
a hierarchy of higher order \syms/ and currents
arise \cite{Poh:survey} 
by repeated replacement of the electromagnetic field
by Poincar\'e and dilation/conformal \sym/ operators 
applied to the field or its dual.
The general classification results obtained in 
\Ref{AncPoh:2004,AncPoh:2001}
state that no other local \syms/ or
local \conslaw/s exist to all orders,
apart from elementary ones of zeroth order
(which are produced through shifting the electromagnetic field
by any particular solution of \Meq/).

Some nonlocal \syms/ and \conslaw/s for \Meq/ were also derived by
Fushchich \& Nikitin \cite{FusNik:1983,FusNik:1987book}, 
in a non-covariant manner using Fourier transform methods.
For the reduction of \Meq/ to 2+1 dimensions,
nonlocal \syms/ and \conslaw/s in covariant form
were obtained by Anco \& Bluman \cite{AncBlu:1997JMP}
through the use of electric and magnetic potentials.
The well-known covariant vector potential for \Meq/
in any number of dimensions gives a Lagrangian system with gauge freedom.
Note that this potential arises from the absence of magnetic charges
and currents in free space 
and hence we refer to it as the magnetic vector potential.
In Lorentz gauge the magnetic potential system reduces
to the vector wave \eq/.
If electric charges and currents are absent, 
as in the source-free \Meq/,
there is an electric \potsys/ analogous
to the magnetic one.
In 3+1 dimensions, the electric and magnetic potentials
are each covariant vector fields that are dual in the sense
that they are exchanged under a \dutr/ on the electromagnetic field.
By comparison, in 2+1 dimensions the electric potential
is a scalar field satisfying the ordinary wave \eq/
and duality is lost.
(In more than three space dimensions, duality is also lost
since the electric potential becomes an antisymmetric tensor field.)

The introduction of these potentials for \Meq/ is an instance of
Bluman's method of \potsys/s 
\cite{BluKumRei:1988,Blu:potsys,BluDor:1995}
and Vinogradov's theory of coverings for PDE systems 
\cite{Vin:covering,KraVin:covering-1,KraVin:covering-2}.
In these two approaches, nonlocal \syms/ of a given PDE system
are realized as local \syms/ of a \potsys/ (or covering system).
\Potsys/s are characterized by the embedding property 
\cite{Blu:potsys} that modulo gauge freedom
their solutions are in one-to-one correspondence with solutions of 
the given PDE system. 
(Gauge freedom, meaning a local \sym/ that depends on an arbitrary
function of all independent variables, 
arises automatically for potentials only
in more than one space dimension.)
It is essential to have a gauge imposed on potentials
in order to obtain nonlocal \syms/,
because as proved by Anco \& Bluman \cite{AncBlu:1997JMP}
when a \potsys/ possesses gauge freedom
then all of its local \syms/ project onto only 
local (gauge-invariant) \syms/ of the original PDE system 
(provided the system is locally well-posed in the sense that 
it is locally solvable \cite{Olv:symmbook}
but has no solutions involving 
an arbitrary function of all independent variables).
Furthermore, an extension of their proof shows that 
all local \conslaw/s of such a \potsys/ 
project onto \conscurr/s whose form is necessarily gauge invariant
modulo terms that are trivially divergence-free,
which presents a severe limitation for obtaining nonlocal \conslaw/s
with essential dependence on potentials. 

In this paper we use a natural joint \potsys/ with Lorentz gauge
imposed to obtain new nonlocal \syms/ and nonlocal \conslaw/s
of \Meq/ in 3+1 dimensions.
This \potsys/ involves the simultaneous introduction of both
electric and magnetic vector potentials.
Thus, it inherits the electric-magnetic duality
invariance of \Meq/
and similarly is non-Lagrangian.
To our knowledge, there has been no previous systematic investigation of
the \syms/ or \conslaw/s of \Meq/ using these joint potentials.

After setting out some preliminaries in \secref{sec:prelim},
we discuss some important inter-relationships among
\syms/, \conslaw/s, and \adjsyms/ of \Meq/
and its various \potsys/s without gauges 
in \secref{sec:cohom}. 
These interrelationships come from the \loccohom{$p$} of \Meq/,
i.e. $p$-forms locally constructed from the spacetime coordinates
and the electromagnetic field and its derivatives (to some finite order)
on all solutions.
We show that the cohomology of closed $p$-forms modulo exact $p$-forms
determines mappings between \syms/ and \adjsyms/.
Since \adjsyms/ generate conserved currents 
through a scaling formula \cite{AncPoh:2001,Anc:2003JPhysA}, 
we obtain a correspondence between \conslaw/s and both \syms/ and \adjsyms/.
Most importantly, this leads to an explicit formula that generates
\conscurr/s directly from any \syms/ of \Meq/ or its \jps/ 
(thus by-passing the absence of a Lagrangian). 
We apply these results to the well-known \geom/ \syms/ of \Meq/
and their counterparts for the \potsys/s without gauges. 
Our results explicitly demonstrate how these \potsys/s yield only
local \syms/ and local \conscurr/s of \Meq/, 
with the exception of one \conscurr/ generated from 
the duality-rotation \sym/ of the \jps/.
This current has an essential dependence on the joint potentials
yet is found to be invariant 
with respect to the gauge freedom in these potentials
modulo trivially conserved terms. 

In \secref{sec:jps-analysis} 
we investigate the \jps/ with Lorentz gauge imposed.
A classification of geometric \syms/ is derived 
by solving the \sym/ determining equations 
using covariant tensorial methods,
from which we obtain new local \syms/ 
along with corresponding new local \conslaw/s
in terms of the potentials. 
As main results, in \secref{sec:nonlocal}
we show that these \syms/ and \conslaw/s are nonlocal
under projection to \Meq/,
and we discuss some of their resulting features.
We make some concluding remarks in \secref{sec:concl}.

\section{Preliminaries}
\label{sec:prelim}

\Meq/ for the electromagnetic field tensor
$\F{}{\mu\nu}(x)=\F{}{[\mu\nu]}(x)$ in \Minksp/
$M\up{4} = (\Rnum^{4},\flat{})$ are given by
\EQ
\der{\mu}\F{\mu\nu}{}(x) = 4\pi \Je{\nu}(x) ,\qquad
\der{\mu}\duF{\mu\nu}{}(x) = 4\pi \Jm{\nu}(x) , 
\endEQ
with electric and magnetic current sources.
Here
\EQ
\duF{}{\mu\nu} = \frac{1}{2} \vol{\mu\nu\sigma\tau}{} \F{\sigma\tau}{}
\label{dualF}
\endEQ
is the dual of $\F{}{\mu\nu}$, 
$\vol{\mu\nu\sigma\tau}{}$ is the spacetime volume form,
$\x{\mu}{}$ are the standard Minkowski coordinates,
and $\der{\mu} = \parder{}{\x{\mu}{}}$ is the coordinate derivative. 
Throughout, indices will be freely lowered or raised using
the spacetime metric $\flat{\mu\nu}$ and its inverse
$\invflat{\mu\nu}$
(with signature $(-+++)$).
When there are no current sources, the field \eq/s
\EQ
\coder{\mu}\F{}{\mu\nu}(x) = 0, \qquad
\coder{\mu}\duF{}{\mu\nu}(x) = 0
\label{ME}
\endEQ
display invariance under the \dutr/
\EQ
\F{}{\mu\nu} \ra \duF{}{\mu\nu}, \qquad
\duF{}{\mu\nu} \ra -\F{}{\mu\nu}.
\label{ME-duality-F}
\endEQ

To introduce potential variables in a covariantly natural way,
we rewrite the source-free \Meq/ \eqref{ME} in the equivalent form
\EQ
\der{[\sigma}\F{}{\mu\nu]}(x) = 0, \qquad
\der{[\sigma}\duF{}{\mu\nu]}(x) = 0.
\label{Meq2}
\endEQ
Since $F(x) = \F{}{\mu\nu}(x) \extder\x{\mu}{} \wedge \extder\x{\nu}{}$
is a closed 2-form and \Minksp/ is topologically trivial, 
we can conclude by Poincar\'{e}'s Lemma that $F(x)$ is exact, \ie/
\EQ
\F{}{\mu\nu}(x) = \der{[\mu} \A{}{\nu]}(x) 
\label{Adefn}
\endEQ
for some 1-form potential $A(x) = \A{}{\nu}(x) \extder\x{\nu}{}$.
The standard {\em \mps/} is given by
\EQ
\coder{\mu} \der{[\mu} \A{}{\nu]}(x) = 0 
\label{MPS}
\endEQ
which is a self-adjoint system and thus arises from a Lagrangian.
This system \eqref{MPS} possesses gauge freedom
\EQ
\A{}{\nu}(x) \ra \A{}{\nu}(x) + \der{\nu}\chi(x),
\label{MPS-gauge-sym}
\endEQ
where $\chi(x)$ is an arbitrary function of $\x{\mu}{}$.

The field \eq/s \eqref{Meq2} also imply that
${*F}(x) = \duF{}{\mu\nu}(x) \extder\x{\mu}{} \wedge \extder\x{\nu}{}$
is a closed 2-form. 
So again by Poincar\'{e}'s Lemma, 
${*F}(x)$ is exact, \ie/
\EQ
\duF{}{\mu\nu}(x) = \der{[\mu} \Ap{}{\nu]}(x) ,
\label{A'defn}
\endEQ
for some 1-form potential $A'(x) = \Ap{}{\nu}(x) \extder\x{\nu}{}$
which satisfies an {\em electric \potsys/} 
analogous to the magnetic \potsys/ for $A(x)$. 
Since by duality 
the electric \potsys/ shares all the same properties 
as the magnetic \potsys/, 
we will omit it in our subsequent discussion and results. 

A further natural \potsys/ of \Meq/ \eqref{ME} is obtained by
introducing both electric and magnetic potentials simultaneously.
Since $\F{}{\mu\nu}(x) = \der{[\mu} \A{}{\nu]}(x)$ 
and $\duF{}{\mu\nu}(x) = \der{[\mu} \Ap{}{\nu]}(x)$ 
must satisfy the duality relation \eqref{dualF}, 
we define their joint {\em electric-magnetic \potsys/} to be
\EQ
\der{[\mu} \Ap{}{\nu]}(x) =
\frac{1}{2} \vol{\mu\nu\sigma\tau}{} \coder{\sigma} \A{\tau}{}(x).
\label{JPS}
\endEQ
The \dutr/ (\ref{ME-duality-F}) on the electromagnetic
field induces a corresponding \dutr/
\EQ
\A{}{\mu} \ra \Ap{}{\mu}, \qquad \Ap{}{\mu} \ra -\A{}{\mu}
\label{JPS-duality}
\endEQ
on the potentials.
(Note that putting $A'=\i A$ would give the self-dual \Meq/.)
The electric-magnetic \potsys/ \eqref{JPS} admits the gauge freedom
\EQs
\A{}{\nu}(x) \ra \A{}{\nu}(x) + \der{\nu}\chi(x), &&
\Ap{}{\nu}(x) \ra \Ap{}{\nu}(x) + \der{\nu}\chi'(x),
\label{JPS-gaugesym}
\endEQs
where $\chi(x)$ and $\chi'(x)$ are arbitrary functions of $\x{\mu}{}$.
Unlike the standard \potsys/ \eqref{MPS}, 
the joint system \eqref{JPS} is not self-adjoint, 
and hence it is a non-Lagrangian system. 

It is important to note that 
via the embedding relations \eqref{Adefn}, \eqref{A'defn}, 
and the duality relation \eqref{dualF}, 
the solutions of the \potsys/s \eqrefs{MPS}{JPS} 
modulo gauge freedom 
are in one-to-one correspondence with the solutions of \Meq/ \eqref{ME}.

Associated with \Meq/ or any of its \potsys/s is 
the respective jet space $\Jsp{q}$ of order $0\leq q \leq \infty$
defined as the coordinate manifold such that each point
($q$-jet) in $\Jsp{q}$ is identified with a spacetime point $x$
and the values of the field or potential(s)
and its partial derivatives up to order $q$ at $x$.
Note here that the jet space $\Jsp{0}$ is identified with $M\up{4} \times E$
where $E$ is the vector space of 2-forms for the case of \Meq/,
1-forms for the case of the magnetic \potsys/,
and pairs of 1-forms in the case of the joint \potsys/.
In this setting we use $\D{\mu}$ to denote 
the total derivative operator with respect to $\x{\mu}{}$
and write a subscript ``$,\mu$'' for coordinates
corresponding to differentiation by $\D{\mu}$ on the field or potential(s)
in the standard way \cite{AncPoh:2001,Poh:survey}.
For \Meq/ the space of solutions is represented by 
the submanifold (solution jet space) $\Rsp{}(F) \subset \Jsp{1}(F)$
whose coordinates quotient out the field equations
on the first-order partial derivatives of the electromagnetic field
in $\Jsp{1}(F)$ \cite{AncPoh:2001,Poh:survey}. 
The $q$-prolonged solution jet space $\Rsp{q}(F) \subset \Jsp{q+1}(F)$
of \Meq/ is defined by a similar quotient 
with respect to the $q$th-order partial derivatives 
of the electromagnetic field equations, $1 \leq q \leq \infty$.
There is an analogous construction of (prolonged) 
solution jet spaces $\Rsp{q}(A,A') \subset \Jsp{q+1}(A,A')$ 
in the case of the joint \potsys/,
and $\Rsp{q}(A) \subset \Jsp{q+2}(A)$ 
in the case of the magnetic \potsys/ \cite{Poh:survey}. 
Note explicit coordinates for $\Rsp{0}(F) :=\Rsp{}(F)$ consist of 
$( \x{\mu}{},\F{}{\mu\nu},\trfr \F{}{\mu(\nu,\sigma)} )$; 
likewise 
$( \x{\mu}{},\A{}{\nu},\Ap{}{\nu},\F{}{\mu\nu},
\A{}{(\nu,\mu)},\Ap{}{(\nu,\mu)} )$
and 
$( \x{\mu}{},\A{}{\nu},\F{}{\mu\nu},\A{}{(\nu,\mu)},
\A{}{(\nu,\mu\sigma)},\trfr \F{}{\mu(\nu,\sigma)} )$
are coordinates for $\Rsp{0}(A,A'):=\Rsp{}(A,A')$ 
and $\Rsp{0}(A):=\Rsp{}(A)$, 
where ``$\trfr$'' on a tensor stands for its totally trace-free part
with respect to the Minkowski metric. 
Similar coordinates can be written down for 
the prolonged solution spaces to all orders, 
representing those components of
the electromagnetic field, potentials, and their partial derivatives 
that are freely specifiable at a spacetime point. 
Throughout, we indicate jet space coordinates by writing 
(derivatives of) $F,A,A'$ without $(x)$ dependence.

It would be typical to proceed by defining \syms/
as infinitesimal transformations 
on the (prolonged) solution jet space $\Rsp{\infty}$.
For our purposes, an equivalent characterization of \syms/
via determining equations \cite{BluAnc:2002book} is better suited.

For \Meq/ or its \potsys/s,
local \syms/ of order $q<\infty$ 
are characterized by $E$-valued functions
on $\Jsp{q} \subset \Jsp{\infty}$ (in the given coordinates)
whose restriction to $\Rsp{\infty}$
satisfies the linearization of the system equations.
Similarly, local \adjsyms/ of order $q$ are characterized by
$\tilde{E}$-valued functions on $\Jsp{q} \subset \Jsp{\infty}$ 
whose restriction to $\Rsp{\infty}$ 
satisfies the adjoint linearization of the system equations,
where $\tilde{E}$ is the vector space of 2-forms
in the case of the joint \potsys/,
1-forms in the case of the magnetic \potsys/,
and pairs of 1-forms in the case of \Meq/.
A \sym/ or \adjsym/ of order $q$ is trivial if it vanishes
when evaluated on $\Rsp{\infty}$; 
two \syms/ or \adjsyms/ that differ by a trivial one 
are considered to be equivalent.

 \mystretch{\STR}
 \begin{table}[h]
 \begin{center}
 \TAB{|c|c|c|} \hline
 System & Symmetry Equations & Adjoint-symmetry Equations
 \\ \hline\hline
 $\AR{c} \mbox{\Meq/}\\
  \coD{\mu} \F{}{\mu\nu} = 0 \\
 \frac{1}{2}\vol{\mu\nu\sigma\tau}{} \coD{\mu} \F{\sigma\tau}{} = 0 
 \endAR$ &
 $\AR{c}
 \coD{\mu} \P{}{\mu\nu} = 0 \\
 \frac{1}{2}\vol{\mu\nu\sigma\tau}{}\coD{\mu} \P{\sigma\tau}{} = 0 \\
 \endAR$ &
 $\D{[\mu} \Qp{}{\nu]} =
 \frac{1}{2} \vol{\mu\nu\sigma\tau}{} \coD{\sigma} \Q{\tau}{}$
 \\ \hline
 $\AR{c}
 \mbox{Magnetic \potsys/}\\
 \coD{\mu} \D{[\mu} \A{}{\nu]} = 0
 \endAR$ &
 $\coD{\mu} \D{[\mu} \Q{}{\nu]} = 0$ &
 $\coD{\mu} \D{[\mu} \Qp{}{\nu]} = 0$
 \\ \hline
 $\AR{c}
 \mbox{Joint \potsys/}\\
 \D{[\mu} \Ap{}{\nu]} =
 \frac{1}{2} \vol{\mu\nu\sigma\tau}{} \coD{\sigma} \A{\tau}{}
 \endAR$ &
 $\D{[\mu} \Qp{}{\nu]} =
 \frac{1}{2} \vol{\mu\nu\sigma\tau}{} \coD{\sigma} \Q{\tau}{}$ &
 $\AR{c}
 \coD{\mu} \P{}{\mu\nu} = 0 \\
 \frac{1}{2}\vol{\mu\nu\sigma\tau}{}\coD{\mu} \P{\sigma\tau}{} = 0 \\
 \endAR$
 \\ \hline
 \endTAB
 \caption{\Sym/ and \adjsym/ \eq/s for \Meq/ and \potsys/s}
 \label{table:sys-sym-adjsym}
 \end{center}
 \end{table}
 \mystretch{1}

Viewed geometrically \cite{AndTor:1996},
a local \sym/ of \Meq/ or its \potsys/s
describes a generalized vector field $\X$ on $M\up{4} \times E$
whose prolongation $\prX$ to $\Jsp{\infty}$
is tangent to the solution jet space $\Rsp{\infty}$
and involves no motion on the spacetime coordinates $\x{\mu}{}$.
In particular, corresponding to 
$P\down{\mu\nu}$, $Q\down{\mu}$, $(Q\down{\mu},Q'\down{\mu})$ 
appearing in the \sym/ equations 
in \tableref{table:sys-sym-adjsym} we have the generators
\EQ
\X = \P{}{\mu\nu} \Parder{\F{}{\mu\nu}}, \qquad
\X = \Q{}{\mu} \Parder{\A{}{\mu}}, \qquad
\X = \Q{}{\mu} \Parder{\A{}{\mu}} + \Qp{}{\mu} \Parder{\Ap{}{\mu}} . 
\endEQ
This definition can be obviously generalized
to allow motion on $\x{\mu}{}$
but it is well-known that every such \sym/ is equivalent to one
(referred to as its evolutionary form) without any motion on $\x{\mu}{}$
\cite{Olv:symmbook,AndTor:1996}. 
In contrast, unlike for \syms/, 
there is no obvious geometrical meaning for \adjsyms/ 
unless the system equations are self-adjoint
(in which case \adjsyms/ coincide with \syms/).

A local conserved current of order $q<\infty$ for \Meq/ or its \potsys/s
is a vector function $\curr{\mu}$ on $\Jsp{q} \subset \Jsp{\infty}$ 
whose divergence vanishes on the solution jet space, namely
\EQ
\D{\mu} \curr{\mu} = 0 \quad \eqtext{ on $\Rsp{\infty}$} .
\endEQ
A current is trivial if it is equal to a curl
$\curr{\mu} = \D{\nu}\curl{\mu\nu}{}$
when evaluated on $\Rsp{\infty}$, 
where $\curl{\mu\nu}{} = \curl{[\mu\nu]}{}$
is some skew-tensor function on $\Jsp{r}$, $q\leq r<\infty$. 
Two currents that differ by a trivial one are considered to be equivalent.
The equivalence class of conserved currents 
containing a current $\curr{\mu}$ 
is called the \conslaw/ associated with $\curr{\mu}$.
A \conslaw/ has order $q$ if the minimum order among all \conscurr/s
in its equivalence class is equal to $q$. 

For \Meq/ and its potential systems,
the explicit coordinates 
introduced earlier for $\Rsp{}$ and its prolongations
can be used to show that 
all nontrivial currents of order $q$ modulo curls
are characterized by multipliers (also called characteristics)
such that the divergence of a current 
$\D{\mu} \curr{\mu}$ on $\Jsp{q+1}$ 
yields, after integration by parts where necessary, 
a contracted product of the multipliers and the system equations. 
It is well-known from general results 
(see for instance \cite{AncBlu:1997PRL})
that multipliers of order $q$ are \adjsyms/
subject to certain conditions on their adjoint-linearization 
on $\Jsp{q}$, 
and there is a homotopy integral formula to recover a current
(modulo a curl) from its multipliers. 
Two main complications arise in dealing with the correspondence
between multipliers and currents for \Meq/. 
Firstly, 
because it is not a PDE system of Cauchy-Kovalevskaya form,
the simple relation \cite{Olv:symmbook,AncBlu:2002EJAM-2}
that two currents are equivalent 
if and only if their multipliers agree on $\Rsp{\infty}$ 
breaks down
and there exist trivial currents
$\curr{\mu} =\D{\nu}( \F{\mu\nu}{}\chi+\duF{\mu\nu}{}\chi' )$
of order $q+1$ whose multipliers are a class of nontrivial \adjsyms/ 
$\Q{}{\nu}=\D{\nu}\chi$, $\Qp{}{\nu}=\D{\nu}\chi'$, 
which do not vanish on $\Rsp{\infty}$,
for any (non-constant) functions $\chi,\chi'$ on $\Jsp{q}$.
Secondly,
due to the linear nature of \Meq/, 
at every order $q>0$ there are other classes of nontrivial \adjsyms/ 
all of which fail to satisfy the adjoint-linearization conditions
even to within the addition of a trivial \adjsym/.  
Similar complications are found to occur for 
the magnetic and joint \potsys/s. 

However, 
an alternative way of generating all nontrivial conserved currents
by-passing these complications
for \Meq/ and its \potsys/s
is provided by a general scaling formula that produces conserved currents
directly from \adjsyms/ 
\cite{AncPoh:2001,Anc:2003JPhysA}. 
This formula is derived from the adjoint relation
between the determining equations for \syms/ and \adjsyms/
of any given PDE system 
\cite{AncBlu:1997PRL}. 
The resulting \conscurr/ formulas for \Meq/ and its \potsys/s
are displayed in \tableref{table:adjrel-conslaw};
the notation $\mid_{\lambda F}$ is used to denote
a scaling of $F$ and derivatives of $F$ in functions on $\Jsp{q}(F)$
by a parameter $\lambda$, 
and likewise for functions on $\Jsp{q}(A)$ or $\Jsp{q}(A,A')$.

 \mystretch{\STR}
 \begin{table}[h]
 \begin{center}
 $\AR{|c|c|c|} \hline
 \mbox{System} & \mbox{Adjoint-symmetry} &
 \mbox{Conserved current formula}\\ \hline\hline
 \AR{c} \mbox{\Meq/} \\
 \coD{\mu} \F{}{\mu\nu}=0 \\
 \coD{\mu} \duF{}{\mu\nu}=0 
 \endAR &
 \Q{}{\nu}, \Qp{}{\nu} \mbox{ on } \Rsp{q}(F) &
 \curr{\mu} = \Int_0^1 \eval{ \l( \Q{}{\nu} \F{\mu\nu}{}
 + \Qp{}{\nu} \duF{\mu\nu}{} \r)}{\lambda F}
 \frac{d\lambda}{\lambda}
 \\ \hline
 \AR{c} \mbox{Magnetic \potsys/} \\
 \coD{\mu} \D{[\mu} \A{}{\nu]}=0
 \endAR &
 \Q{}{\nu} \mbox{ on } \Rsp{q}(A) &
 \AR{l}
 \curr{\mu} =
 \Int_0^1 \eval{ \l( \Q{}{\nu} \coD{[\mu} \A{\nu]}{}
 - \coD{[\mu} \Q{\nu]}{} \A{}{\nu} \r) }{\lambda A}
 \frac{d\lambda}{\lambda}
 \endAR
 \\ \hline
 \AR{c} \mbox{Joint \potsys/} \\
 \D{[\mu} \Ap{}{\nu]} = {*\D{[\mu} \A{}{\nu]}}  
 \endAR
 &
 \P{}{\mu\nu} \mbox{ on } \Rsp{q}(A,A') &
 \curr{\mu} =
 \Int_0^1 \eval{\l( \P{\mu\nu}{} \Ap{}{\nu}
 - \duP{\mu\nu}{} \A{}{\nu} \r)}{\lambda A,\lambda A'}
 \frac{d\lambda}{\lambda}
 \\ \hline
 \endAR$
\caption{Conserved current formulas for \Meq/ and \potsys/s}
\label{table:adjrel-conslaw}
 \end{center}
 \end{table}
 \mystretch{1}

General results in \Ref{Anc:2003JPhysA} 
establish a key property of these \conscurr/ formulas. 
A direct proof in the case of \Meq/ 
was given in \Ref{AncPoh:2001}.

\begin{prop}
\label{prop:adjsym-conscur}
For an \adjsym/ that agrees with a multiplier on the solution jet space,
the conserved current scaling formula generates 
an equivalent nontrivial current;
otherwise for an \adjsym/ differing from any multiplier
on the solution jet space,
it yields a trivial \conscurr/.
\end{prop}

In the case of the magnetic \potsys/,
which is self-adjoint and thus arises from a Lagrangian,
\adjsyms/ are the same as \syms/, 
and multipliers correspond to those \syms/ under which 
the Lagrangian of the system is invariant (to within a divergence).
The scaling formula in the Lagrangian case
produces a current equivalent to the one
that comes from Noether's theorem applied to such \syms/.

\section{Cohomology, local \syms/ and local \conslaw/s}
\label{sec:cohom}

In this section we present a unified account
of interrelationships (i.e. mappings) among 
local \syms/, local \adjsyms/, and local \conslaw/s of \Meq/, 
and the magnetic and joint \potsys/s on \Minksp/. 
The cohomology of differential 1-forms and 2-forms on
the solution jet space of these systems together with a
locality-projection theorem for local \syms/ will be the main tools
in the derivation of these results.

We will consistently use $P$ to denote a \horiz/ 2-form,
$Q$ a \horiz/ 1-form, $\chi$ a scalar function (0-form),
e.g. on $J^q(F)$, 
 \EQ
 P = P\down{\mu\nu}[F] \extder\x{\mu}{} \wedge \extder\x{\nu}{}, \qquad
 Q = Q\down{\mu}[F] \extder\x{\mu}{}, \qquad
 \chi = \chi[F],
 \endEQ
where dependence on jet space coordinates to some finite order 
is denoted by $[F]$. 
A similar notation will be used for differential forms on $J^q(A)$ and
$J^q(A,A')$ and when these variables carry primes, tildes, etc. 
The total differential will be denoted by $\totder$, e.g.
 \EQ
 \totder{P} = \D{[\sigma} P\down{\mu\nu]}[F] 
 \extder\x{\sigma}{} \wedge \extder\x{\mu}{} \wedge \extder\x{\nu}{}, \qquad
 \totder{Q} = \D{[\mu} Q\down{\nu]}[F] 
 \extder\x{\mu}{} \wedge \extder\x{\nu}{}, \qquad
 \totder{\chi} = \D{\mu} \chi[F] 
 \extder\x{\mu}{}.
 \endEQ
In differential form notation, 
the system equations and determining equations for \syms/ and \adjsyms/ 
are summarized in \tableref{table:sym-adjsym-diff-form}.

 \mystretch{\STR}
 \begin{table}[h]
 \begin{center}
 $\AR{|c|c|c|} \hline
 \mbox{System} & \mbox{\Sym/ Equations} &
    \mbox{\Adjsym/ Equations} \\ \hline\hline
 \AR{c} \totder F=0 \\ \totder{*F}=0  \endAR & 
 \AR{c} \totder P[F]=0 \\ \totder{*P}[F]=0 \endAR
 & \totder Q'[F] = *\totder Q[F] \\ \hline
 \totder{*\totder A} = 0 & \totder{*\totder Q}[A] = 0 &
 \totder{*\totder Q'}[A] = 0 \\ \hline
 \totder A' = {*\totder A} & \totder Q'[A,A'] = *\totder Q[A,A'] &
 \AR{c}  \totder P[A,A']=0 \\ \totder{*P}[A,A']=0 \endAR \\ \hline
 \endAR$
 \caption{\Sym/ and \adjsym/ \eq/s for \Meq/ and \potsys/s}
 \label{table:sym-adjsym-diff-form}
 \end{center}
 \end{table}
 \mystretch{1}

Associated with the magnetic and joint \potsys/s, 
there is a natural embedding of the respective jet spaces 
$\Jsp{q+1}(A)$ and $\Jsp{q+1}(A,A')$
into the jet space $\Jsp{q}(F)$ 
under the total differential mapping given by 
$F=\totder{A}=-{*\totder{A'}}$
and its obvious prolongation $(0 \leq q \leq \infty)$. 
This embedding extends to the prolonged solution jet spaces
$\Rsp{\infty} \subset \Jsp{\infty}$ 
and is linear and one-to-one 
modulo the gauge freedom on the potentials,
namely $F=0$ if and only if $A=\totder\chi$, $A'=\totder\chi'$,
where $\chi,\chi'$ are any scalar functions on $\Jsp{q}$.

We note the following consequences of gauge freedom
in the \potsys/s here.
The (adjoint-) \sym/ determining equations of the standard \potsys/ 
have solutions of the form $Q[A] = \totder\chi[A]$
(i.e. gauge symmetries on $A$) 
for an {\it arbitrary} scalar function $\chi$ on $\Rsp{q}(A)$.
Such solutions can be also viewed as representing
a gauge freedom in the form of $Q[A]$.
Similarly, the \sym/ determining equations of the joint \potsys/ 
have solutions 
$Q[A,A']=\totder\chi[A,A']$, $Q'[A,A']=\totder\chi'[A,A']$
(i.e. gauge \syms/ on $A,A'$)
for a pair of {\em arbitrary} scalar functions
$\chi,\chi'$ on $\Rsp{q}(A,A')$; 
the \adjsym/ determining equations of this system 
can be shown to have no gauge freedom in the form of $P[A,A']$
(basically, the latter equations are locally well-posed 
as a PDE system for $P$). 
By comparison with the corresponding determining equations
for \syms/ and \adjsyms/ of \Meq/,
it follows that $Q[F]=\totder\chi[F]$, $Q'[F]=\totder\chi'[F]$
represent gauge freedom in the form of \adjsyms/,
for pairs of {\it arbitrary} scalar functions $\chi,\chi'$ 
on $\Rsp{q}(F)$, 
while there is no gauge freedom in the form
of the \syms/ (since \Meq/ are a locally well-posed PDE system).

The total differential $\totder$ obviously satisfies $\totder^2 = 0$ 
and hence defines a complex of \horiz/ forms on $\Jsp{\infty}$,
which is a generalization of the de Rham complex on \Minksp/
to the jet space setting \cite{And:1992}.
What we will refer to as \horiz/ forms on $\Rsp{q}$ can be defined
more precisely as the pullback to $\Rsp{\infty}$ (via the inclusion map)
of \horiz/ forms on $J^\infty$ 
whose coefficients depend on the jet coordinates up to a finite order 
$q+1$ in the case of \Meq/ and the joint \potsys/; 
$q+2$ in the case of the magnetic \potsys/.
For notational convenience, we will also denote the induced
total differential mapping on $\Rsp{\infty}$ by $\totder$ and
hence obtain the $\totder$-complex on $\Rsp{\infty}$.
Results concerning the local cohomology of this complex, 
i.e. $\totder$-closed modulo $\totder$-exact \horiz/ forms,
will underlie our study of the local \syms/
and \adjsyms/ of \Meq/ and its \potsys/s.
It is worth remarking that, in contrast, 
off $\Rsp{\infty}$ 
the local cohomology of the (free) $\totder$-complex is trivial,
since (see \cite{And:1992})
all $\totder$-closed \horiz/ $p$-forms on $\Jsp{\infty}$ 
for $1\leq p<4$ are $\totder$-exact. 

\subsection{Cohomology and locality projection}

We now state the cohomology and locality-projection theorems
and then give their proofs afterwards.

\begin{thm} 
{\bf (Local Cohomology)}

1-form cohomology: Let $Q$ be a \horiz/ 1-form on $\Rsp{q}$,
$0 \leq q < \infty$.
If $Q$ is closed, \ie/ $\totder Q=0$, then
\EQs
{\rm (i)} 
& \totder F=\totder {*F}=0 & \Lra \qquad
Q[F] = \totder\chi[F] ;
\label{ME-1form}\\
{\rm (ii)} 
& \totder{*\totder A}=0 & \Lra \qquad
Q[A] = \totder\chi[A] ; 
\label{MPS-1form}\\
{\rm (iii)} 
& \totder A' = *\totder A & \Lra \qquad
Q[A,A'] = \totder\chi[A,A'] . 
\label{JPS-1form}
\endEQs
Thus, the \loccohom{1} is trivial for each of these systems.

2-form cohomology: Let $P$ be a \horiz/ 2-form on $\Rsp{q}$,
$0 \leq q < \infty$.
If $P$ is closed, \ie/ $\totder P=0$, then
\EQs
{\rm (i)} 
& \totder F = \totder{*F}=0 & \Lra \qquad
P[F] = \c{1} F + \c{2} {*F} + \totder Q[F] ;
\label{ME-2form}\\
{\rm (ii)} 
& \totder{*\totder A}=0 & \Lra \qquad
P[A] = c\, {*F} + \totder Q[A] ,\quad 
F= \totder{A} ;
\label{MPS-2form}\\
{\rm (iii)}
& \totder A' = *\totder A & \Lra \qquad
P[A,A'] = \totder Q[A,A'] ;
\label{JPS-2form}
\endEQs
for some constants $c,c_1,c_2$. 
Thus, $F$ and $\duF{}{}$ represent the only nontrivial \loccohom{2}, 
and this cohomology is killed by 
the introduction of a corresponding potential.
\end{thm}

We remark that local 1-form and 2-form cohomology has
a field-theoretic interpretation in terms of conserved charges
\cite{Tor:lecturenotes}, 
describing electromagnetic fluxes through closed loops and surfaces 
in spacetime. 
(Namely, the integral of $*F(x),F(x)$ over any closed surface 
yields the total magnetic and electric charge 
enclosed within the surface;
in Minkowski space, these charges vanish 
for all smooth solutions $F(x)$ of \Meq/.)

\begin{thm} 
{\bf (Locality Projection)}

Let $Q$ be a \horiz/ 1-form on $\Rsp{q}$, $0 \leq q < \infty$.
If the 2-form $*\totder Q$ is closed, \ie/ $\totder{*\totder Q}=0$, 
then
\EQs
{\rm (i)} 
& \totder{*\totder A}=0 & \Lra \qquad
\totder Q[A] = P[F] ,\eqtext{ where $F=\totder A$} ;
\\
{\rm (ii)}
& \totder A'={*\totder A} & \Lra \qquad
\totder Q[A,A']=P[F] ,\eqtext{ where $F = \totder A = -{*\totder A'}$} .
\endEQs
Thus, any essential dependence on potentials in $Q$ is killed
under total exterior differentiation.
\end{thm}

A proof of the \loccohom{2} theorem for \Meq/
in the case where $P$ is a linear function on $\Rsp{q}(F)$ 
is given in \Ref{The:MSc} 
using tensorial methods.
This proof amounts to showing that 
the vanishing of the cohomology equation \eqref{ME-2form} 
has no nontrivial linear solutions.
The nonlinear case can be reduced to the linear case 
by standard linearization techniques 
and the whole computation is especially tractable in spinor form 
with the methods used in \Ref{AncPoh:2001}.
The proof for the \potsys/s 
and the \loccohom{1} theorem 
can be done by the same techniques.

The locality-projection theorem is a consequence of applying
a general result proved by Anco \& Bluman \cite{AncBlu:1997JMP}:
for a locally well-posed PDE system 
(i.e. if it is locally solvable \cite{Olv:symmbook}
such that no solutions depend on an arbitrary function of 
all independent variables),
the local \syms/ of any \potsys/ with gauge freedom
project onto only local \syms/ under the embedding
of the solution space of the \potsys/ into the solution space
of the given PDE system.
In particular, no projected \syms/ have any essential dependence
on the potentials.
If a \horiz/ 1-form $Q$ on $\Rsp{q}(A)$ 
satisfying $\totder{*\totder Q} =0$ is viewed as 
a local \sym/ of the magnetic \potsys/, 
then since \Meq/ is a locally well-posed system
we immediately conclude that the projected \sym/ $P=\totder Q$ for \Meq/ 
must be a \horiz/ 2-form on $\Rsp{q-1}(F)$. 
A similar argument applies to a \horiz/ 1-form $Q$ on $\Rsp{q}(A,A')$.
Because $\totder{*\totder Q} =0$ implies that 
the \horiz/ 2-form $P'=*\totder Q$ is closed, 
the equation $*\totder Q=\totder Q'$ 
holds for some 1-form $Q'$ on $\Rsp{q}(A,A')$,
by the cohomology equation \eqref{JPS-2form}. 
If we then view the pair $(Q,Q')$ as a local \sym/ of the joint \potsys/, 
we again conclude 
$P=\totder Q$ must be a 2-form on $\Rsp{q-1}(F)$. 

It is crucial that both the cohomology and locality-projection theorems
are formulated on the solution jet spaces $\Rsp{q}$ 
of finite order $q<\infty$.
Indeed, on the infinite-order solution jet space $\Rsp{\infty}$,
these theorems break down in the following manner.

\begin{prop}
On $\Rsp{\infty}(F)$, the \loccohom{2} becomes formally trivial:
\EQ
F = \totder A_F \qquad \mbox{and} \qquad \duF{}{} = \totder A'_F
\endEQ
where
\EQs
A_F &=& 
\sum_{k\geq 0} \frac{(-1)^k}{(k+1)!} x\hook (\dilop^k F) , 
\label{A-F}\\
A'_F &=& 
\sum_{k\geq 0} \frac{(-1)^k}{(k+1)!} x\hook (\dilop^k {*F}) , 
\label{A'-F}
\endEQs
and $\dilop=\x{\sigma}{}\D{\sigma}$ 
denotes the dilation operator.
Furthermore, these 1-forms satisfy Cronstrom's gauge
\EQ
x \hook A_F = x \hook A'_F = 0 . 
\endEQ
\end{prop}

The proof amounts to an explicit computation and will be omitted.
We note that the formal series \eqrefs{A-F}{A'-F} arise from 
integration by parts of the Poincar\'e homotopy formula 
for the de Rham cohomology of differential forms on \Minksp/ 
\cite{Olv:symmbook}.
Thus, the notion of nonlocality or nontrivial cohomology
associated with \Meq/ and its \potsys/s
is meaningful only when we work in finite-order jet spaces.

\subsection{Mappings and duality}

We now give the statements of our main results
which are consequences of the cohomology and
locality-projection theorems.
Throughout this section, it is understood that we work on
finite-order solution jet spaces $\Rsp{q}$, $0\leq q < \infty$.

\begin{thm}
\label{thm:PQdecomp}
The local \syms/ and local \adjsyms/ of \Meq/ and its \potsys/s
have the following decompositions:
\Items
\item[{\rm (i)}] $\totder F=\totder{*F}=0 \Lra$
\EQs
&&
P[F] = c\, F + c'\, {*F} + \totder\t{Q}[F] ,\quad
{*P}[F] = c\, {*F} -c'\, F + \totder\t{Q}'[F] , 
\label{ME:Pdecomp}\\
&&
Q[F] = \t{Q}[F] + \totder\chi[F] ,\quad
Q'[F] = \t{Q}'[F] + \totder\chi'[F] ; 
\label{ME:QQ'decomp}
\endEQs
\item[{\rm (ii)}] $\totder{*\totder A}=0 \Lra$
\EQs
&&
Q[A] = c\, A + \t{Q}[F] + \totder\chi[A] ,
\label{PS:Qdecomp}\\
&&
Q'[A] = c\, A + \t{Q}[F] + \totder\chi[A] ,
\label{PS:Q'decomp}
\endEQs
with $F=\totder A$; 
\item[{\rm (iii)}] $\totder A'=*\totder A \Lra$
\EQs
&&
Q[A,A'] = c\, A + c'\, A' + \t{Q}[F] + \totder\chi[A,A'] ,\quad
Q'[A,A'] = c\, A' -c'\, A + \t{Q}'[F] + \totder\chi'[A,A'] ,
\nonumber\\&&
\label{JPS:QQ'decomp}\\
&&
P[A,A']= c\, F + c'\, {*F} + \totder\t{Q}[F] ,\quad
{*P}[A,A']= c\, {*F} -c'\, F + \totder\t{Q}'[F] , 
\label{JPS:Pdecomp}
\endEQs
with $F=\totder A, {*F}=\totder A'$;
\endItems
for some constants $c,c'$,
where 
\EQ
\totder\t{Q}'[F] =*\totder\t{Q}[F]
\label{QQ'relation}
\endEQ
in all cases. 
These decompositions 
are unique up to addition of arbitrary gradients 
$\totder\chi[F]$ to $\t{Q}[F]$, $\totder\chi'[F]$ to $\t{Q}'[F]$, 
and are stable under the duality invariance 
\EQ
P \ra {*P} ,\quad (Q,Q') \ra (Q',-Q)
\label{PQduality}
\endEQ
of the \sym/ equations and \adjsym/ equations. 
\end{thm}

We mention that, moreover, 
the 1-forms $\t{Q}[F],\t{Q}'[F]$ here are 
canonically related under the duality transformation 
\eqref{ME-duality-F} on $F$, 
as will be discussed in Proposition~\ref{prop:dualitydecomp} later. 


In the \sym/ decompositions 
\eqref{ME:Pdecomp}, \eqref{PS:Qdecomp}, \eqref{JPS:QQ'decomp}, 
the cohomology terms correspond to
infinitesimal scaling and duality-rotation transformations
\EQs
\X_{\it scal} &=& 
F\Parder{F} ,\quad 
A\Parder{A} ,\quad 
A\Parder{A} + A'\Parder{A'} , 
\label{Xscaling}\\
\X_{\it dual} &=& 
*F\Parder{F} ,\quad 
A'\Parder{A} - A\Parder{A'} .
\label{Xduality}
\endEQs
Note the duality-rotation is not realized as a local \sym/
of the standard \potsys/.
The \sym/-decomposition terms that involve $\chi$ or $\chi'$ 
correspond to a gauge freedom in the \syms/, 
namely the infinitesimal transformations
\EQ
\X_{\it gauge} = 
\totder\chi\Parder{A} ,\quad 
\totder\chi\Parder{A} + \totder\chi'\Parder{A'}
\label{Xgauge}
\endEQ
for arbitrary functions $\chi,\chi'$ on $\Jsp{q}$ are \syms/. 
The remaining terms in the \sym/ decompositions 
represent gauge-invariant infinitesimal transformations
\EQ
\X = 
\totder\t{Q}[F] \Parder{F} ,\quad
\t{Q}[F] \Parder{A} ,\quad
\t{Q}[F] \Parder{A} + \t{Q}'[F] \Parder{A'} 
\endEQ
which are themselves \syms/.
Note these terms are well-defined only up to gauge freedom \eqref{Xgauge}
such that $\chi,\chi'$ are functions on $\Jsp{q}(F)$. 

These results establish that the vector spaces of \syms/ and \adjsyms/
of each system are a direct sum of 
cohomology subspaces 
spanned by the separate terms proportional to $c,c'$,
and a complementary subspace
identified with the gauge-invariant non-cohomology terms
involving $\t{Q},\t{Q}'$, 
up to gradient terms $\totder\chi,\totder\chi'$. 
Hereafter we let 
$X$ denote a vector space of \syms/,
$Y$ a vector space of \adjsyms/,
and we use superscripts $c,c',0$ to distinguish
the respective subspaces in the direct sum decomposition;
superscripts $\chi,\chi'$ will denote the
vector subspace defined by all $\totder\chi,\totder\chi'$ terms
while a tilde will stand for the quotient with respect to this subspace. 
Thus we have the following vector space decompositions:
\EQs
&{\rm (i)} &
X_F = X^c_F \oplus X^{c'}_F \oplus X^0_F ,\quad
Y_F = Y^0_F ,
\\
& {\rm (ii)} &
X_A = X^c_A \oplus X^0_A ,\quad
Y_A = Y^c_A \oplus Y^0_A ,
\\
& {\rm (iii)} &
X_{A,A'}
= X^c_{A,A'} \oplus X^{c'}_{A,A'} \oplus X^0_{A,A'} ,\quad 
Y_{A,A'} = Y^c_{A,A'} \oplus Y^{c'}_{A,A'} \oplus Y^0_{A,A'} , 
\endEQs 
where the $X^0,Y^0$ subspaces in the case of \horiz/ 1-forms 
naturally partition into equivalence classes 
modulo gradients
\EQs
& {\rm (iv)} &
\tilde Y^0_F = Y^0_F/ Y^{\chi,\chi'}_F ,\quad
\tilde X^0_A = X^0_A/X^{\chi}_A  ,\quad
\tilde Y^0_A = Y^0_A/Y^{\chi}_A  ,\quad
\tilde X^0_{A,A'} = X^0_{A,A'}/X^{\chi,\chi'}_{A,A'} .
\label{gaugeinvXY}
\endEQs
It will be convenient to introduce a linear map $*'$ 
on \horiz/ 1-forms $\t{Q}[F] \mod{\totder\chi[F]}$ 
by the equation 
\EQ
\totder{*'\t{Q}}[F] = *\totder\t{Q}[F] 
\quad\eqtext{ on $\Rsp{q}(F)$, }
\label{Qduality}
\endEQ
which defines an automorphism of each vector space \eqref{gaugeinvXY}.
Note the relation \eqref{QQ'relation} can be simply written
$\t{Q}' = {*'\t{Q}}$. 

Importantly, interrelationships hold between any two decompositions
\eqsref{ME:Pdecomp}{JPS:Pdecomp}, 
given by linear maps summarized in the following four theorems.

\begin{thm} 
{\bf (Self-correspondences related to the systems
$\totder{F}=\totder{*F}=0$, 
$\totder{A'}=*\totder{A}$, 
$\totder{*\totder{A}}=0$)}
\label{thm:self-mappings}

There is a linear mapping between:
\Items
\item[{\rm (i)}] 
local \syms/ on $\Rsp{q}(F)$ and  local \adjsyms/ on $\Rsp{q-1}(F)$, 
given by
\EQs
&&
P[F] \mod{ F,{*F} } \llra \totder(\t{Q}[F] \mod{ \totder\chi[F] } ) ,
\\
&&
*P[F] \mod{ F,{*F} } \llra \totder( \t{Q}'[F] \mod{ \totder\chi'[F] } ) , 
\endEQs
corresponding to the isomorphism of vector spaces
\EQ
X^0_F \cong \tilde Y^0_F . 
\endEQ
\item[{\rm (ii)}] 
local \syms/ on $\Rsp{q}(A,A')$ and local \adjsyms/ on $\Rsp{q+1}(A,A')$,
given by
\EQs
&&
\totder( Q[A,A'] \mod{\totder\chi[A,A']} ) \llra P[A,A'] ,
\label{JPS-QtoPmap}\\
&&
\totder( Q'[A,A'] \mod{\totder\chi'[A,A']} ) \llra *P[A,A'] , 
\label{JPS-Q'toPmap}
\endEQs
corresponding to the isomorphism of vector spaces
\EQ
X^c_{A,A'} \cong Y^c_{A,A'} ,\quad
X^{c'}_{A,A'} \cong Y^{c'}_{A,A'} ,\quad
\tilde X^0_{A,A'} \cong Y^0_{A,A'} . 
\endEQ
\item[{\rm (iii)}] 
local \syms/ on $\Rsp{q}(A)$ and local \adjsyms/ on $\Rsp{q}(A)$,
given by the direct identification 
\EQ
Q[A] \llra Q'[A] ,
\label{PS:QtoQ'id}
\endEQ
namely $X_A = Y_A$, 
as well as a dual identification 
\EQ
{*'}( Q[A] \mod{A,\totder\chi[A]} ) \llra Q'[A] \mod{A,\totder\chi[A]} ,
\label{PS:QtoQ'}
\endEQ
corresponding to a nontrivial duality (isomorphism) of vector spaces
$\tilde X^0_A \cong \tilde Y^0_A$. 
\endItems
\end{thm}

Note the composition of the maps \eqrefs{PS:QtoQ'id}{PS:QtoQ'} 
yields a linear mapping of local (adjoint-) \syms/ 
modulo $A,\totder\chi[A]$ on $\Rsp{q}(A)$ 
into themselves,
corresponding to the vector space automorphism 
$*': \tilde X^0_A \lra \tilde X^0_A$
(and correspondingly $\tilde Y^0_A \lra \tilde Y^0_A$). 

\begin{thm} 
{\bf (Correspondences related to the systems 
$\totder{*\totder A}=0$ and $\totder A'=*\totder A$)}
\label{thm:std-joint}

There is a linear mapping between:
\Items
\item[{\rm (i)}] 
local (adjoint-) \syms/ on $\Rsp{q}(A)$ and local \syms/ on $\Rsp{q}(A,A')$, 
given by
\EQs
Q[A] \mod{\totder\chi[A]} &\llra& Q[A,A'] \mod{A',\totder\chi[A,A']} , 
\\
*'Q[A] \mod{\totder\chi[A]} &\llra& Q'[A,A'] \mod{A, \totder\chi'[A,A']} , 
\endEQs
where $*'$ is extended by $*'A := A'$ and linearity 
so it is well-defined 
on all local (adjoint-) \syms/ $Q[A]$ modulo gradients $\totder\chi[A]$, 
corresponding to the isomorphism of vector spaces
\EQ
X^c_A \cong X^c_{A,A'} ,\quad
\tilde X^0_A \cong \tilde X^0_{A,A'} .
\endEQ
\item[{\rm (ii)}] 
local (adjoint-) \syms/ on $\Rsp{q}(A)$ 
and local \adjsyms/ on $\Rsp{q+1}(A,A')$, 
given by
\EQ
\totder (Q[A] \mod{\totder\chi[A]}) \llra P[A,A'] \mod *F , 
\endEQ
corresponding to the isomorphism of vector spaces
 \EQ
X^c_A \cong Y^c_{A,A'} ,\quad
\tilde X^0_A \cong Y^0_{A,A'} .
\endEQ
\endItems
\end{thm}

\begin{thm} 
{\bf (Correspondences related to the systems 
$\totder F= \totder{*F}=0$ and $\totder A'=*\totder A$)}
\label{thm:ME-joint}

There is a linear mapping between:
\Items
\item[{\rm (i)}] 
local \syms/ on $\Rsp{q}(F)$ and local \adjsyms/ on $\Rsp{q+1}(A,A')$, 
given by
\EQ
P[F] \llra P[A,A']
\endEQ
which is an isomorphism of vector spaces 
\EQ 
X^c_F \cong Y^c_{A,A'} ,\quad
X^{c'}_F \cong Y^{c'}_{A,A'} ,\quad
X^0_F \cong Y^0_{A,A'} . 
\endEQ
\item[{\rm (ii)}] 
local \adjsyms/ on $\Rsp{q}(F)$ and local \syms/ on $\Rsp{q+1}(A,A')$, 
given by
\EQs
&&
Q[F] \mod{\totder\chi[F]} \llra Q[A,A'] \mod{A,A',\totder\chi[A,A']} ,
\\
&&
Q'[F] \mod{\totder\chi'[F]} \llra Q'[A,A'] \mod{A,A',\totder\chi'[A,A']} , 
\endEQs
corresponding to the isomorphism of vector spaces
\EQ
\tilde Y^0_F \cong \tilde X^0_{A,A'} . 
\endEQ
\item[{\rm (iii)}] 
local \syms/ on $\Rsp{q+1}(A,A')$ and local \syms/ on $\Rsp{q}(F)$, 
given by
\EQs
&&
\totder( Q[A,A'] \mod{\totder\chi[A,A']} ) \llra P[F] ,
\\
&&
\totder( Q'[A,A'] \mod{\totder\chi'[A,A']} ) \llra *P[F] , 
\endEQs
corresponding to the isomorphism of vector spaces
\EQ
X^c_{A,A'} \cong X^c_F ,\quad
X^{c'}_{A,A'} \cong X^{c'}_F ,\quad
\tilde X^0_{A,A'} \cong X^0_F . 
\endEQ
\item[{\rm (iv)}] 
local \adjsyms/ on $\Rsp{q}(F)$ and local \adjsyms/ on $\Rsp{q+1}(A,A')$,
given by
\EQs
&&
\totder( Q[F] \mod{\totder\chi[F] }) \llra P[A,A'] \mod{F,*F} ,
\\
&&
\totder( Q'[F] \mod{\totder\chi'[F] }) \llra *P[A,A'] \mod{F,*F} , 
\endEQs
corresponding to the isomorphism of vector spaces
\EQ
\tilde Y^0_F \cong Y^0_{A,A'} . 
\endEQ
\endItems
\end{thm}

\begin{thm} 
{\bf (Correspondences related to the systems
$\totder F= \totder{*F}=0$ and ${\totder{*\totder A}}=0$)}
\label{thm:ME-std}

There is a linear mapping between:
\Items
\item[{\rm (i)}] 
local \syms/ on $\Rsp{q}(F)$ and local \syms/ on $\Rsp{q+1}(A)$, 
given by
\EQ
P[F] \mod{*F} \llra \totder( Q[A] \mod{\totder\chi[A]} ) , 
\endEQ
corresponding to the isomorphism of vector spaces
\EQ
X^c_F \cong X^c_A ,\quad
X^0_F \cong \tilde X^0_A . 
\endEQ
\item[{\rm (ii)}] 
local \adjsyms/ of $\Rsp{q}(F)$ and local \syms/ of $\Rsp{q+1}(A)$,
given by
\EQs
Q[F] \mod{\totder\chi[F]} &\llra & Q[A] \mod{A,\totder\chi[A]},
\\
Q'[F] \mod{\totder\chi'[F]} &\llra & *'Q[A] \mod{A,\totder\chi[A]},
\endEQs
corresponding to the isomorphism of vector spaces
\EQ
\tilde Y^0_F \cong \tilde X^0_A . 
\endEQ
\endItems
\end{thm}

\Proof{}
We give the proof of Theorem~\ref{thm:self-mappings} in detail.
The proofs of Theorems \ref{thm:std-joint}--\ref{thm:ME-std}
are readily derived in a similar manner from 
the decompositions \eqsref{ME:Pdecomp}{JPS:Pdecomp}
combined with the cohomology equations \eqsref{ME-1form}{JPS-2form}
and Theorem~\ref{thm:self-mappings}.

For part (i) of Theorem~\ref{thm:self-mappings}, 
by the \sym/ decomposition \eqref{ME:Pdecomp} with $c=c'=0$
(i.e. quotienting out the scaling and duality-rotation terms),
the pair of \horiz/ 1-forms $(\t{Q},\t{Q}')$ satisfy
the \adjsym/ equation on $\Rsp{q}(F)$.
Conversely, in the \adjsym/ decomposition \eqref{ME:QQ'decomp}
we see that the \horiz/ 2-forms $\totder\t{Q}$, $\totder\t{Q}'$, 
and their duals are closed due to the \adjsym/ equation
and hence they directly satisfy 
the \sym/ equations on $\Rsp{q}(F)$.

Part (ii) follows analogously from the 
\sym/ and \adjsym/ decompositions \eqrefs{JPS:QQ'decomp}{JPS:Pdecomp},
including the cases $c\neq 0,c'\neq 0$ via the obvious mappings
$F \llra \totder{A}$, ${*F} \llra \totder{A'}$.

Finally for part (iii) of the theorem, 
we apply the 2-form cohomology equation \eqref{MPS-2form}
to the \sym/ equation 
$\totder{*\totder{Q}}=0$ for the \horiz/ 1-form $Q$ on $\Rsp{q}(A)$,
which shows that the 2-form $*\totder(Q+c A)$ is exact
for some constant $c$. 
Hence the equation $*\totder(Q+c A) = \totder{Q'}$ holds 
for some \horiz/ 1-form $Q'$ on $\Rsp{q}(A)$,
and we see that the dual 2-form $*\totder{Q'}$ is exact
and therefore is closed. 
Thus, $Q'$ satisfies the (adjoint-) \sym/ equation $\totder{*\totder{Q'}}=0$
on $\Rsp{q}(A)$. 
\endProof

We now prove the \sym/ and \adjsym/ decompositions 
\eqsref{ME:Pdecomp}{JPS:Pdecomp}. 

\begin{prop} 
\label{prop:MEsym} 
Every local \sym/ $P=P[F]$ 
of order $q$ of $\totder{F}=\totder{*F}=0$
has the form \eqref{ME:Pdecomp}
for some constants $c,c'$,
where $(\t{Q}[F],\t{Q}'[F])$ is a local \adjsym/ 
of order $q-1$ of $\totder{F}=\totder{*F}=0$, 
and conversely. 
\end{prop}

\Proof{} 
Since $P[F]$ and ${*P}[F]$ are closed \horiz/ 2-forms on $\Rsp{q}(F)$ 
for finite $q$
then by the 2-form cohomology equation \eqref{ME-2form} we have 
\EQs
P[F] &=& \c{1} F + \c{2} {*F} + \totder\t{Q}[F], 
\label{ME-Pform}\\
{*P}[F] &=& \c{3} F + \c{4} {*F} + \totder\t{Q}'[F], 
\label{ME-*Pform}
\endEQs
for some constants $\c{1},\c{2},\c{3},\c{4}$ 
and some \horiz/ 1-forms $\t{Q}[F],\t{Q}'[F]$. 
Applying $*$ to \eqref{ME-Pform} and equating it to \eqref{ME-*Pform},
we obtain
 \EQ
 (\c{1} - \c{4}) {*F} - (\c{2} + \c{3}) F =
 \totder\t{Q}'[F] - *\totder\t{Q}[F].
 \label{Fcohomeq}
 \endEQ
The differential order of the right side is at least one
while the left side is of differential order zero,
hence on $\Rsp{q}(F)$ a descent argument 
(similar to the ones used in the classification results 
in \Ref{AncPoh:2001,AncPoh:2002}) 
shows that both sides of \eqref{Fcohomeq} must vanish
and so $\totder\t{Q}'[F] = *\totder\t{Q}[F]$.
Since $F$, ${*F}$ are \loccohom{2} elements
(which are linearly independent), 
then we must have $\c{1}=\c{4}\equiv c$, $\c{2}=-\c{3}\equiv c'$,
which establishes \eqref{ME:Pdecomp}. 
The converse is immediate.
\endProof

\begin{prop} 
\label{prop:JPSsym}
Every local \sym/ $Q=Q[A,A']$, $Q'=Q'[A,A']$
of order $q$ of $\totder{A'}={*\totder{A}}$ 
has the form \eqref{JPS:QQ'decomp}
for some constants $c,c'$, 
and some scalar functions $\chi[A,A'],\chi'[A,A']$, 
with $F=\totder A = -{*\totder A'}$, 
where $(\t{Q}[F], \t{Q}'[F])$ is a local \adjsym/ 
of order $q-1$ of $\totder{F} = \totder{*F}=0$, 
and conversely. 
\end{prop}

\Proof{} 
By the locality-projection theorem, both $\totder{Q}=P[F]$,
$\totder{Q'}={*P}[F]$ are local \syms/ of $\totder{F}=\totder{*F}=0$.
Using Proposition~\ref{prop:MEsym}
and writing $F=\totder{A},{*F}=\totder{A'}$, 
we have
\EQ
\totder( Q[A,A'] - c A - c' A' - \t{Q}[F] ) = 0 ,\quad
\totder( Q'[A,A'] - c A' + c' A - \t{Q}'[F] ) = 0 , 
\endEQ
where $(\t{Q}[F]$, $\t{Q}'[F])$ is a local \adjsym/ 
of $\totder{F}=\totder{*F}=0$. 
Since the \loccohom{1} on $\Rsp{q}(A,A')$ for finite $q$ is trivial,
we obtain \eqref{JPS:QQ'decomp}. 
The converse is immediate.
\endProof

\begin{prop} 
\label{cor:JPS-adjsym}
Every local \adjsym/ $P=P[A,A']$ 
of order $q$ of $\totder{A'}={*\totder{A}}$ 
has the form \eqref{JPS:Pdecomp}
for some constants $c,c'$, 
with $F=\totder{A} = -{*\totder{A'}}$,
where $(\t{Q}[F],\t{Q}'[F])$ is a local \adjsym/ 
of order $q-2$ of $\totder F = \totder{*F}=0$, 
and conversely. 
\end{prop}

\Proof{}
Since the \loccohom{2} on $\Rsp{q}(A,A')$ for finite $q$ is trivial, 
we have $P=\totder Q$, ${*P}=\totder Q'$
for some \horiz/ 1-forms $Q=Q[A,A']$, $Q'=Q'[A,A']$. 
Hence the pair $(Q,Q')$ is a local \sym/ of $\totder{A'}=*\totder{A}$. 
By Proposition~\ref{prop:JPSsym}, 
$\totder$ applied to \eqref{JPS:QQ'decomp} yields \eqref{JPS:Pdecomp}. 
The converse is immediate.
\endProof

\begin{prop} 
Every local (adjoint-) \sym/ $Q=Q[A]$ 
of order $q$ of $\totder{*\totder{A}}=0$
has the form \eqref{PS:Qdecomp}
for some constant $c$ and some scalar function $\chi[A]$, 
with $F = \totder A$, 
where $(\t{Q}[F],\t{Q}'[F])$ is a local \adjsym/ 
of order $q-1$ of $\totder F = \totder{*F}=0$, 
and conversely. 
\end{prop}

\Proof{}
Regard $Q[A]$ as a \horiz/ 1-form on $\Jsp{q}(A,A') \supset \Jsp{q}(A)$
with no dependence on the coordinates involving $A'$. 
Since $\totder{A'}=*\totder{A}$ implies $\totder{*\totder A}=0$, 
the 2-form $*\totder Q[A]$ is closed 
on $\Rsp{q}(A,A')$ whose \loccohom{2} for finite $q$ is trivial. 
Hence $*\totder Q[A] = \totder Q'[A,A']$ 
holds for some \horiz/ 1-form $Q'[A,A']$
and thus the pair $(Q,Q')$ is a local \sym/ of $\totder{A'}={*\totder{A}}$. 
Then by Proposition~\ref{prop:JPSsym}, we have 
\EQ
Q[A] = cA + c' A' + \t{Q}[F] + \totder\chi[A,A'] . 
\endEQ
But since $Q[A]$ is independent of $A'$
we must have $c' = 0$ and $\chi = \chi[A]$. 
Consequently, we obtain \eqref{PS:Qdecomp}. 
The converse is immediate.
\endProof

Finally, we point out the effect of the \dutr/
\EQ
F \ra {*F} ,\quad
(A,A') \ra (A',-A)
\label{FAA'duality}
\endEQ
on \syms/ and \adjsyms/ of \Meq/ and its \potsys/s,
which stems from
the explicit classification of \adjsyms/ 
$(Q[F],Q'[F])$ on $\Rsp{\infty}(F)$ 
given in \Ref{AncPoh:2001}.
This classification shows that 
the \horiz/ 1-forms 
$Q[F],Q'[F]$, modulo gradients $\totder\chi[F],\totder\chi'[F]$,
are {\it linear} functions of the jet space coordinates
and hence have well-defined parity (eigenvalue) decompositions 
with respect to the combined \dutr/s \eqrefs{FAA'duality}{PQduality}. 
Consider more specifically the transformation \eqref{FAA'duality}
composed with the inverse of the transformation \eqref{PQduality}
\EQ
(\t{Q}[F],\t{Q}'[F]) \ra (-\t{Q}'[*F],\t{Q}[*F])
 \label{combined-duality}
\endEQ
acting on the vector space $\tilde Y^0_F$. 
(Here the notation $[*F]$ denotes
the \dutr/ \eqref{FAA'duality} 
on the jet space coordinates $[F]$ in $\Jsp{q}(F)$; 
likewise $[A',-A]$ will denote 
the duality transformation on the coordinates $[A,A']$ in $\Jsp{q}(A,A')$.
Note that the \dutr/ is well-defined on the
corresponding solution jet spaces $\Rsp{q-1}(F)$ and $\Rsp{q-1}(A,A')$.
However, it does not exist on either $\Jsp{q}(A)$ or $\Rsp{q-2}(A)$.)
Due to the linearity of $\t{Q}[F]$ and $\t{Q}'[F]$, 
the square of the transformation \eqref{combined-duality} is the identity
and hence its eigenvalues are $\pm 1$. 
The eigenspaces of even/odd parity corresponding to these eigenvalues
are then given by 
\EQ
\t{Q}_{\pm}[*F] = \pm\t{Q}'_{\pm}[F] ,\quad
\t{Q}'_{\pm}[*F] = \mp\t{Q}_{\pm}[F] \quad\eqtext{ on $\Rsp{q}(F)$. }
\label{QQ'dualityrelation}
\endEQ
Equivalently, we can write this canonical relation as
\EQ
{*'\t{Q}}_{\pm}[F] = \pm \t{Q}_{\pm}[*F]
\endEQ
and likewise for $\t{Q}'_{\pm}[F]$,
where $*'$ is the linear map \eqref{Qduality}. 
This parity decomposition result \eqref{QQ'dualityrelation} 
extends to all solutions of
the \sym/ equations and \adjsym/ equations 
on $\Rsp{q}(F)$ and $\Rsp{q}(A,A')$
through the (cohomology and locality-projection) 
decompositions in Theorem~\ref{thm:PQdecomp}. 

\begin{prop}
\label{prop:dualitydecomp}
The \dutr/ \eqref{FAA'duality}
on the electromagnetic field and its joint potentials 
induces a corresponding \dutr/ \eqref{PQduality} 
on local \syms/ and local \adjsyms/ via
\EQs
&&
{*P_{\pm}}[F] = \pm {P_{\pm}[*F]} ,\quad
{*P_{\pm}}[A,A'] = \pm {P_{\pm}[A',-A]} , 
\\
&&
Q'_{\pm}[F] \mod{\totder\chi'[F]} = \pm Q_{\pm}[*F] \mod{\totder\chi[F]} ,
\\
&&
Q'_{\pm}[A,A'] \mod{\totder\chi'[A,A']} = 
\pm Q_{\pm}[A',-A] \mod{\totder\chi[A,A']} ,
\endEQs
where $\pm$ denotes even/odd parity parts 
with respect to the combined duality transformations
\EQs
&&
P[F] \ra -{*P[*F]} 
\quad\eqtext{ on $X^c_F \oplus X^{c'}_F \oplus X^0_F$, }
\label{combined-dutr1}\\
&&
P[A,A'] \ra -{*P}[A',-A]
\quad\eqtext{ on $Y^c_{A,A'} \oplus Y^{c'}_{A,A'} \oplus Y^0_{A,A'}$, }
\\
&&
(Q[F],Q'[F]) \ra (-Q'[*F],Q[*F]) 
\quad\eqtext{ on $\tilde Y^0_F$, }
\label{combined-dutr3}\\
&&
(Q[A,A'],Q'[A,A']) \ra (-Q'[A',-A],Q[A',-A]) 
\quad\eqtext{ on $X^c_{A,A'} \oplus X^{c'}_{A,A'} \oplus \tilde X^0_{A,A'}$. }
\label{combined-dutr4}
\endEQs
\end{prop}

In addition we remark that a complete and explicit classification of
all local \syms/ and \adjsyms/ of the magnetic and joint \potsys/s 
is now available by combining the decompositions in Theorem~\ref{thm:PQdecomp}
with the classification of local \adjsyms/ of \Meq/ from \Ref{AncPoh:2001}.
An illustration of this result applied to 
a geometric class of \syms/ and \adjsyms/
will be given later.

\subsection{Conservation law formulas}

There is an important application of the preceding results
to local \conslaw/s of \Meq/ and its \potsys/s.
We begin with a few preliminaries. 
In differential form notation 
a conserved current of order $q<\infty$ 
given by a vector function $\curr{\mu}$ 
corresponds to a \horiz/ 3-form 
$*\curr{}\down{\mu\nu\sigma} = \vol{\mu\nu\sigma\tau}{} \curr{\tau}$
on $\Jsp{q}$
that is closed on $\Rsp{\infty}$. 
A current is trivial if and only if the corresponding \horiz/ 3-form is exact
(\ie/ $*\curr{}\down{\mu\nu\sigma}
= \D{[\mu} {*\curl{}{\nu\sigma]}}
= \frac{1}{3} \vol{\mu\nu\sigma\tau}{} \D{\alpha}\curl{\alpha\tau}{}$)
on $\Rsp{\infty}$. 
Thus, the \loccohom{3} of \Meq/ and its potential systems
describes the nontrivial \conslaw/s of these systems. 
(Note this cohomology is clearly quite rich, 
in contrast to the local 1-form and 2-form cohomology.
See \Ref{Tor:lecturenotes} for a general discussion of \conslaw/s 
of field equations from this perspective.)
The formulas in \tableref{table:adjrel-conslaw}
that generate all conserved currents (modulo curls)
in terms of \adjsyms/ are summarized 
in \tableref{table:conscur-diff-form} 
in this notation. 

 \mystretch{\STR}
 \begin{table}[h]
 \begin{center}
 $\AR{|c|c|c|c|} \hline
 \mbox{System} & \mbox{Adjoint-symmetry} & \mbox{Conserved 3-form current} 
 \\ \hline \hline
 \AR{c}
 \mbox{\Meq/}\\
 \totder F=0 \\ \totder{*F}=0
 \endAR  & \totder{Q'}[F] =*\totder{Q}[F] &
 *\curr{} = 
 \Int_0^1 \eval{\l( Q' \wedge F - Q \wedge *F \r)}{\lambda F}
 \frac{d\lambda}{\lambda} 
 \\ \hline
 \AR{c}
 \mbox{Magnetic \potsys/}\\
 \totder{*\totder A}=0
 \endAR  & \totder{*\totder{Q}}[A] = 0 & 
 *\curr{} = 
 \Int_0^1 \eval{\l( A \wedge *\totder Q - Q \wedge *\totder A \r)}{\lambda A}
 \frac{d\lambda}{\lambda} 
 \\ \hline
 \AR{c}
 \mbox{Joint \potsys/}\\
 \totder A'=*\totder A
 \endAR & \AR{c} \totder{P}[A,A'] =0 \\ \totder{*P}[A,A']=0 \endAR &
 *\curr{} = 
 \Int_0^1 \eval{\l( -A \wedge P - A' \wedge {*P} \r)} {\lambda A,\lambda A'}
 \frac{d\lambda}{\lambda} 
 \\ \hline
 \endAR$
 \caption{Conserved current formulas in differential form notation}
\label{table:conscur-diff-form}
 \end{center}
 \end{table}
 \mystretch{1}

Now, through the \adjsym/ decompositions in Theorem~\ref{thm:PQdecomp}, 
the conserved current formulas for the \potsys/s can be simplified
to remove inessential dependence on the potentials.

\begin{prop}
For the \potsys/ $\totder{*\totder A}=0$,
all nontrivial local conserved 3-form currents of order $q$ 
are generated by
\EQ
 *\curr{} = 
\int_0^1 \eval{\l( \t{Q}'\wedge F - \t{Q} \wedge *F \r)}{\lambda F}
\frac{d\lambda}{\lambda} 
\mod{ \totder{*\curl{}{}} }
\label{MPS-current}
\endEQ
for $q\geq 1$, 
where $\t{Q}[F]$ is the gauge-invariant part of the (adjoint-) \sym/
$Q[A]$ in the decomposition \eqref{PS:Qdecomp} 
and $\t{Q}'[F] = {*'\t{Q}}[F]$ is its image under the map \eqref{Qduality}, 
on $\Rsp{q-1}(F)$. 
\end{prop}

\Proof{}
We substitute the decomposition \eqref{PS:Qdecomp} 
into the integrand of $*\curr{}$ in \tableref{table:conscur-diff-form}
to obtain
\EQ
A\wedge *\totder Q - Q\wedge *\totder A =
A\wedge *\totder\t{Q} - \t{Q}\wedge *F
+ \chi \totder\duF{}{} - \totder(\chi {*F}),
\label{stdPS-integrand}
\endEQ
where the second last term vanishes on $\Rsp{}(A)$,
and the last term is an exact 3-form which we may drop.
Next we apply the relation 
$\totder{*'\t{Q}}=*\totder\t{Q}$ to the first term on the right 
in \eqref{stdPS-integrand},
where, note, $*'\t{Q}$ is determined only to within a gradient $\totder\chi'$.
This yields
\EQ
A \wedge *\totder\t{Q} = A \wedge \totder{*'}\t{Q}
 = F \wedge {*'}\t{Q}- \chi'\totder{F} + \totder(\chi' F -A \wedge {*'}\t{Q}),
\endEQ
where we again drop the last two terms as before. 
Hence, the integrand of $*\curr{}$ modulo an exact 3-form on $\Rsp{}(A)$
is simply ${*'\t{Q}} \wedge F - \t{Q} \wedge *F$.
\endProof

\begin{prop}
\label{prop:JPScurrentformula}
For the joint \potsys/ $\totder A' = *\totder A$, 
all nontrivial local conserved 3-form currents of order $q$ 
are generated by linear combinations of
\EQ
 *\curr{} = 
\int_0^1 \eval{\l( -\t{Q} \wedge F  - \t{Q}' \wedge *F \r)}{\lambda F}
\frac{d\lambda}{\lambda} 
\mod{ \totder{*\curl{}{}} }
\quad \eqtext{ on $\Rsp{q-1}(A,A')$}
\label{JPS-current}
\endEQ
for $q\geq 1$, 
and when $q=1$,
\EQ
 *\curr{} = -\frac{1}{2}( A \wedge F + A' \wedge \duF{}{} )
\quad \eqtext{ on $\Rsp{}(A,A')$, } 
\label{JPS-dualityrot-current}
\endEQ
where $\t{Q}[F], \t{Q}'[F]$ are determined 
by the gauge-invariant parts of the \adjsym/ $P[A,A']$
and its dual ${*P}[A,A']$ in the decomposition \eqref{JPS:Pdecomp},
satisfying the relation $\t{Q}' = {*'\t{Q}}$.
\end{prop}

\Proof{}
As in the previous proof, we first substitute the decompositions 
in \eqref{JPS:Pdecomp}
into the integrand of $*\curr{}$
in \tableref{table:conscur-diff-form}, 
obtaining 
\EQ
-A \wedge P - A' \wedge {*P} = c( -A\wedge F - A' \wedge *F)
+ c' (-A \wedge *F + A' \wedge F) -A \wedge \totder\t{Q}
- A' \wedge \totder\t{Q}' . 
\endEQ
Note that on $\Rsp{q-1}(A,A')$ we have
\EQs
 A' \wedge F- A \wedge *F &=& A' \wedge \totder A - A \wedge \totder A'
 = \totder( A \wedge A') ,
\label{simplify1}\\
 -A \wedge \totder\t{Q} - A' \wedge \totder\t{Q}' &=& -F \wedge \t{Q}
 - *F \wedge \t{Q}'
 +\totder( A \wedge \t{Q} + A' \wedge \t{Q}') .
\label{simplify2}
\endEQs
The last terms in both \eqrefs{simplify1}{simplify2} 
are exact 3-forms which we drop.
Hence, 
the integrand of $*\curr{}$ modulo an exact 3-form on $\Rsp{q-1}(A,A')$ is 
$c( -A\wedge F - A' \wedge *F) - \t{Q} \wedge F - \t{Q}' \wedge *F$.
\endProof

These 3-form formulas \eqref{MPS-current}, 
\eqref{JPS-current}, \eqref{JPS-dualityrot-current}
now lead to our main results pertaining to \conslaw/s.
First, we see that with the exception of 
one conserved current \eqref{JPS-dualityrot-current}
coming from the joint \potsys/,
all other local conserved currents of the two \potsys/s
are equivalent to local conserved currents of \Meq/.
In particular, by the \adjsym/ decompositions in Theorem~\ref{thm:PQdecomp}, 
the \horiz/ 1-forms $\t{Q}[F]$ and $\t{Q}'[F]$
in the currents \eqrefs{MPS-current}{JPS-current}
can be directly identified with \adjsyms/
$(Q[F],Q'[F])$ of \Meq/
by $Q=\t{Q}[F]$ and $Q'=\t{Q}'[F]$. 
There is also a dual identification given by
$Q = \t{Q}'[F]$, $Q' = -\t{Q}[F]$
which reflects the duality invariance \eqref{PQduality}
of the \adjsym/ equation $\totder Q'=*\totder Q$ on $\Rsp{q}(F)$.
Note that the gauge freedom in the form of the \adjsyms/ 
given by
\EQ
Q \ra Q + \totder\chi[F], \qquad 
Q' \ra Q' + \totder\chi'[F] , 
\endEQ
for arbitrary functions $\chi,\chi'$ on $\Jsp{q}(F)$
yields only trivial currents, 
since
\EQ
\totder\chi'[F] \wedge F - \totder\chi[F] \wedge {*F} 
= \totder (\chi'[F] F - \chi[F] {*F})
\quad\eqtext{ on $\Rsp{}(F)$ }
\label{gaugeinvcurr}
\endEQ
is an exact 3-form.
This result \eqref{gaugeinvcurr} applies as well to 
the exceptional current \eqref{JPS-dualityrot-current},
coming from the identification of the \adjsyms/ $Q=A'$, $Q'=-A$
with an essential dependence on the potentials. 
Therefore, all \conslaw/s given by the currents 
\eqrefs{MPS-current}{JPS-current}
as well as the current 
\eqref{JPS-dualityrot-current}
are gauge invariant. 

Second, 
through the mappings relating local \adjsyms/ of \Meq/
to local \syms/ of \Meq/ and its \potsys/s 
in Theorems \ref{thm:ME-joint} and~\ref{thm:ME-std}, 
we obtain a direct correspondence between local conserved currents
and local \syms/ of each system.
This is especially interesting because \Meq/ and the joint \potsys/
are not self-adjoint systems,
i.e. Noether's theorem relating \conslaw/s and \syms/
through a Lagrangian is inapplicable.

\begin{thm}
\label{thm:MEcurrentformula}
All \syms/ of \Meq/ induced through its \jps/
directly generate conserved currents
via the formula
\EQ
*\curr{} 
= \int_0^1 \eval{\l( Q' \wedge F - Q \wedge *F \r)}{\lambda A,\lambda A'}
 \frac{d\lambda}{\lambda} 
=  \int_0^1 \eval{\l( A \wedge {*P} - A' \wedge P \r)}{\lambda A,\lambda A'}
\frac{d\lambda}{\lambda}
\mod{ \totder{*\curl{}{}} }
\label{MEcurrentformula}
\endEQ
where 
the pair of \horiz/ 1-forms $(Q[A,A'],Q'[A,A'])$ is 
identified with any local \sym/ $\X$ of the joint \potsys/,
and the \horiz/ 2-form $P[A,A']$ is 
identified with any corresponding \sym/ of \Meq/,
written in terms of the joint potentials
via the relations $\totder{Q[A,A']}=P[A,A'],\totder{Q'}[A,A']={*P}[A,A']$. 
For the scaling symmetry \eqref{Xscaling}, 
the current \eqref{MEcurrentformula} is trivial.
\end{thm}

The proof of this theorem is similar to that of 
Proposition~\ref{prop:JPScurrentformula}
and will be omitted. 

We note that this scaling formula \eqref{MEcurrentformula} 
generates the exceptional current
\eqref{JPS-dualityrot-current}
naturally from the duality-rotation \sym/ \eqref{Xduality}
of the \jps/ and \Meq/. 
The implications of the explicit dependence 
on the potentials in this current 
will be dealt with in more detail in \secref{sec:nonlocal}.
More generally, the scaling formula \eqref{MEcurrentformula} 
yields \conscurr/s for any \sym/, {\it local or nonlocal},
$\X = P \Parder{F}$ admitted by \Meq/. 

Finally, we emphasize a main aspect of the interrelations 
we have derived between the \sym/ structure and the \conslaw/ structure 
of \Meq/ and its \potsys/s. 

\begin{cor} 
No \syms/ of $\totder{F} =\totder{*F}=0$
with essential dependence on a potential
arise from projection of local \syms/ of the \potsys/s
$\totder{*\totder A}=0$ and $\totder A' = *\totder A$
under the mapping $F=\totder A = -{*\totder A'}$, 
due to their gauge freedom
$A \ra A +\totder\chi$, $A' \ra A' +\totder\chi'$. 
There is only one \conslaw/ of $\totder{F} =\totder{*F}=0$ 
with explicit dependence on a potential 
arising from a projection of the local conserved currents of these \potsys/s,
namely, the duality-rotation current \eqref{JPS-dualityrot-current}.
This exceptional \conslaw/ nevertheless is gauge invariant. 
\end{cor}

Thus the locality projection results proven in \Ref{AncBlu:1997JMP} 
for \syms/ of well-posed PDE systems 
when applied to \Meq/ 
extend in a natural sense to gauge-invariant \conslaw/s. 
Indeed, a gauge-invariance projection result 
can be established directly for \conslaw/s of 
any locally regular PDE system
by an extension of the proof of locality projection for \syms/,
which we now outline. 

By a locally regular PDE system 
we mean it and all its prolongations are locally solvable \cite{Olv:symmbook} 
and of constant rank \cite{Olv:symmbook},
so all \conslaw/s arise from multipliers. 
Consider a conserved 3-form current $*\curr{}$ 
of any \potsys/ with gauge freedom.
Since conservation of $*\curr{}$ holds for all solutions of the \potsys/, 
it must remain conserved under 
infinitesimal gauge transformations $\X_{\it gauge}$ on the potentials. 
Hence the 3-form current $*\bar\Psi := \X_{\it gauge}{*\curr{}}$
is also conserved and consequently projects to a \conscurr/ of
the original PDE system through the embedding property of 
the respective solution spaces \cite{Blu:potsys}. 
The projected current $*\bar\Psi$ necessarily depends on 
an arbitrary function of $x$ 
and therefore so does its associated multiplier. 
Treating this function as an auxiliary dependent variable
and applying the corresponding Euler-Lagrange operator,
we obtain a divergence identity holding on the PDE system. 
An integration-by-parts method can then be used to reconstruct
an equivalent current from the divergence identity,
directly resulting in $*\bar\Psi = \totder{*\curl{}{}}$
holding for all solutions of the system. 
Hence we conclude that the 3-form current $*\curr{}$ 
projects to a gauge-invariant \conslaw/ of the given PDE system, 
namely $*\curr{}$ is invariant with respect to the gauge freedom 
on the potentials modulo trivially conserved (exact 3-form) terms.

\subsection{Geometric \syms/ and \conslaw/s}
\label{geometricPQsolutions}

To conclude this section, we consider 
the geometric \syms/, \adjsyms/, and conserved currents of 
\Meq/ and its \potsys/s
in light of our main results. 

\begin{defn}
\label{geom-pform}
A \horiz/ $p$-form on $\Jsp{1}$ will be called {\em geometric}
if it is locally constructed from $F$ in the case of \Meq/
and from $A$ or $A'$ in the case of the \potsys/s, 
using the Minkowski metric, volume form, spacetime coordinates,
and exterior derivatives. 
\end{defn}

To proceed we will make use of the following Lie derivative identities
holding for $p$-forms $\omega$ 
on any 4-dimensional spacetime manifold:
$\Lie{\xi} \omega = \xi \hook \extder\omega + \extder(\xi \hook \omega)$, 
and for $p=2$, 
$*\Lie{\xi} \omega = \Lie{\xi} {*\omega}$. 
These identities extend in an obvious way to a definition of 
the Lie derivative on \horiz/ forms 
in a jet space setting, 
for example,
$\Lie{\xi} \omega[F] 
= \xi \hook \totder\omega[F] + \totder(\xi \hook \omega[F])$, 
and 
$*\Lie{\xi} \omega[F] 
= \Lie{\xi} {*\omega[F]}$.
Thus, it follows that 
any Lie derivatives of $F$, $A$, or $A'$ produce geometric \horiz/ forms
in the sense of Definition~\ref{geom-pform}. 

Accordingly, 
a \sym/, \adjsym/, or \conscurr/ will be called {\it geometric}
if it is described by a geometric \horiz/ form. 
(Higher-order \syms/, \adjsyms/, and \conscurr/s of an analogous form 
arise from geometric ones in an explicit manner 
by the repeated application of
Lie derivatives on $F,A,A'$, 
with respect to general \CKvec/s, 
\ie/ general conformal symmetry operators.)
Note the classes of geometric \syms/, \adjsyms/, and \conscurr/s 
are each invariant under the duality transformations 
\eqrefs{PQduality}{FAA'duality}. 

The geometric \syms/ of \Meq/ and its \potsys/s consist of
the general conformal \syms/
\EQ
\X_{\it conf} = 
\Lie{\xi} F\Parder{F} ,\quad  
\Lie{\xi} A\Parder{A} ,\quad 
\Lie{\xi} A\Parder{A} + \Lie{\xi} A'\Parder{A'} ,
\label{confsymm}
\endEQ
and their corresponding dual \syms/
\EQ
\X'_{\it conf} = 
\Lie{\xi}{*F}\Parder{F} ,\quad 
\Lie{\xi} A'\Parder{A} - \Lie{\xi} A\Parder{A'} ,
\endEQ
in addition to the obvious scaling and duality-rotation \syms/ 
\eqrefs{Xscaling}{Xduality}. 
Here $\xi$ denotes a general conformal \Kvec/ of the Minkowski metric, 
\ie/ $\Lie{\xi}\flat{} = \Omega\flat{}$, $\Omega = \frac{1}{2}\div\xi$, 
representing 
four translation \syms/, 
six \rb/ \syms/, 
a dilation \sym/,
and four conformal (inversion) \syms/.
(Note that among these infinitesimal transformations,
$\X_{\it scal}$, $\X_{\it dual}$, $\X_{\it conf}$ are of point-type 
\cite{BluAnc:2002book}
while $\X'_{\it conf}$ is not, \ie/ its type is first-order.)
It is well known that the \syms/ $\X_{\it conf}$ 
comprise the 15-dimensional conformal Lie algebra $\SO(4,2)$,
while $\X_{\it conf}$ and $\X'_{\it conf}$ together form 
a 30-dimensional Lie algebra isomorphic to the complexification of $\SO(4,2)$;
the scaling and duality-rotation \syms/ $\X_{\it scal},\X_{\it dual}$
commute with both $\X_{\it conf}$ and $\X'_{\it conf}$.
Thus we note the Lie algebra of geometric \syms/ of 
\Meq/ and its \potsys/s 
has the structure $\U(1)^2 \times \SO(4,2)\otimes\Cnum$.

The \horiz/ 1-forms and 2-forms associated with these \syms/
and the corresponding \adjsyms/ obtained through
the mappings in Theorems~\ref{thm:self-mappings} to~\ref{thm:ME-std}
are summarized in \tableref{table:geometric-forms}.
All these pairs of \horiz/ forms $(P,*P)$ and $(Q,Q')$ have even-parity 
under the respective combined \dutr/s 
\eqsref{combined-dutr1}{combined-dutr4}.
(We remark that, in contrast, the analogous pairs
associated with the chiral \syms/ \cite{AncPoh:2004,AncPoh:2001} 
of \Meq/ possess odd-parity and exhibit a non-geometric form
that involves symmetrized derivatives of $F$.)

 \begin{table}[h]
 \begin{center}
 $\AR{|c|c|c|c|} \hline
 Q & Q' & P=\totder Q & {*P}=\totder Q' \\ \hline\hline
 A & A' & F & {*F} \\ \hline
 A' & -A & {*F} & -F \\ \hline
 \xi\hook F & \xi \hook {*F} & \Lie{\xi} F & \Lie{\xi} {*F} \\ \hline
 \xi \hook {*F} & -\xi \hook F & \Lie{\xi} {*F} & -\Lie{\xi} F \\ \hline
 \endAR$
 \caption{1-forms and 2-forms associated with geometric \syms/
 and \adjsyms/ for \Meq/ and \potsys/s}
 \label{table:geometric-forms}
 \end{center}
 \end{table}

Substitution of the 1-forms and 2-forms 
in \tableref{table:geometric-forms} 
into the conserved current formulas
in \tableref{table:conscur-diff-form} 
generates four currents, two of which are trivial.
The two other currents generated are the duality-rotation one 
\eqref{JPS-dualityrot-current} discussed earlier 
and one which is equivalent to the well-known 
geometric stress-energy currents of \Meq/
\EQs
*\curr{} 
&=& \frac{1}{2} \l( (\xi \hook {*F}) \wedge F - (\xi \hook F) \wedge *F \r) 
\quad \eqtext{ on $\Rsp{}(F)$}
\label{ME:stressenergy}
\\
&=& \frac{1}{2} (A \wedge \Lie{\xi}{*F} - A' \wedge \Lie{\xi} F)
\mod{ \totder{*\curl{}{}} } \eqtext{ on $\Rsp{1}(A,A')$} , 
\nonumber
\endEQs
where $\xi$ is a general \CKvec/.

\section{Analysis of the \jps/ in Lorentz gauge}
\label{sec:jps-analysis}

In light of the main results obtained in \secref{sec:cohom},
we must investigate the imposition of gauges in order
to obtain new \syms/ or new \conslaw/s
of the \potsys/s for \Meq/.
In the case of the \mps/ \eqref{MPS},
some well-known gauges are shown in \tableref{MPS-common-gauges}.
Note that $\x{\mu}{} =\{ x^0, x^i \}$ denotes time and space
coordinates in \Minksp/.

 \mystretch{\STR}
 \begin{table}[h]
 \begin{center}
 \TAB{|c|c|} \hline
 Gauge Name & Description \\ \hline\hline
 Lorentz & $\coder{\mu} \A{}{\mu}(x) =0$, \ie/ 
 $\der{0}\A{}{0}(x) = \coder{i} \A{}{i}(x)$ \\ \hline
 Coulomb & $\coder{i} \A{}{i}(x) =0$ \\ \hline
 Temporal & $\A{}{0}(x) =0$ \\ \hline
 Axial & $n\up{i} \A{}{i}(x)=0$, $n^i =\const$ \\ \hline
 Cronstrom & $\x{\mu}{} \A{}{\mu}(x) = 0$, \ie/ 
 $\x{i}{}\A{}{i}(x) = -\x{0}{}\A{}{0}(x)$ \\ \hline
 \endTAB
 \caption{Well-known gauges for the \mps/
 \label{MPS-common-gauges}}
 \end{center}
 \end{table}
 \mystretch{1}

Since \Meq/ \eqref{ME} are manifestly covariant and coordinate-independent,
natural gauges to investigate for \sym/ purposes should
also have these properties.
It should be noted that in 2+1 spacetime dimensions,
when temporal gauge or axial gauge is imposed on the \mps/,
no new local geometrical \syms/ arise \cite{AncBlu:1997JMP};
moreover, some of the Poincar\'e and conformal \syms/ 
are manifestly lost.

A standard covariant gauge choice for the \mps/ \eqref{MPS}
is the Lorentz gauge.
Cronstrom's gauge is also covariant,
but it is explicitly coordinate-dependent
and so we will not consider this gauge here.
With Lorentz gauge imposed on the magnetic potential 
we have the augmented system
\EQ
 \coder{\mu} \der{[\mu} \A{}{\nu]}(x) = 0, \qquad
 \coder{\mu} \A{}{\mu}(x) = 0,
\label{A-MPS}
\endEQ
which is no longer self-adjoint, and hence is a non-Lagrangian system. 
A well-known feature of the \potsys/ \eqref{A-MPS}
compared to the system without Lorentz gauge is that 
the conformal \syms/ \eqref{confsymm} 
on $\A{}{\mu}(x)$ are lost.
On the other hand, no new geometric \syms/ are gained,
as shown in the linear case in \Ref{The:MSc}.

For the \jps/ \eqref{JPS}, 
duality between the electric and magnetic potentials 
motivates the choice of Lorentz gauge on
both $\A{}{\mu}(x)$ and $\Ap{}{\mu}(x)$. 
Hence, we consider the augmented \potsys/
\EQ
\der{[\mu} \Ap{}{\nu]}(x) =
\frac{1}{2} \vol{\mu\nu\sigma\tau}{} \coder{\sigma} \A{\tau}{}(x),
\qquad \coder{\mu} \A{}{\mu}(x) = 0,
\qquad \coder{\mu} \Ap{}{\mu}(x) = 0.
\label{JPS-L}
\endEQ
Like the \jps/ itself, the augmented system is not self-adjoint
and thus is not a Lagrangian system. 
Note, importantly, 
the duality invariance \eqref{JPS-duality} on the potentials 
is retained.

Due to the imposition of Lorentz gauge,
the \potsys/ \eqref{JPS-L} becomes a locally well-posed PDE system,
with no gauge freedom on the potentials. 
As is the case for the standard \potsys/ in Lorentz gauge,
the potentials still admit a residual freedom given by 
gradients \eqref{JPS-gaugesym} involving scalar functions $\chi(x),\chi'(x)$, 
but in order to preserve Lorentz gauge 
these functions are restricted to satisfy the wave equation
$\coder{\mu}\der{\mu}\chi(x)=0$, 
$\coder{\mu}\der{\mu}\chi'(x)=0$.
(Hence the transformation \eqref{JPS-gaugesym} no longer
involves {\it arbitrary} functions of $\x{\mu}{}$.)
Modulo this residual freedom,
the solutions of the \potsys/ \eqref{JPS-L} in Lorentz gauge
are in one-to-one correspondence with the solutions of \Meq/ \eqref{ME}.

In this section, we will classify geometric \syms/
and corresponding conserved currents 
admitted by this \potsys/ \eqref{JPS-L}. 
We will denote the solution jet space of the system 
by $\divfrRsp{}(A,A')$. 
Note the coordinates of $\divfrRsp{}(A,A')$ are related to 
those of $\Rsp{}(A,A')$ by quotienting out 
the Lorentz gauge equations 
on derivatives of the potentials.

\subsection{Symmetry analysis}

The determining \eq/s for local symmetries 
$\X = \Q{}{\mu} \Parder{\A{}{\mu}} + \Qp{}{\mu} \Parder{\Ap{}{\mu}}$
of order $q <\infty$ 
of the \potsys/ \eqref{JPS-L} are given by 
\EQs
&&
\D{[\mu} \Qp{}{\nu]} 
= *\D{[\mu} \Q{}{\nu]} , 
\label{JPS-L:Qdeteq1}\\
&&
\coD{\mu} \Q{}{\mu} = 0, \quad
\coD{\mu} \Qp{}{\mu} = 0 , 
\label{JPS-L:Qdeteq2}
\endEQs
on $\divfrRsp{q}(A,A') \subset \Jsp{q+1}(A,A')$. 
These equations retain the duality invariance 
\EQ
(Q,Q') \lra (Q',-Q) . 
\label{QQ'duality}
\endEQ
Note that the loss of gauge freedom in this \potsys/
under Lorentz gauge implies there is no gradient freedom 
in the form of \syms/ $\Q{}{\mu},\Qp{}{\mu}$. 
We therefore refer to \eqref{JPS-L:Qdeteq2} 
as Lorentz gauge equations on $(Q,Q')$. 

Anco \& Pohjanpelto have shown that 
any local \sym/ of order $0\leq q<\infty$ of \Meq/ 
is {\it linear} in the field $\F{}{\mu\nu}$ and its derivatives 
on $\Rsp{q}(F)$ \cite{AncPoh:2004,AncPoh:2002}.
The same method can be expected to establish an analogous result 
for local \syms/ of the \potsys/ \eqref{JPS-L}. 
Moreover, the geometric \syms/ of Maxwell's equations and
the joint \potsys/ have even-parity 
under the respective combined \dutr/s 
\eqrefs{combined-dutr1}{combined-dutr4}.
This motivates a classification using 
a linear homogeneous even-parity ansatz for geometric \syms/:
\EQs
&& 
\Q{}{\mu}[A,A'] = 
\Lie{\xi} \A{}{\mu} + \Lie{\xi'} \Ap{}{\mu}
+ \a{\mu}{\nu}(x) \A{}{\nu} + \ap{\mu}{\nu}(x) \Ap{}{\nu},
\label{JPS-L:XA}\\
&&
\Qp{}{\mu}[A,A'] = 
\Lie{\xi} \Ap{}{\mu} - \Lie{\xi'} \A{}{\mu}
+ \a{\mu}{\nu}(x) \Ap{}{\nu} - \ap{\mu}{\nu}(x) \A{}{\nu},
\label{JPS-L:XA'}
\endEQs
where $\kv{\nu}{}$, $\kvp{\nu}{}$, $\a{\mu}{\nu}$, $\ap{\mu}{\nu}$,
are functions of $x$ to be determined.
Note the even-parity condition characterizing 
this class of \syms/ is expressed by 
\EQ
\Qp{}{\mu}[A,A'] = \Q{}{\mu}[A',-A] . 
\label{JPS-L-symmduality}
\endEQ

\begin{thm} 
\label{thm:JPS-L-sym}
The (even-parity) class of geometric \syms/ $\X$ of the form 
\eqsref{JPS-L:XA}{JPS-L:XA'}
admitted by the \potsys/ \eqref{JPS-L} 
consists of:
\Items
\item[{\rm (i)}]
the scaling and duality-rotation transformations
\EQs
&&
\sX = 
\A{}{\mu} \Parder{\A{}{\mu}}
 + \Ap{}{\mu} \Parder{\Ap{}{\mu}} , 
\label{JPS-L-Xscaling}\\
&&
\dX = 
\Ap{}{\mu} \Parder{\A{}{\mu}}
 - \A{}{\mu} \Parder{\Ap{}{\mu}} ,
\label{JPS-L-Xduality}
\endEQs
\item[{\rm (ii)}]
the \rbvec/ transformation 
\EQ
\rbX{}= 
\l( \gam{\mu}{\nu} \A{}{\nu} + \dugam{\mu}{\nu} \Ap{}{\nu} \r) 
\Parder{\A{}{\mu}} 
+ \l( \gam{\mu}{\nu} \Ap{}{\nu} - \dugam{\mu}{\nu} \A{}{\nu} \r)  
\Parder{\Ap{}{\mu}} ,
\label{JPS-L-Xrbvec}
\endEQ
with 
\EQ
\gam{}{\mu\nu} = \gam{}{[\mu\nu]}=\const , 
\label{skewmatr}
\endEQ
\item[{\rm (iii)}]
the \HKV/ transformation and its dual 
\EQs
&&
\hX{}=    
\Lie{\xi} \A{}{\mu} \Parder{\A{}{\mu}}
+ \Lie{\xi} \Ap{}{\mu} \Parder{\Ap{}{\mu}} ,
\label{JPS-L-XHKvec}\\
&&
\hX{}'=    
\Lie{\xi} \Ap{}{\mu} \Parder{\A{}{\mu}}
- \Lie{\xi} \A{}{\mu} \Parder{\Ap{}{\mu}} , 
\label{JPS-L-XHKvecdual}
\endEQs
with 
\EQ
\kv{\mu}{} = 
\k{1}{\mu}{} + \k{2}{\mu\nu}{} \x{}{\nu} + \k{3}{}{} \x{\mu}{} ,\quad
\k{1}{\mu}{},\k{2}{\mu\nu}{}=\k{2}{[\mu\nu]}{},\k{3}{}{}=\const, 
\label{HKV}
\endEQ
\item[{\rm (iv)}]
the \CKV/ transformation and its dual 
\EQs
&& \cX{} =
\l( \wLie{\xi} \A{}{\mu} + \z{\mu}{\nu} \A{}{\nu}
+ \duz{\mu}{\nu} \Ap{}{\nu} \r) \Parder{\A{}{\mu}}
+ \l( \wLie{\xi} \Ap{}{\mu} + \z{\mu}{\nu} \Ap{}{\nu}
- \duz{\mu}{\nu} \A{}{\nu} \r) 
\Parder{\Ap{}{\mu}} , 
\label{JPS-L-XCKvec}\\
&& \cX{}' =
\l( \wLie{\xi} \Ap{}{\mu} + \z{\mu}{\nu} \Ap{}{\nu}
- \duz{\mu}{\nu} \A{}{\nu} \r) 
\Parder{\A{}{\mu}}
- \l( \wLie{\xi} \A{}{\mu} + \z{\mu}{\nu} \A{}{\nu}
+ \duz{\mu}{\nu} \Ap{}{\nu} \r) 
\Parder{\Ap{}{\mu}} , 
\label{JPS-L-XCKvecdual}
\endEQs
where 
$\wLie{\xi} := \Lie{\xi} + \frac{1}{4} \Omega$
and 
$\z{}{\mu\nu} := -\frac{1}{2} \coder{[\mu} \kv{\nu]}{}$, 
$\Omega:= \frac{1}{2} \der{\mu}\kv{\mu}{}$, 
with 
\EQ
\kv{\mu}{} = 
\k{4}{\sigma}{} \x{}{\sigma} \x{\mu}{}
- \frac{1}{2} \k{4}{\mu}{} \x{\sigma}{} \x{}{\sigma} ,\quad
\k{4}{\mu}{}=\const .
\label{CKV}
\endEQ
\endItems
This class is closed under the \dutr/ \eqref{QQ'duality}. 
\end{thm}

Before outlining the proof of this theorem,
we first discuss geometrical features of these \syms/
and the structure of their Lie algebra.

The parameters 
$\k{1}{\mu}{}$, $\k{2}{\mu\nu}{}$, $\k{3}{}{}$, $\k{4}{\mu}{}$ 
appearing in the homothetic/conformal \Kvec/s $\kv{\mu}{}$ 
correspond respectively to 
four translations, three rotations and three boosts, 
one dilation, and four inversions of $M^4$,
which are conformal isometries determined by 
the conformal Killing equation
\EQ
\Lie{\xi} \flat{\mu\nu} = \Omega \flat{\mu\nu} . 
\label{conformalisom}
\endEQ
The isometries \eqref{conformalisom}
generated by translations, \rbs/, and a dilation
comprise the homothetic \Kvec/s \eqref{HKV} on $M^4$
and induce a corresponding Lie derivative action on the potentials,
i.e. infinitesimal Poincar\'e and dilation transformations.
In contrast to the situation 
for the standard \potsys/ in Lorentz gauge \eqref{A-MPS}, 
the infinitesimal conformal transformation
associated with genuine conformal isometries \eqref{conformalisom} 
given by inversions on $M^4$ 
are not lost.
However, the transformation is modified through 
combining a weighted Lie derivative 
with respect to a \CKvec/ \eqref{CKV}
and an internal rotation on the potentials 
via the coefficients $\z{}{\mu\nu}$ and $\duz{}{\mu\nu}$.

The infinitesimal \rbvec/ transformation 
with parameters $\gam{}{\mu\nu} = \gam{}{[\mu\nu]}$ 
is genuinely new as there is no local counterpart of it
in \Meq/ or the standard \potsys/ with or without Lorentz gauge.
Along with the infintesimal scaling and duality-rotation transformations, 
these transformations are internal (non-spacetime) \syms/ 
in the sense that there is no associated motion on spacetime.
Note they comprise three internal rotations and three internal boosts
in addition to the one scaling and one duality-rotation. 

We give an alternative representation for the internal \syms/
as follows.
At any point in spacetime,
identify $\Rnum^{4} \times \Rnum^{4} \cong \Rnum^{4} \otimes \Rnum^{2}$,
and introduce basis elements 
$e\down{1},e\down{2}$ for $\Rnum^{2}$. 
We will identify the jet space coordinates of the potentials
$\A{}{\mu},\Ap{}{\mu}$ with 
$\A{}{\mu} \otimes e\down{1}$, $\A{}{\mu} \otimes e\down{2}$.
Let ${\rm id} \in {\rm Hom}(\Rnum^{4})$ be the identity map,
and ${\rm R} \in {\rm Hom}(\Rnum^{2})$ a standard rotation on $\Rnum^{2}$,
\ie/ ${\rm id}(v\down\mu) = v\down{\mu}$,
${\rm R}(e\down{1}) = e\down{2}$,
${\rm R}(e\down{2}) = -e\down{1}$,
for $v\down{\mu} \in \Rnum^{4}$.
Then the duality-rotation is given by the transformation 
\EQ
{\rm id} \otimes {\rm R} \in {\rm Hom}(\Rnum^{4} \otimes \Rnum^{2})
\endEQ
which acts non-trivially only on the $\Rnum^{2}$ factor,
or off-diagonally on the space $\Rnum^{4} \otimes \Rnum^{2}$,
with matrix representation
\EQ
 \l( \AR{c|c} 0 & -\mathbb{I} \\ \hline \mathbb{I} & 0 \endAR \r).
\endEQ
The \rbsvec/ combine a nontrivial action on
the $\Rnum^{2}$ and $\Rnum^{4}$ factors.
Define ${\rm R}\down{\gamma} \in {\rm Hom}(\Rnum^{4})$
to be the standard \rb/ operator with parameters
$\gam{}{\mu\nu} = \gam{}{[\mu\nu]}$.
Recall that the \rbs/ act infinitesimally
by ${\rm R}\down{\gamma}(v\down{\mu}) = \gam{\mu}{\nu} v\down{\nu}$
for $v\down{\mu} \in \Rnum^{4}$.
Here the constant parameters $\gam{}{\mu\nu}$ 
geometrically determine a 2-dimensional plane in $\Rnum^{4}$ 
on which the \rb/ takes place
(namely, the cokernel of ${\rm R}\down{\gamma}$). 
With this notation, the internal \rbs/ act via
\EQ
{\rm R}\down{\gamma}\otimes {\rm id}
+ {\rm R}\down{*\gamma}\otimes {\rm R}
 \in {\rm Hom}(\Rnum^{4} \otimes \Rnum^{2}).
\label{IRB}
\endEQ
We then see that the transformation \eqref{IRB}
is a sum of diagonal and off-diagonal
\rbs/ on $\Rnum^{4}\otimes \Rnum^{2}$:
the diagonal action involves a standard \rb/ on $\Rnum^{4}$,
and the plane for the off-diagonal \rb/
is the dual of the plane for the diagonal \rb/.

The geometric \syms/ \eqsref{JPS-L-Xscaling}{JPS-L-XCKvecdual}
comprise a 38-dimensional \sym/ algebra. 
To describe its commutator structure, 
we first extend the transformation $\cX{}$ and its dual $\cX{}'$ 
to general \CKvec/s \eqref{conformalisom}. 


\begin{prop}
\label{JPS-L-combinedsymm}
For a \HKvec/ 
$\kv{\mu}{} = 
\k{1}{\mu}{} + \k{2}{\mu\nu}{} \x{}{\nu} + \k{3}{}{} \x{\mu}{}$, 
\EQ
\cX{} = \hX{} +\rbX{} + \frac{1}{4}\Omega \sX ,\quad
\cX{}' = \hX{}' -\durbX{} + \frac{1}{4}\Omega \dX ,\quad
\gam{}{} = \z{}{}
\endEQ
is a geometric \sym/ of the \jps/ in Lorentz gauge,
where $\z{}{\mu\nu} = \frac{1}{2} \k{2}{\mu\nu}{}=\const$, 
$\Omega=2\k{3}{}{}=\const$ 
are the (scaled) curl and divergence of $\kv{\mu}{}$. 
\end{prop}

Let 
$[\gamma_1,\gamma_2]\up{\mu\nu} 
= 2 \gamma_1\up{\sigma[\mu} \gamma_2\updown{\nu]}{\sigma}$
denote the commutator of two skew-tensors 
$\gamma_1\up{\mu\nu},\gamma_2\up{\mu\nu}$
viewed as matrices, 
and 
$[\xi_1,\xi_2]\up{\mu} 
= \Lie{\xi_1}\xi_2\up{\mu} = -\Lie{\xi_2}\xi_1\up{\mu}$
denote the commutator of two general conformal \Kvec/s 
$\xi_1\up{\mu},\xi_2\up{\mu}$. 
Note a general \CKvec/ $\xi\up{\mu}$ has a decomposition into 
a sum of a \HKvec/ \eqref{HKV} and a \CKvec/ \eqref{CKV}
whose curl and divergence are given by 
\EQ
\z{}{\mu\nu} = -\frac{1}{2} \coder{[\mu} \kv{\nu]}{} 
= \frac{1}{2} \k{2}{\mu\nu}{} -\k{4}{[\mu}{} \x{\nu]}{} ,\quad
\Omega = \frac{1}{2} \der{\mu} \kv{\mu}{}
= 2\k{3}{}{} + 2\k{4}{\mu}{} \x{}{\mu} ,
\endEQ
where $\z{}{\mu\nu}$ and $\Omega$ are constant only in the case of \HKvec/. 
Recall, any two \CKvec/s \eqref{CKV} commute,
while the commutator of any two \HKvec/s \eqref{HKV} is again a \HKvec/, 
which is given by the well-known Poincar\'e Lie algebra 
\cite{Bat:1909,Cun:1909,Ibr:1968}
enlarged by the one-dimensional Lie algebra of dilations. 
(In particular, dilations commute with all \HKvec/s except translations,
whose commutator is again a translation \ie/ 
$[x,k_1]\up{\mu} = -\k{1}{\mu}{}$; 
the Poincar\'e Lie algebra is the semidirect product of
the Lie algebra $\SO(3,1)$ of rotations/boosts
and the abelian Lie algebra $\U(1)^4 \cong \Rnum^4$ of translations.)
The commutator of a \CKvec/ \eqref{CKV} and a \HKvec/ \eqref{HKV} 
is given by the parameters
\EQ
\kb{1}{\mu}{} =
0 ,\quad
\kb{2}{\mu\nu}{} = 
2 \k{4}{[\mu}{} \k{1}{\nu]}{} ,\quad
\kb{3}{}{} = 
-\k{4}{}{\nu} \k{1}{\nu}{} ,\quad
\kb{4}{\mu}{} = 
- \k{4}{\mu}{} \k{3}{}{} - \k{4}{}{\nu} \k{2}{\nu\mu}{} . 
\endEQ

\begin{thm} {\bf (Geometric symmetry algebra) }
\label{JPS-L-symalg}

The nonzero commutators of the geometric \syms/ 
\eqsref{JPS-L-Xscaling}{JPS-L-XCKvecdual} of the \jps/ in Lorentz gauge
have the algebraic structure 
\EQs
&&
[\rbX{1},\rbX{2}] = 2\rbX{3} ,\quad\mbox{ where } 
\gamma_3 = [\gamma_2,\gamma_1] ,
\\
&&
[\cX{1},\cX{2}] = -[\cX{1}',\cX{2}'] = \cX{3} ,\quad
[\cX{1},\cX{2}'] = \cX{3}' ,
\mbox{ where } 
\xi_3 = [\xi_2,\xi_1] . 
\endEQs
Moreover, 
$\sX,\dX$ span a two-dimensional abelian Lie algebra, 
$\rbX{}$ spans the six-dimensional \rb/ Lie algebra $\SO(3,1)$, 
$\cX{}$ spans the 15-dimensional conformal Lie algebra $\SO(4,2)$;
the Lie algebra spanned by $\cX{},\cX{}'$ is isomorphic to 
the complexification of $\SO(4,2)$. 
Thus the geometric \syms/ together form a 
38-dimensional Lie algebra 
$\U(1)^2 \times \SO(3,1) \times \SO(4,2)\otimes\Cnum$.
\end{thm}

The computation of this \sym/ algebra is straightforward and will be omitted.
To conclude the \sym/ analysis, 
we now derive the geometric \sym/ classification
by solving the \sym/ determining equations 
\eqrefs{JPS-L:Qdeteq1}{JPS-L:Qdeteq2}. 

\Proof{ of Theorem \ref{thm:JPS-L-sym}}
Explicit coordinates for the solution jet space $\divfrRsp{}(A,A')$
are given by 
$(\x{\mu}{},\A{}{\nu},\Ap{}{\nu},
\F{}{\mu\nu},\trfr\A{}{\mu\nu},\trfr\Ap{}{\mu\nu})$
where
\EQ
\F{}{\mu\nu} = \A{}{[\nu,\mu]} 
= -\frac{1}{2} \vol{\mu\nu}{\sigma\tau} \Ap{}{\tau,\sigma} 
,\quad
\A{}{\mu\nu} = \A{}{(\nu,\mu)} ,\quad
\Ap{}{\mu\nu} = \Ap{}{(\nu,\mu)} . 
\endEQ
These components represent the linearly independent parts of
the potentials and their first-order derivatives
at a point in spacetime, subject to the system equations \eqref{JPS-L}. 

To proceed, we substitute \eqrefs{JPS-L:XA}{JPS-L:XA'}
into the determining equations \eqrefs{JPS-L:Qdeteq1}{JPS-L:Qdeteq2}.
The second-order derivative terms in $A$ and $A'$ 
are found to vanish on $\divfrRsp{1}(A,A') \subset \Jsp{2}(A,A')$. 
For the remaining terms, we extract the coefficients of 
the linearly independent coordinates 
on $\divfrRsp{}(A,A')$ as follows:
(i) $\A{}{\nu},\Ap{}{\nu}$,
(ii) $\F{}{\mu\nu}$,
(iii) $\trfr\A{}{\mu\nu},\trfr\Ap{}{\mu\nu}$. 
By setting the coefficients to vanish,
we obtain the equations
\EQs
&& 
\der{[\alpha} \a{\beta]\sigma}{} = 
\frac{1}{2} \vol{\alpha\beta}{\mu\nu} \der{\mu} \ap{\nu\sigma}{} ,
\label{JPS-L:coeff-eqn-1}\\
&& 
\der{\sigma} \ba{}{\sigma\alpha} = 0 ,\quad
\der{\sigma} \bap{}{\sigma\alpha} = 0 ,
\label{JPS-L:coeff-eqn-2}\\
&& 
\a{[\alpha}{(\mu} \id{\nu)}{\beta]} + \frac{1}{2} \ap{}{\sigma(\mu}
\vol{}{\nu)}\down{\alpha\beta\sigma} =
\frac{1}{4} \invflat{\mu\nu} (
\a{[\alpha\beta]}{} 
- \frac{1}{2} \vol{\alpha\beta\sigma\tau}{} \ap{}{\sigma\tau} ) ,
\label{JPS-L:coeff-eqn-3}\\
&& 
\ta{(\alpha\beta)}{} = 
\frac{1}{4} \flat{\alpha\beta} \ta{\sigma}{\sigma} , \quad
\tap{(\alpha\beta)}{} = 
\frac{1}{4} \flat{\alpha\beta} \tap{\sigma}{\sigma} ,
\label{JPS-L:coeff-eqn-4}\\
&& 
\ta{[\alpha}{\sigma} \vol{\beta]\sigma}{\mu\nu}
 - \ta{}{\sigma[\mu} \epsilon\up{\nu]}\down{\alpha\beta\sigma}
+ 2(\tap{[\alpha}{[\mu}+ \tap{}{[\mu}\down{[\alpha}) \id{\nu]}{\beta]}
- \tap{\sigma}{\sigma}\id{[\mu}{\alpha} \id{\nu]}{\beta}
=0 , 
\label{JPS-L:coeff-eqn-5}\\
&& 
\ap{[\alpha\beta]}{} = 
\frac{1}{2} \vol{\alpha\beta\sigma\tau}{} \a{}{\sigma\tau} , 
\label{JPS-L:coeff-eqn-6}
\endEQs
where, for notational convenience, we have defined
\EQs
&&
\ba{\alpha\beta}{} := \a{\alpha\beta}{} + \der{\alpha} \xi\down{\beta} ,
\quad
\bap{\alpha\beta}{} := \ap{\alpha\beta}{} + \der{\alpha} \xi'\down{\beta} ,
\\
&&
\ta{\alpha\beta}{} := \a{\alpha\beta}{} + 2\der{\alpha} \xi\down{\beta} ,
\quad
\tap{\alpha\beta}{} := \ap{\alpha\beta}{} + 2\der{\alpha} \xi'\down{\beta}.
\endEQs

From \eqref{JPS-L:coeff-eqn-3},
contracting on $\nu,\beta$, 
and then symmetrizing on $\alpha,\mu$,
we find
\EQ
\a{(\alpha\beta)}{} =
\frac{1}{4} \flat{\alpha\beta} \a{\sigma}{\sigma} . 
\label{symmproof:1}
\endEQ
Similarly, multiplication of \eqref{JPS-L:coeff-eqn-3}
by $\vol{}{\alpha\beta\sigma\tau}$ followed by a similar contraction
and symmetrization leads to 
\EQ
\ap{(\alpha\beta)}{} =
\frac{1}{4} \flat{\alpha\beta} \ap{\sigma}{\sigma} . 
\label{symmproof:2}
\endEQ
Combining \eqrefs{symmproof:1}{symmproof:2}
with \eqref{JPS-L:coeff-eqn-4}, we obtain
\EQ
\der{(\alpha} \kv{}{\beta)} 
=\frac{1}{4} \flat{\alpha\beta} \der{\sigma} \kv{\sigma}{} ,\quad
\der{(\alpha} \kvp{}{\beta)} 
= \frac{1}{4} \flat{\alpha\beta} \der{\sigma} \kvp{\sigma}{} .
\label{symmproof:KVeq}
\endEQ
Through \eqsref{symmproof:1}{symmproof:KVeq}
and \eqref{JPS-L:coeff-eqn-6}, 
the equations \eqref{JPS-L:coeff-eqn-3}--\eqref{JPS-L:coeff-eqn-5}
reduce to identities.

The differential equation \eqref{symmproof:KVeq}
is equivalent to the conformal Killing equation \eqref{conformalisom}
and hence both $\kv{\mu}{}(x)$ and $\kvp{\mu}{}(x)$
are general \CKvec/s:
\EQs
\kv{\mu}{} &=& 
\k{1}{\mu}{} + \k{2}{\mu\nu}{} \x{}{\nu}
+ \k{3}{}{} \x{\mu}{}
+ \k{4}{\nu}{} \x{}{\nu} \x{\mu}{}
- \frac{1}{2} \k{4}{\mu}{} \x{\nu}{} \x{}{\nu} , 
\label{xi-CKV}\\
\kvp{\mu}{} &=& 
\kp{1}{\mu}{}
+ \kp{2}{\mu\nu}{} \x{}{\nu}
+ \kp{3}{}{} \x{\mu}{}
+ \kp{4}{\nu}{} \x{}{\nu} \x{\mu}{}
- \frac{1}{2} \kp{4}{\mu}{} \x{\nu}{} \x{}{\nu} , 
\label{xi'-CKV}
\endEQs
where
$\k{1}{\mu}{}$, 
$\k{2}{\mu\nu}{} = \k{2}{[\mu\nu]}{}$, 
$\k{3}{}{}$, $\k{4}{\mu}{}$, 
and
$\kp{1}{\mu}{}$, 
$\kp{2}{\mu\nu}{} = \kp{2}{[\mu\nu]}{}$, 
$\kp{3}{}{}$, 
$\kp{4}{\mu}{}$ are constants.

We decompose $\a{\nu\sigma}{}$ and $\ap{\nu\sigma}{}$ 
into their symmetric and antisymmetric parts 
using \eqrefs{symmproof:1}{symmproof:2}, 
with the notation
\EQ
\ah{\nu\sigma}{} := \a{[\nu\sigma]}{} ,\quad
\aph{\nu\sigma}{} := \ap{[\nu\sigma]}{} .
\endEQ
Elimination of $\aph{\nu\sigma}{}$ in \eqref{JPS-L:coeff-eqn-1} 
using \eqref{JPS-L:coeff-eqn-6}
then gives the equation 
\EQ
\der{[\alpha} \ah{\beta]\sigma}{} 
-\frac{1}{2} \der{\sigma} \ah{\alpha\beta}{} 
= \flat{\sigma[\alpha} \coder{\mu} \ah{\beta]\mu}{} 
+ \frac{1}{4} \flat{\sigma[\alpha} \der{\beta]} \a{\nu}{\nu} 
+\frac{1}{8} \vol{\alpha\beta\mu\sigma}{} \coder{\mu} \ap{\nu}{\nu}.
\label{trproof:1}
\endEQ
By contracting \eqref{trproof:1} on $\beta,\sigma$, 
followed by using \eqrefs{JPS-L:coeff-eqn-2}{xi-CKV}, 
we obtain
\EQ
\der{\alpha} \a{\sigma}{\sigma} 
= \frac{4}{3} \coder{\sigma} \ah{\sigma\alpha}{}
= -\coder{\sigma} \der{\sigma} \xi\down{\alpha} = 2 \k{4}{}{\alpha}.
\label{divpart}
\endEQ
Consequently, the trace-part of $\a{\alpha\beta}{}$ is
\EQ
 \a{\sigma}{\sigma} = 4\lambda + 2\k{4}{\sigma}{} \x{}{\sigma} ,
\label{trproof:2}
\endEQ
where $\lambda$ is a constant.
Then, 
a similar elimination of $\ah{\nu\sigma}{}$ in \eqref{JPS-L:coeff-eqn-1} 
leads to the trace-part of $\ap{\alpha\beta}{}$,
\EQ
 \ap{\sigma}{\sigma} = 4\lambda' + 2\kp{4}{\sigma}{} \x{}{\sigma} ,
\label{trproof:3}
\endEQ
where $\lambda'$ is a constant.

Hence \eqref{trproof:1} becomes
\EQ
\der{[\alpha} \ah{\beta]\sigma}{} 
- \frac{1}{2} \der{\sigma} \ah{\alpha\beta}{} 
= 
\k{4}{}{[\alpha} \flat{\beta]\sigma} 
- \frac{1}{4} \vol{\alpha\beta\sigma\tau}{} \kp{4}{\tau}{} 
\label{asymproof:1}
\endEQ
through \eqsref{divpart}{trproof:3}.
Antisymmetrization of \eqref{asymproof:1} on $\alpha,\beta,\sigma$ 
leads to 
\EQ
\der{[\alpha} \ah{\beta]\sigma}{}
+ \frac{1}{2} \der{\sigma} \ah{\alpha\beta}{} 
= -\frac{3}{4} \vol{\alpha\beta\sigma\tau}{} \kp{4}{\tau}{} .
\label{asymproof:2}
\endEQ
Combining \eqrefs{asymproof:1}{asymproof:2} we obtain
\EQ
\der{\sigma} \ah{\alpha\beta}{} =
-\k{4}{}{[\alpha} \flat{\beta]\sigma}
- \frac{1}{2} \vol{\alpha\beta\sigma\tau}{} \kp{4}{\tau}{} ,
\label{asymproof:3}
\endEQ
which yields 
\EQ
 \ah{\alpha\beta}{} = 
\gam{\alpha\beta}{} + \z{\alpha\beta}{} - \duzp{\alpha\beta}{} ,
\label{asymproof:4}
\endEQ
where $\gam{\alpha\beta}{} = \gam{[\alpha\beta]}{}=\const$, 
and
$\z{\alpha\beta}{} = 
-\frac{1}{2} \der{[\alpha} \kv{}{\beta]} 
= -\k{4}{}{[\alpha} \x{}{\beta]}$,
$\zp{\alpha\beta}{} = 
-\frac{1}{2} \der{[\alpha} \kvp{}{\beta]} 
= -\kp{4}{}{[\alpha} \x{}{\beta]}$.
Finally, using \eqref{JPS-L:coeff-eqn-6}
we have 
\EQ
\aph{\alpha\beta}{} = 
\dugam{\alpha\beta}{} + \duz{\alpha\beta}{} + \zp{\alpha\beta}{} . 
\label{asymproof:5}
\endEQ
All equations \eqref{JPS-L:coeff-eqn-1}--\eqref{JPS-L:coeff-eqn-6}
now reduce to identities.
\endProof

We remark that if the ansatz \eqrefs{JPS-L:XA}{JPS-L:XA'} is
generalized to include odd parity terms of zeroth order
on $\Jsp{1}(A,A')$, 
then the proof goes through and no new \syms/ are obtained. 
It would be a natural generalization to next include 
odd parity first-order terms, 
but we have not investigated the outcome.

\subsection{Conservation law analysis}

Since the solution space of the \jps/ in Lorentz gauge
has a natural embedding into 
the solution space of the unaugmented \potsys/, 
both the duality-rotation current 
\eqref{JPS-dualityrot-current}
connected with duality rotations on the potentials
and the stress-energy currents \eqref{ME:stressenergy}
associated with general conformal \Kvec/s, 
discussed in \secref{geometricPQsolutions},  
continue to be admitted when Lorentz gauge is imposed.
New conserved currents are suggested by 
the appearance of the new local \syms/ admitted under Lorentz gauge. 

The \jps/ remains non-self-adjoint in Lorentz
gauge and hence it is a non-Lagrangian system. Consequently, conserved
currents arise through \adjsyms/ via a scaling formula similar
to the one in \tableref{table:adjrel-conslaw} 
for the system without Lorentz gauge. 

Local \adjsyms/ of system \eqref{JPS-L} consist of 
a \horiz/ 2-form $\P{}{\mu\nu}[A,A']$ 
as in the case without Lorentz gauge,
plus a pair of differential scalar functions $\chi[A,A'],\chi'[A,A']$,
which arise respectively from the adjoint of the \sym/ equations 
\eqrefs{JPS-L:Qdeteq1}{JPS-L:Qdeteq2}. 
In particular, 
the determining equations for local \adjsyms/ of order $q<\infty$ 
take the form
\EQs
&& 
\coD{\mu} \P{}{\mu\nu} + \D{\nu} \chi'  =0 ,
\label{JPS-L-Pdeteq1}\\
&&
\coD{\mu} \duP{}{\mu\nu} - \D{\nu} \chi =0 , 
\label{JPS-L-Pdeteq2}
\endEQs
on $\Rsp{q}(A,A')$.
Note these equations have the duality invariance
\EQ
(\P{}{\mu\nu},\chi,\chi') \ra (\duP{}{\mu\nu},\chi',-\chi) . 
\label{JPS-L-Pduality}
\endEQ

\begin{prop}
All nontrivial local conserved currents of the \potsys/ \eqref{JPS-L}
are generated from local \adjsyms/ 
by the conserved current formula 
\EQ
\curr{\mu} = 
\int_0^1 \eval{\l(\P{\mu\nu}{} \Ap{}{\nu}
- \duP{\mu\nu}{} \A{}{\nu} + \chi \A{\mu}{}
+ \chi'\Ap{\mu}{}\r)}{\lambda A, \lambda A'} 
\frac{d\lambda}{\lambda} . 
\label{JPS-L-currentformula}
\endEQ
\end{prop}

To apply this result, 
rather than solve the \adjsym/ equations 
we will now utilize our \sym/ classification results
and derive a mapping from local \syms/ to local \adjsyms/,
extending the mapping \eqsref{JPS-QtoPmap}{JPS-Q'toPmap}
obtained in Theorem \ref{thm:self-mappings} for the \jps/ without gauges.

Since this map \eqsref{JPS-QtoPmap}{JPS-Q'toPmap} 
relied only on the adjoint relation between 
the respective determining equations
for local \syms/ 
$\Q{}{\mu}[A,A'],\Qp{}{\mu}[A,A']$ 
and local \adjsyms/
$\P{}{\mu\nu}[A,A']$ 
of the unaugmented \potsys/, 
it carries over to the \potsys/ in Lorentz gauge
if we project out the functions 
$\chi[A,A'],\chi'[A,A']$ 
associated with the Lorentz gauge equations \eqref{JPS-L:Qdeteq2}. 

\begin{thm}
For the joint \potsys/ in Lorentz gauge,
there is a linear mapping from local \syms/ into local \adjsyms/ 
given by
\EQ
(\Q{}{\nu}, \Qp{}{\nu}) \lra 
(\P{}{\mu\nu}= \D{[\mu} \Q{}{\nu]},
*\P{}{\mu\nu} = \D{[\mu} \Qp{}{\nu]},\chi=0,\chi'=0) 
\label{JPS-L:sym-adjsym1}
\endEQ
as well as a dual mapping
\EQ
(\Q{}{\nu}, \Qp{}{\nu}) \lra 
(\P{}{\mu\nu} = -\D{[\mu}\Qp{}{\nu]}, 
*\P{}{\mu\nu} = \D{[\mu} \Q{}{\nu]},
\chi=0,\chi'=0)
\label{JPS-L:sym-adjsym2}
\endEQ
coming from the duality invariance \eqref{JPS-L-Pduality}.
Associated to these correspondences 
\eqrefs{JPS-L:sym-adjsym1}{JPS-L:sym-adjsym2}
are respective formulas \eqref{JPS-L-currentformula}
that directly generate local \conscurr/s 
\EQ
\curr{\mu} = 
\int_0^1 \eval{(\Qp{}{\nu} \F{\mu\nu}{}
- \Q{}{\nu} \duF{\mu\nu}{} )}{\lambda A, \lambda A'}
\frac{d\lambda}{\lambda}
\label{JPS-L:sym-curr1}
\endEQ
and
\EQ
\curr{\mu} = 
\int_0^1 \eval{(\Q{}{\nu} \F{\mu\nu}{}
+ \Qp{}{\nu} \duF{\mu\nu}{} )}{\lambda A, \lambda A'}
\frac{d\lambda}{\lambda}
\label{JPS-L:sym-curr2}
\endEQ
from local \syms/ of this \potsys/. 
\end{thm}

\Proof{}
The \adjsym/ equations \eqrefs{JPS-L-Pdeteq1}{JPS-L-Pdeteq2}
reduce directly to the \sym/ equations 
\eqrefs{JPS-L:Qdeteq1}{JPS-L:Qdeteq2}
through substitution of the mappings 
\eqrefs{JPS-L:sym-adjsym1}{JPS-L:sym-adjsym2}. 
Similarly, 
the \conscurr/ formula \eqref{JPS-L-currentformula}
reduces to formulas \eqrefs{JPS-L:sym-curr1}{JPS-L:sym-curr2}
modulo curls, since
$\A{}{\nu} \coD{[\mu}\Qp{\nu]}{}
= \A{}{\nu} {*\coD{[\mu}\Q{\nu]}{}}
= \frac{1}{2} \vol{}{\mu\nu\alpha\beta} \l(
\Q{}{\beta} \D{\nu} \A{}{\alpha} 
+ \D{\alpha}( \Q{}{\beta}\A{}{\nu} ) \r)$
and likewise for the other term 
$\Ap{}{\nu} \coD{[\mu}\Q{\nu]}{} 
= -\Ap{}{\nu} {*\coD{[\mu}\Qp{\nu]}{}}$. 
\endProof

We remark that a converse for either correspondence 
\eqref{JPS-L:sym-adjsym1} or \eqref{JPS-L:sym-adjsym2} 
would rely on generalizing the \loccohom{2} theorem to the
\jps/ in Lorentz gauge,
which is unnecessary for the purpose of generating conserved currents.

Of the two \conscurr/ formulas,
the latter one \eqref{JPS-L:sym-curr2} is distinguished 
by the property that it generates a trivial current 
from the scaling \sym/ \eqref{JPS-L-Xscaling}
and a nontrivial current from 
the duality-rotation \sym/ \eqref{JPS-L-Xduality}
(while this correspondence is reversed by the dual formula
\eqref{JPS-L:sym-curr1}). 
We emphasize that both formulas generate the same \conscurr/s
when applied to the class of geometric \syms/
\eqsref{JPS-L-Xscaling}{JPS-L-XCKvecdual},
since this class exhibits duality-invariance \eqref{QQ'duality}. 

We now list in \tableref{table:JPS-L-sym-curr} 
the \conscurr/s that arise through formula \eqref{JPS-L:sym-curr2} 
from the geometric \syms/ 
\eqsref{JPS-L-Xscaling}{JPS-L-XCKvecdual} 
classified in Theorem~\ref{thm:JPS-L-sym}.
For comparison,
we write out the stress-energy currents, 
\EQ
\curr{\mu} = 
\kv{\sigma}{} ( 
 \F{}{\sigma\nu} \F{\mu\nu}{} + \duF{}{\sigma\nu} \duF{\mu\nu}{} ) 
:= \consT{\mu}{\sigma} \kv{\sigma}{}
\label{hKVcurr}
\endEQ
where $\kv{\sigma}{}$ is a general \CKvec/
and $\consT{\mu}{\sigma}$ denotes the conserved stress-energy tensor of \Meq/. 

 \begin{table}[h]
 \begin{center}
 $\AR{|c|c|c|} \hline
 & \mbox{Geometric Symmetry} & \mbox{Conserved current}
 \\ \hline\hline
 \mbox{scaling} & 
 \AR{c} 
 \Q{}{\mu} = \A{}{\mu} \\
 \Qp{}{\mu} = \Ap{}{\mu} \endAR &
 \AR{rl}
 \scurr{\mu} =& 
 \frac{1}{2} (\A{}{\nu} \F{\mu\nu}{} + \Ap{}{\nu} \duF{\mu\nu}{} ) \\
 = & \D{\nu} \l( 
 \frac{1}{4} \vol{}{\mu\nu\sigma\tau} \A{}{\sigma} \Ap{}{\tau} \r) 
  \endAR
 \\ \hline
 \mbox{duality-rotation} & 
 \AR{c} 
 \Q{}{\mu} = \Ap{}{\mu} \\ 
 \Qp{}{\mu} = -\A{}{\mu} \endAR &
 \dcurr{\mu} = 
 \frac{1}{2} (\Ap{}{\nu} \F{\mu\nu}{} - \A{}{\nu} \duF{\mu\nu}{} )
 \\ \hline
 \AR{c} \mbox{internal} \\ \mbox{\rbs/} \\ 
 \gam{\mu\nu}{} \mbox{ given by \eqref{skewmatr} } \endAR & 
 \AR{rl} 
 \Q{}{\mu} &= 
 \gam{\mu}{\nu} \A{}{\nu} + \dugam{\mu}{\nu} \Ap{}{\nu} \\
 \Qp{}{\mu} &= 
 \gam{\mu}{\nu} \Ap{}{\nu} - \dugam{\mu}{\nu} \A{}{\nu} \endAR &
 \AR{rl}
 \rbcurr{\mu} = &
 \frac{1}{2} \gam{\nu}{\sigma} (
 \A{}{\sigma} \F{\mu\nu}{} + \Ap{}{\sigma} \duF{\mu\nu}{} ) \\
 & + \frac{1}{2} \dugam{\nu}{\sigma} (
 \Ap{}{\sigma} \F{\mu\nu}{} - \A{}{\sigma} \duF{\mu\nu}{} )
 \endAR
 \\ \hline
 \AR{c} \mbox{general conformal} \\ \mbox{\Kvec/} \\ 
 \kv{\mu}{}, \z{}{\mu\nu} \mbox{ given by \eqref{CKV} } \endAR &
 \AR{rl} 
 \Q{}{\mu} &= 
 \wLie{\xi} \A{}{\mu} + \z{\mu}{\nu} \A{}{\nu} + \duz{\mu}{\nu} \Ap{}{\nu} \\
 \Qp{}{\mu} &= 
 \wLie{\xi} \Ap{}{\mu} + \z{\mu}{\nu} \Ap{}{\nu} - \duz{\mu}{\nu} \A{}{\nu} 
 \endAR &
 \AR{rl}
 \ccurr{\mu} 
 =& 
 \frac{1}{2}(\wLie{\xi}\A{}{\nu}) \F{\mu\nu}{} 
 + \frac{1}{2}(\wLie{\xi}\Ap{}{\nu}) \duF{\mu\nu}{} \\
 & + \frac{1}{2} \z{\nu}{\sigma}( \A{}{\sigma} \F{\mu\nu}{} 
 + \Ap{}{\sigma} \duF{\mu\nu}{} ) \\
 & + \frac{1}{2} \duz{\nu}{\sigma}( \Ap{}{\sigma} \F{\mu\nu}{} 
 - \A{}{\sigma} \duF{\mu\nu}{} ) 
 \endAR
 \\ \hline
 \AR{c} \mbox{general conformal} \\ \mbox{\Kvec/ dual} \endAR &
 \AR{rl} 
 \Q{}{\mu} &= 
 \wLie{\xi} \Ap{}{\mu} + \z{\mu}{\nu} \Ap{}{\nu} -\duz{\mu}{\nu} \A{}{\nu} \\
 \Qp{}{\mu} &= 
 -\wLie{\xi} \A{}{\mu} - \z{\mu}{\nu} \A{}{\nu} -\duz{\mu}{\nu} \Ap{}{\nu} 
 \endAR &
 \AR{rl}
 \cducurr{\mu} =& 
 \frac{1}{2} (\wLie{\xi}\Ap{}{\nu}) \F{\mu\nu}{} 
 - \frac{1}{2} (\wLie{\xi}\A{}{\nu}) \duF{\mu\nu}{} \\
 & + \frac{1}{2} \z{\nu}{\sigma}( \Ap{}{\sigma} \F{\mu\nu}{} 
 - \A{}{\sigma} \duF{\mu\nu}{} ) \\
 & - \frac{1}{2} \duz{\nu}{\sigma}( \A{}{\sigma} \F{\mu\nu}{} 
 + \Ap{}{\sigma} \duF{\mu\nu}{} )
 \endAR
 \\ \hline
 \AR{c} \mbox{Lie derivative terms} \\ 
 \mbox{in $\Q{}{\mu},\Qp{}{\mu}$} \\ \\
 \wLie{\xi} = \Lie{\xi} +\frac{1}{4} \Omega \\
 \Omega \mbox{ given by \eqref{CKV} } \endAR & &
 \AR{rl} 
 & \frac{1}{2} (\Lie{\xi} \A{}{\nu}) \F{\mu\nu}{} 
 + \frac{1}{2} (\Lie{\xi} \Ap{}{\nu}) \duF{\mu\nu}{} \\
 &= \D{\nu}\l( \frac{1}{2} \kv{\sigma}{} (
 \A{}{\sigma} \F{\mu\nu}{} + \Ap{}{\sigma} \duF{\mu\nu}{} ) \r) \\
 &\qquad + \consT{\mu}{\sigma}\kv{\sigma}{} \\ \hline 
 & \frac{1}{2} (\Lie{\xi} \Ap{}{\nu}) \F{\mu\nu}{}
 - \frac{1}{2} (\Lie{\xi} \A{}{\nu}) \duF{\mu\nu}{} \\
 &= \D{\nu}\l( \frac{1}{2} \kv{\sigma}{} (
 \Ap{}{\sigma} \F{\mu\nu}{} - \A{}{\sigma} \duF{\mu\nu}{} ) \r)
 \endAR  
 \\ \hline
 \endAR$
\caption{Conserved currents derived from geometric symmetries of 
the joint \potsys/ in Lorentz gauge}
\label{table:JPS-L-sym-curr}
 \end{center}
 \end{table}

Trivial currents (i.e. curls) are produced by 
both the scaling \sym/ \eqref{JPS-L-Xscaling}
and the \HKV/ dual \syms/ \eqref{JPS-L-XHKvecdual}. 
The \HKV/ \syms/ \eqref{JPS-L-XHKvec}
themselves reproduce
the same stress-energy currents (modulo trivial currents)
as admitted by the unaugmented \potsys/.

In contrast the currents associated with 
the new \rbvec/ symmetries \eqref{JPS-L-Xrbvec}
as well as the \CKV/ symmetries \eqref{JPS-L-XCKvec}
and their dual symmetries \eqref{JPS-L-XCKvecdual}
are genuinely new local \conscurr/s. 
Nontriviality of these currents is established
in \secref{sec:nonlocal}, 
where we will simplify all inessential dependence on potentials 
through the embedding of the solution space of the \potsys/ \eqref{JPS-L}
into the solution space of \Meq/.

Finally, we remark that the 
duality-rotation current, \rbvec/ currents, 
\CKV/ currents and their dual currents
are related by considering a general \CKvec/
as was noted for the corresponding \syms/ 
in Proposition~\ref{JPS-L-combinedsymm}. 

\begin{prop}
\label{JPS-L-combinedcurr}
For a \HKvec/ 
$\kv{\mu}{} = 
\k{1}{\mu}{} + \k{2}{\mu\nu}{} \x{}{\nu} + \k{3}{}{} \x{\mu}{}$, 
\EQ
\ccurr{\mu} 
= \consT{\mu}{\sigma}\kv{\sigma}{} +\rbcurr{\mu} \modcurl,\quad
\cducurr{\mu} 
= -\rbducurr{\mu} +\frac{1}{4}\Omega \dcurr{\mu} \modcurl,\quad
\gam{}{} = \z{}{}
\endEQ
is a geometric current of the \jps/ in Lorentz gauge,
where $\z{}{\mu\nu} = \frac{1}{2} \k{2}{\mu\nu}{}=\const$, 
$\Omega=2\k{3}{}{}=\const$. 
\end{prop}

\section{New nonlocal \syms/ and \conslaw/s of \Meq/}
\label{sec:nonlocal}

In this section 
we derive nonlocal symmetries and nonlocal conserved currents of \Meq/
through projection of 
the new geometric symmetries and new geometric currents of 
the \jps/ in Lorentz gauge found in \secref{sec:jps-analysis}. 
The projection is defined on $\Jsp{1}(A,A')$
using the natural embedding of 
the solution jet space $\divfrRsp{}(A,A')$ of 
the potentials in Lorentz gauge
into the solution jet space of \Meq/ as follows:
We note first that, with the introduction of potentials, 
points in $\Jsp{q}(F)$ are identified with 
equivalence classes of points in $\Jsp{q+1}(A,A')$ 
under (prolonged) gauge transformations 
$\A{}{\mu} \ra \A{}{\mu} + \D{\mu}\chi$, 
$\Ap{}{\mu} \ra \Ap{}{\mu} + \D{\mu}\chi'$, 
for any functions $\chi,\chi'$ on $\Jsp{q}(A,A')$. 
Any representative in each equivalence class satisfying the equations
\EQ
\coD{\mu}\D{\mu}\chi(A) = \invflat{\mu\nu}\A{}{\mu,\nu} ,\quad
\coD{\mu}\D{\mu}\chi'(A') = \invflat{\mu\nu}\Ap{}{\mu,\nu}
\label{JPS-L:waveeq}
\endEQ
corresponds to the choice of Lorentz gauge being imposed on the potentials.
(Note $\chi,\chi'$ are unique to within the addition of 
any solution of the source-free wave equation.)
Hence we have an embedding of $\divfrRsp{}(A,A')$ into $\Rsp{}(A,A')$
given by the explicit jet space coordinates 
$(\x{\mu}{},\tA{}{\nu},\tAp{}{\nu},\F{}{\mu\nu},\tA{}{\mu\nu},\tAp{}{\mu\nu})$
such that 
\EQ
\tA{}{\nu} = \A{}{\nu} -\D{\nu}\chi(A) ,\quad
\tAp{}{\nu} = \Ap{}{\nu} -\D{\nu}\chi'(A')
\label{tAA'coords}
\endEQ
satisfy Lorentz gauge 
\EQ
\invflat{\mu\nu} \tA{}{\nu,\mu} = \invflat{\mu\nu} \tAp{}{\nu,\mu} = 0 . 
\endEQ
Note here 
\EQ
\F{}{\mu\nu} = \tA{}{[\nu,\mu]} = -{*\tAp{}{[\nu,\mu]}} ,
\endEQ
and 
\EQs
&&
\tA{}{\mu\nu} = \trfr( \A{}{\mu\nu} -\D{\mu}\D{\nu}\chi(A) )
= \A{}{(\nu,\mu)} -\D{\mu}\D{\nu}\chi(A) ,
\\
&&
\tAp{}{\mu\nu} = \trfr( \Ap{}{\mu\nu} -\D{\mu}\D{\nu}\chi'(A') )
= \Ap{}{(\nu,\mu)} -\D{\mu}\D{\nu}\chi'(A') ,
\label{dertAA'coords}
\endEQs
where, we recall, 
$\A{}{\mu\nu} = \A{}{(\nu,\mu)}$, 
$\Ap{}{\mu\nu} = \Ap{}{(\nu,\mu)}$.
Hereafter, $\Rsp{}(\divfrA,\divfrA') \subset \Jsp{1}(A,A')$
will denote $\divfrRsp{}(A,A') \subset \Jsp{1}(A,A')$
under this embedding of the solution jet spaces.

\subsection{Induced \syms/ of \Meq/}

\begin{defn}
\label{prop:JPS-L-projsymm}
Any local symmetry 
$\X\tA{}{\mu} = \Q{}{\mu}[\divfrA,\divfrA']$, 
$\X\tAp{}{\mu} = \Qp{}{\mu}[\divfrA,\divfrA']$ 
of the \jps/ on $\Rsp{}(\divfrA,\divfrA')$
projects to a symmetry 
$\X\F{}{\mu\nu} = \D{[\mu}\Q{}{\nu]}[A,A']$
of \Meq/ on $\Rsp{}(A,A')$
via the transformation \eqsref{tAA'coords}{dertAA'coords}. 
A projected symmetry is {\em local} in the electromagnetic field
iff
$\X\F{}{\mu\nu}$ has {\it no} essential dependence on the potentials
$\A{}{\nu},\Ap{}{\nu},\A{}{\mu\nu},\Ap{}{\mu\nu}$,
so $\D{[\mu}\Q{}{\nu]}$ is a \horiz/ 2-form on $\Rsp{}(F)$;
and otherwise a projected symmetry $\X\F{}{\mu\nu}[A,A']$ is {\em nonlocal}. 
\end{defn}

We now list in \tableref{JPS-L-ME-sym} 
the projected geometric symmetries 
obtained from Theorem~\ref{thm:JPS-L-sym}.
The projections are a straightforward computation 
using Definition~\ref{prop:JPS-L-projsymm}. 

 \mystretch{\STR}
 \begin{table}[h]
 \begin{center}
 \TAB{|c|c|} \hline
 Geometric \sym/ type &
 \TAB{c} Induced \sym/ of \\
 $\coD{\mu}\F{}{\mu\nu} = 0$,\quad $\coD{\mu}\duF{}{\mu\nu} = 0$ 
 \endTAB \\ \hline\hline
 \TAB{c} 
 scaling \eqref{JPS-L-Xscaling} \\
 and duality-rotation \eqref{JPS-L-Xduality} 
 \endTAB &
 $\AR{rl}
 \sX= &
 \F{}{\mu\nu} \parder{\F{}{\mu\nu}} \\ 
 \dX= &
 \duF{}{\mu\nu} \parder{\F{}{\mu\nu}} 
 \endAR$ \\ \hline
 \TAB{c} 
 internal \rb/ \eqref{JPS-L-Xrbvec} \\
 $\gam{\mu\nu}{}$ given by \eqref{skewmatr} \endTAB & 
 $\rbX{}= 
 ( \tA{\sigma}{[\mu} \gam{\nu]\sigma}{} 
 + \tAp{\sigma}{[\mu} \dugam{\nu]\sigma}{} ) \parder{\F{}{\mu\nu}}$ 
 \\ \hline
 \TAB{c} 
 translation, \rb/, dilation \eqref{JPS-L-XHKvec} \\
 and dual \eqref{JPS-L-XHKvecdual} \\
 $\kv{\mu}{}$ given by \eqref{HKV} \endTAB &
 $\AR{rl}
 \hX{}= & 
 (\Lie{\xi} \F{}{\mu\nu}) \parder{\F{}{\mu\nu}}  \\ 
 \hX{}'= &
 (\Lie{\xi} \duF{}{\mu\nu}) \parder{\F{}{\mu\nu}} 
 \endAR$ \\ \hline
 \TAB{c} 
 conformal \eqref{JPS-L-XCKvec} \\
 and dual \eqref{JPS-L-XCKvecdual} \\
 $\kv{\mu}{}$, $\z{}{\mu\nu}$, $\Omega$ given by \eqref{CKV} \\
 $\wLie{\xi} := \Lie{\xi} + \frac{1}{4} \Omega$ 
 \endTAB &
 $\AR{rl} 
 \cX{}= &
 ( \wLie{\xi} \F{}{\mu\nu}
 +\tA{\sigma}{[\mu} \z{\nu]\sigma}{} 
 + \tAp{\sigma}{[\mu} \duz{\nu]\sigma}{} \\ 
 & - \tA{\sigma}{}\der{\sigma}\z{\mu\nu}{} 
 + \tAp{\sigma}{}\der{\sigma}\duz{\mu\nu}{} ) 
 \parder{\F{}{\mu\nu}} \\
 \cX{}'= &
 ( \wLie{\xi} \duF{}{\mu\nu}
 +\tAp{\sigma}{[\mu} \z{\nu]\sigma}{} 
 - \tA{\sigma}{[\mu} \duz{\nu]\sigma}{} \\ 
 & - \tAp{\sigma}{}\der{\sigma}\z{\mu\nu}{} 
 - \tA{\sigma}{}\der{\sigma}\duz{\mu\nu}{} ) \parder{\F{}{\mu\nu}} 
 \endAR$ \\ \hline
 \endTAB
\caption{Symmetries of \Meq/ induced by geometric \syms/
of the \jps/ in Lorentz gauge
 \label{JPS-L-ME-sym}}
 \end{center}
 \end{table}

The projected symmetries with essential dependence on the potentials
are simply the \CKV/ \syms/ and their dual \syms/, 
and the \rbvec/ \syms/,
which comprise 
the new geometric symmetries found for the \jps/ in Lorentz gauge.
Note the Lie derivative term in the \CKV/ \syms/
corresponds itself precisely to 
the local conformal \syms/ of \Meq/, 
and likewise for the Lie derivative term in the dual \syms/.
By subtraction of these local symmetries 
\EQ
\X_{\it conf} = (\Lie{\xi} \F{}{\mu\nu}) \Parder{\F{}{\mu\nu}} ,\quad
\X_{\it conf}' = (\Lie{\xi} \duF{}{\mu\nu}) \Parder{\F{}{\mu\nu}} ,
\label{ME:localsymm}
\endEQ
we are left with new symmetries $\tX$ 
of a similar form to the \rbvec/ \syms/, 
involving no dependence on derivatives of 
the electromagnetic field $\F{}{\mu\nu}$ or its dual, 
as given by the following transformations: 

\Items
\item[{\rm (i)}] 
\rbvec/ transformation 
\EQ
\rbtX{} = \Big( 
\A{\sigma}{[\mu} \gam{\nu]\sigma}{} + \Ap{\sigma}{[\mu} \dugam{\nu]\sigma}{} 
+ \gam{[\mu}{\sigma} \D{\nu]}\D{\sigma}\chi(A)
+ \dugam{[\mu}{\sigma} \D{\nu]}\D{\sigma}\chi'(A')
\Big) \Parder{\F{}{\mu\nu}} ,  
\label{ME:rbvecsymm}
\endEQ
with $\gam{}{\mu\nu} = \gam{}{[\mu\nu]}=\const$,
\item[{\rm (ii)}] 
\cvec/ transformation 
\EQs
\ctX{} =  && \Big( 
\frac{1}{4} \Omega \F{}{\mu\nu}
+ \A{\sigma}{[\mu} \z{\nu]\sigma}{} + \Ap{\sigma}{[\mu} \duz{\nu]\sigma}{} 
- \A{\sigma}{} \der{\sigma}\z{\mu\nu}{} 
+ \Ap{\sigma}{}\der{\sigma}\duz{\mu\nu}{} 
\nonumber\\&&
- \D{[\mu}( \z{\nu]}{\sigma} \D{\sigma}\chi(A) 
+ \duz{\nu]}{\sigma} \D{\sigma}\chi'(A') 
+\frac{1}{3} \coder{\sigma} \z{\nu]\sigma}{} \chi(A) )
\Big) \Parder{\F{}{\mu\nu}} , 
\label{ME:Cvecsymm}
\endEQs
and dual transformation 
\EQs
\ctX{}' =  && \Big( 
\frac{1}{4} \Omega \duF{}{\mu\nu}
+ \Ap{\sigma}{[\mu} \z{\nu]\sigma}{} 
-\A{\sigma}{[\mu} \duz{\nu]\sigma}{} 
- \Ap{\sigma}{} \der{\sigma}\z{\mu\nu}{} 
- \A{\sigma}{}\der{\sigma}\duz{\mu\nu}{} 
\nonumber\\&&
- \D{[\mu}( \z{\nu]}{\sigma} \D{\sigma}\chi'(A') 
- \duz{\nu]}{\sigma} \D{\sigma}\chi(A) 
+\frac{1}{3} \coder{\sigma} \z{\nu]\sigma}{} \chi'(A') )
\Big) \Parder{\F{}{\mu\nu}} , 
\label{ME:Cvecdualsymm}
\endEQs
where 
$\z{}{\mu\nu} := -\frac{1}{2} \coder{[\mu} \kv{\nu]}{}$
and $\Omega:= \frac{1}{2} \der{\mu} \kv{\mu}{}$, 
with 
$\kv{\mu}{} = 
\k{4}{\sigma}{} \x{}{\sigma} \x{\mu}{}
- \frac{1}{2} \k{4}{\mu}{} \x{\sigma}{} \x{}{\sigma}$,
$\k{4}{\mu}{}=\const$;
$\chi(A),\chi'(A')$ are scalar functions satisfying 
the wave equation \eqref{JPS-L:waveeq}. 
\endItems
Their explicit dependence on the potentials means that these symmetries
are nonlocal and nontrivial. 

We remark that 
the transformations \eqrefs{ME:Cvecsymm}{ME:Cvecdualsymm} 
continue to be admitted as \syms/ 
if the \CKvec/ $\kv{\mu}{}$ is replaced by a \HKvec/ \eqref{HKV}
as seen from Proposition~\ref{JPS-L-combinedsymm}. 
Indeed, 
in the case of a \rb/ $\kv{\mu}{} = \k{2}{\mu\nu}{} \x{}{\nu}$, 
these transformations respectively reduce to 
the \rbvec/ \syms/ \eqref{ME:rbvecsymm} 
with parameters 
$\gam{}{\mu\nu} = 
\frac{1}{2} \k{2}{\mu\nu}{}, 
-\frac{1}{4} \invvol{\mu\nu}{\sigma\tau}\k{2}{\sigma\tau}{}$;
in contrast these transformations yield 
a multiple $\frac{1}{2} \k{3}{}{}$ of 
the scaling and duality-rotation \syms/
in the case of a dilation $\kv{\mu}{} = \k{3}{}{}\x{\mu}{}$,
and trivial \syms/
in the case of a translation $\kv{\mu}{} =\k{1}{\mu}{}$. 

Thus, the nonlocal symmetries we have found for \Meq/ 
arise from new symmetries of the form \eqrefs{ME:Cvecsymm}{ME:Cvecdualsymm}
for a general \CKvec/ 
\ie/ $\kv{\mu}{}$ is any generator of 
a conformal isometry \eqref{conformalisom} of \Minksp/. 

There is a deeper unity between 
the \rbvec/ \syms/ and the \cvec/ and dual \syms/.
Consider the transformations \eqrefs{ME:Cvecsymm}{ME:Cvecdualsymm} 
using the sum of a \rb/ \Kvec/ and a \CKvec/
\EQ
\kv{\mu}{} = 
\k{2}{\mu\nu}{} \x{}{\nu} 
+ \k{4}{\sigma}{} \x{}{\sigma} \x{\mu}{}
- \frac{1}{2} \k{4}{\mu}{} \x{\sigma}{} \x{}{\sigma} . 
\label{unifyKV}
\endEQ
We observe that only the (scaled) curl $\z{}{\mu\nu}$ and divergence $\Omega$
of this \Kvec/ enter these transformations,
where $\Omega$ is related to $\z{}{\mu\nu}$ by
\EQ
\z{}{\mu\nu} 
= \frac{1}{2} \k{2}{\mu\nu}{} -\k{4}{[\mu}{} \x{\nu]}{} ,\quad
\Omega 
= 2\k{4}{\mu}{} \x{}{\mu} = \frac{4}{3} \x{}{\nu} \der{\mu} \z{}{\mu\nu} .
\endEQ
Moreover, $\z{}{\mu\nu}$ has precisely the form of the dual of 
a \KY/ tensor, 
namely $\ky{}{\mu\nu}(x) := \duz{}{\mu\nu}$ satisfies the
\KY/ equation \cite{KYtensors:1,KYtensors:2}
\EQ
\coder{(\sigma} \ky{}{\mu)\nu} =0 . 
\label{KYeq}
\endEQ
(More general \KY/ tensors of conformal type 
are parameters for local chiral \syms/ of \Meq/
\cite{FusNik:1983,FusNik:1987book,AncPoh:2002,AncPoh:2004}
and first arose in the study of integrals of the geodesic equations 
for light rays in the curved Kerr metric;
they are also connected with separation of variables of \Meq/ 
in that metric.)

Thus the nonlocal internal \syms/ 
\eqref{ME:rbvecsymm}, \eqref{ME:Cvecsymm}, \eqref{ME:Cvecdualsymm}
have the following geometric form:
\EQs
\kyX 
=&& \Big( 
\frac{1}{3} \x{}{\sigma} \der{\tau} \duky{}{\sigma\tau} \F{}{\mu\nu}
-\tA{\sigma}{[\mu} \duky{\nu]\sigma}{} 
+ \tAp{\sigma}{[\mu} \ky{\nu]\sigma}{} 
+ \tA{}{\sigma} \coder{\sigma}\duky{\mu\nu}{} 
+ \tAp{}{\sigma}\coder{\sigma}\ky{\mu\nu}{} 
\Big) \Parder{\F{}{\mu\nu}} , 
\nonumber\\
= && \Big( 
\frac{1}{3} \x{}{\sigma} \der{\tau} \duky{}{\sigma\tau} \F{}{\mu\nu}
-\A{\sigma}{[\mu} \duky{\nu]\sigma}{} 
+ \Ap{\sigma}{[\mu} \ky{\nu]\sigma}{} 
+ \A{}{\sigma} \coder{\sigma}\duky{\mu\nu}{} 
+ \Ap{}{\sigma}\coder{\sigma}\ky{\mu\nu}{} 
\nonumber\\&&
+\D{[\mu}( \duky{\nu]}{\sigma} \D{\sigma}\chi(A) 
- \ky{\nu]}{\sigma} \D{\sigma}\chi'(A') 
+\frac{1}{3} \coder{\sigma} \duky{\nu]\sigma}{} \chi(A) )
\Big) \Parder{\F{}{\mu\nu}} , 
\label{ME:KYsymm}\\
\kyX' 
=  && \Big( 
\frac{1}{3} \x{}{\sigma} \der{\tau} \duky{}{\sigma\tau} \duF{}{\mu\nu}
-\tAp{\sigma}{[\mu} \duky{\nu]\sigma}{} 
- \tA{\sigma}{[\mu} \ky{\nu]\sigma}{} 
+ \tAp{}{\sigma} \coder{\sigma}\duky{\mu\nu}{} 
- \tA{}{\sigma}\coder{\sigma}\ky{\mu\nu}{} 
\Big) \Parder{\F{}{\mu\nu}} , 
\nonumber\\
= && \Big( 
\frac{1}{3} \x{}{\sigma} \der{\tau} \duky{}{\sigma\tau} \duF{}{\mu\nu}
-\Ap{\sigma}{[\mu} \duky{\nu]\sigma}{} 
- \A{\sigma}{[\mu} \ky{\nu]\sigma}{} 
+ \Ap{}{\sigma} \coder{\sigma}\duky{\mu\nu}{} 
- \A{}{\sigma}\coder{\sigma}\ky{\mu\nu}{} 
\nonumber\\&&
+\D{[\mu}( \duky{\nu]}{\sigma} \D{\sigma}\chi'(A') 
+ \ky{\nu]}{\sigma} \D{\sigma}\chi(A) 
+\frac{1}{3} \coder{\sigma} \duky{\nu]\sigma}{} \chi'(A') )
\Big) \Parder{\F{}{\mu\nu}} , 
\label{ME:KYdualsymm}
\endEQs
where 
\EQ
\ky{}{\mu\nu} = 
\kb{1}{\mu\nu}{} + \invvol{\mu\nu}{\sigma\tau} \kb{2}{\sigma}{} \x{\tau}{} ,
\quad
\kb{1}{\mu\nu}{} = \kb{1}{[\mu\nu]}{}, \kb{2}{\sigma}{} =\const
\label{KYT}
\endEQ
is a \KY/ tensor. 
Note that only the set of constant \KY/ tensors is preserved 
under duality $\ky{}{\mu\nu} \ra \duky{}{\mu\nu}$. 

\begin{thm}
\label{ME-nonlocalsymm}
\Meq/ admits the (nontrivial) nonlocal \syms/
\eqrefs{ME:KYsymm}{ME:KYdualsymm}
depending on an arbitrary \KY/ tensor \eqref{KYT}. 
Under the \dutr/ \eqref{JPS-duality} on the potentials, 
these \syms/ are interchanged,
and in the case of a constant \KY/ tensor 
they are related through directly replacing 
this tensor with its dual.
Thus the \syms/ comprise a 14-dimensional vector space. 
\end{thm}

The vector space structure of these nonlocal \syms/ 
has a basis consisting of 
six of \rbvec/ type \eqref{ME:rbvecsymm}, 
four of \cvec/ type \eqref{ME:Cvecsymm}
and four of dual type \eqref{ME:Cvecdualsymm}.
However, their commutator structure is not closed. 
Note the \rbvec/ \syms/ themselves comprise a $\SO(3,1)$ Lie algebra,
which follows from projection of the commutator structure 
stated in Theorem~\ref{JPS-L-symalg}
for the corresponding local \syms/
on the solution jet space $\divfrRsp{}(A,A')$ of the \jps/ in Lorentz gauge.
In contrast the \cvec/ \syms/ and their duals
do not arise from projection of any local \syms/ of this \potsys/,
and as a consequence,
their commutators do not have the form of 
local transformations on $\divfrRsp{}(A,A')$, 
\ie/ they necessarily involve a nonlocal dependence on the potentials
and hence there is no natural Lie algebra structure for them. 
The same conclusion holds for the commutators of
the \rbvec/ \syms/ with the \cvec/ \syms/ or their duals. 
These commutators are still themselves 
nonlocal \syms/ of the \potsys/ and hence of \Meq/. 
More generally, 
there is an enveloping (nonlocal) \sym/ algebra 
generated by the span of the \syms/
\eqref{ME:rbvecsymm}, \eqref{ME:Cvecsymm}, \eqref{ME:Cvecdualsymm},
and their repeated commutators. 

If instead we consider the respective superpositions
$\ctX{} +\X_{\it conf}$, $\ctX{}' +\X'_{\it conf}$, 
of the nonlocal \syms/ \eqref{ME:Cvecsymm}, \eqref{ME:Cvecdualsymm} 
of \Meq/ 
and the corresponding local \syms/ \eqref{ME:localsymm} 
given by \rb/ \Kvec/s and \CKvec/s, 
their collective enveloping algebra collapses to 
the Lie algebra of \rbs/ and complexified inversions,
namely the natural semidirect product of 
$\SO(3,1)$ with $\U(1)^4 \otimes \Cnum$, 
as follows from Theorem~\ref{JPS-L-symalg}.

\subsection{Induced conservation laws}

\begin{defn}
\label{JPS-L-projcurr}
Any local \conscurr/ $\curr{\mu}[\divfrA,\divfrA']$ 
on $\Rsp{}(\divfrA,\divfrA')$
directly projects to a \conscurr/ 
$\curr{\mu}[A,A']$ of \Meq/ on $\Rsp{}(A,A')$
via the transformation \eqsref{tAA'coords}{dertAA'coords}. 
A projected current is {\em local} in the electromagnetic field
iff,
up to addition of a curl, 
it has {\it no} essential dependence on the potentials
$\A{}{\nu},\Ap{}{\nu},\A{}{\mu\nu},\Ap{}{\mu\nu}$,
so that 
$\curr{\mu}[A,A']+ \triv{\nu}{\mu\nu}[A,A']$
for some antisymmetric tensor function $\curl{\mu\nu}{}[A,A']$
is a vector function on $\Rsp{}(F)$; 
and otherwise a projected current $\curr{\mu}[A,A']$
is {\em nonlocal}. 
\end{defn}

For the geometric \conscurr/s 
listed in \tableref{table:JPS-L-sym-curr} 
for the \jps/ in Lorentz gauge, 
the Lie derivative terms in the general \CKV/ currents
simply project to the corresponding stress-energy currents \eqref{hKVcurr},
while the Lie derivative terms in the dual currents project to a curl. 
The remaining internal terms in those currents 
as well as the \rbvec/ currents
and duality-rotation current 
separately project to new \conscurr/s $\tcurr\up{\mu}$ of \Meq/: 
\Items
\item[{\rm (i)}] 
duality-rotation \conslaw/ 
\EQ
\dtcurr{\mu}
= \tAp{}{\nu} \F{\mu\nu}{} - \tA{}{\nu} \duF{\mu\nu}{} 
= \Ap{}{\nu} \F{\mu\nu}{} - \A{}{\nu} \duF{\mu\nu}{} \modcurl, 
\label{ME:dualitycurr}
\endEQ
\item[{\rm (ii)}] 
\rbvec/ \conslaw/ 
\EQs
\rbtcurr{\mu} 
= &&
\gam{\nu}{\sigma}( 
\tA{}{\sigma} \F{\mu\nu}{} + \tAp{}{\sigma} \duF{\mu\nu}{} )
+ \dugam{\nu}{\sigma}( 
\tAp{}{\sigma} \F{\mu\nu}{} - \tA{}{\sigma} \duF{\mu\nu}{} )
\nonumber\\
= &&
-\frac{1}{2} \gam{}{\nu\sigma}( 
\A{\mu}{} \F{}{\nu\sigma} + \Ap{\mu}{} \duF{}{\nu\sigma} )
+ 2 \gam{}{\nu[\sigma} ( 
\A{}{\sigma} \F{\mu]}{\nu} + \Ap{}{\sigma} \duF{\mu]}{\nu} )
\nonumber\\&&
+\wavecurr{\mu}(\chi(A),\gam{}{\nu\sigma}\F{}{\nu\sigma})
+\wavecurr{\mu}(\chi'(A'),\gam{}{\nu\sigma}\duF{}{\nu\sigma})
\label{ME:rbveccurr}\\&&
\modcurl, 
\nonumber
\endEQs
with $\gam{}{\mu\nu} = \gam{}{[\mu\nu]}=\const$,
\item[{\rm (iii)}] 
\cvec/ \conslaw/ 
\EQs
\ctcurr{\mu}
= &&
\z{\nu}{\sigma}( 
\tA{}{\sigma} \F{\mu\nu}{} + \tAp{}{\sigma} \duF{\mu\nu}{} )
+ \duz{\nu}{\sigma}( 
\tAp{}{\sigma} \F{\mu\nu}{} - \tA{}{\sigma} \duF{\mu\nu}{} )
+ \frac{1}{4} \Omega ( \tA{}{\nu} \F{\mu\nu}{} + \tAp{}{\nu} \duF{\mu\nu}{} )
\nonumber\\
= &&
-\frac{1}{2} \z{}{\nu\sigma} ( 
\A{\mu}{} \F{}{\nu\sigma} + \Ap{\mu}{} \duF{}{\nu\sigma} )
+ 2 \z{}{\nu[\sigma} (
\A{}{\sigma} \F{\mu]}{\nu} + \Ap{}{\sigma} \duF{\mu]}{\nu} )
-\frac{1}{8} \vol{}{\mu\sigma\alpha\beta} 
\der{\sigma}\Omega\ \A{}{\alpha} \Ap{}{\beta} 
\nonumber\\&&
+\wavecurr{\mu}(\chi(A),\z{}{\nu\sigma}\F{}{\nu\sigma})
+\wavecurr{\mu}(\chi'(A'),\z{}{\nu\sigma}\duF{}{\nu\sigma})
\label{ME:Cveccurr}\\&&
\modcurl, 
\nonumber
\endEQs
and dual \conslaw/ 
\EQs
\ductcurr{\mu}
= &&
\z{\nu}{\sigma}( 
\tAp{}{\sigma} \F{\mu\nu}{} - \tA{}{\sigma} \duF{\mu\nu}{} )
- \duz{\nu}{\sigma}( 
\tA{}{\sigma} \F{\mu\nu}{} + \tAp{}{\sigma} \duF{\mu\nu}{} )
+ \frac{1}{4} \Omega ( \tAp{}{\nu} \F{\mu\nu}{} - \tA{}{\nu} \duF{\mu\nu}{} )
\nonumber\\
= &&
-\frac{1}{2} \z{}{\nu\sigma} ( 
\Ap{\mu}{} \F{}{\nu\sigma} - \A{\mu}{} \duF{}{\nu\sigma} )
+ 2 \z{}{\nu[\sigma} (
\Ap{}{\sigma} \F{\mu]}{\nu} - \A{}{\sigma} \duF{\mu]}{\nu} )
\nonumber\\&&
+ \frac{1}{4} \Omega ( \Ap{}{\nu} \F{\mu\nu}{} -\A{}{\nu} \duF{\mu\nu}{} )
-\wavecurr{\mu}(\chi(A),\z{}{\nu\sigma}\duF{}{\nu\sigma})
+\wavecurr{\mu}(\chi'(A'),\z{}{\nu\sigma}\F{}{\nu\sigma})
\label{ME:Cvecdualcurr}\\&&
\modcurl, 
\nonumber
\endEQs
where $\z{}{\mu\nu} := -\frac{1}{2} \coder{[\mu} \kv{\nu]}{}$
and $\Omega:= \frac{1}{2} \der{\mu} \kv{\mu}{}$, 
with 
$\kv{\mu}{} = 
\k{4}{\sigma}{} \x{}{\sigma} \x{\mu}{}
- \frac{1}{2} \k{4}{\mu}{} \x{\sigma}{} \x{}{\sigma}$,
$\k{4}{\mu}{}=\const$; 
here $\chi(A),\chi'(A')$ are scalar functions satisfying 
the wave equation \eqref{JPS-L:waveeq},
and $\wavecurr{\mu}(f,g) := \frac{1}{2}( g\coD{\mu}f - f\coD{\mu}g )$
is a skew-bilinear vector function depending on any scalar expressions $f,g$. 
\endItems
In writing downs these currents 
we have simplified some terms through integration by parts
and multiplied by an overall factor of $2$. 
We remark that $\wavecurr{\mu}(f,g)$ has the form of a conserved current
formula for the ordinary scalar wave equation \cite{AncBlu:1996JMP},
taking $f,g$ to be a pair of symmetries. 
These terms are not separately conserved in the currents here. 

Note that if the \CKvec/ $\kv{\mu}{}$ in the currents \eqref{ME:Cveccurr}
is replaced by a \HKvec/ \eqref{HKV}, 
we obtain 
the \rbvec/ currents \eqref{ME:rbveccurr}
with $\gam{}{\mu\nu} = \frac{1}{2} \k{2}{\mu\nu}{}$
in the case of a \rb/ 
$\kv{\mu}{} = \k{2}{\mu\nu}{} \x{}{\nu}$, 
and trivial currents 
in the case of a translation $\kv{\mu}{} =\k{1}{\mu}{}$
or a dilation $\kv{\mu}{} = \x{\mu}{}$. 
Similarly, from the dual currents \eqref{ME:Cvecdualcurr} 
we obtain 
the \rbvec/ current \eqref{ME:rbveccurr}
with $\gam{}{\mu\nu} 
= -\frac{1}{4} \invvol{\mu\nu}{\alpha\beta} \k{2}{\alpha\beta}{}$
in the case of a \rb/ 
$\kv{\mu}{} = \k{2}{\mu\nu}{} \x{}{\nu}$,
the duality-rotation current \eqref{ME:dualitycurr} 
in the case of a dilation $\kv{\mu}{} = \x{\mu}{}$,
and a trivial current 
in the case of translation $\kv{\mu}{} =\k{1}{\mu}{}$. 

Thus, the new currents we have found for \Meq/ 
come from new \conslaw/s of the form \eqrefs{ME:Cveccurr}{ME:Cvecdualcurr}
that exist for a general \CKvec/, 
\ie/ $\kv{\mu}{}$ is any generator of 
a conformal isometry \eqref{conformalisom} of \Minksp/. 

We now prove these \conslaw/s \eqsref{ME:dualitycurr}{ME:Cvecdualcurr}
are nontrivial and nonlocal.

Firstly, it is useful to define the weight of 
a quadratic current $\curr{\mu}$ on $\Jsp{q}(A,A')$ 
to be the maximum of the weights of all monomial terms in $\curr{\mu}$,
given by counting the total number of derivatives that appear on 
the potentials:
\ie/ $\A{}{\mu},\Ap{}{\mu}$ have weight $0$, 
$\F{}{\mu\nu}$ has weight $1$, 
while $\chi(A),\chi'(A')$ are counted as weight $-1$
through equation \eqref{JPS-L:waveeq}. 
The lowest weight nontrivial {\it local} \conscurr/s of \Meq/ 
as shown in \Ref{AncPoh:2001} 
are the stress-energy currents \eqref{ME:stressenergy},
which have weight $2$ on $\Jsp{}(F) \subset \Jsp{1}(A,A')$.
In comparison, the currents \eqsref{ME:dualitycurr}{ME:Cvecdualcurr}
each have weight $1$ and cannot be equivalent consequently 
to any nontrivial local current of \Meq/.
This establishes, moreover, that these currents 
\eqsref{ME:dualitycurr}{ME:Cvecdualcurr}
are nonlocal. 
Thus it remains to show only that they are nontrivial
when restricted to the solution jet space $\Rsp{}(\divfrA,\divfrA')$.

\begin{lem}
Suppose 
\EQ
\curr{\mu} = 
( \k{1}{\mu}{\alpha\beta\sigma} \tA{\sigma}{} 
+ \k{2}{\mu}{\alpha\beta\sigma} \tAp{\sigma}{} )\F{\alpha\beta}{} 
\label{currLHS}
\endEQ
is a \conscurr/ on $\Rsp{}(\divfrA,\divfrA')$.
Then this current is trivial iff 
\EQ
\k{2}{\mu\alpha\beta\sigma}{}
= \frac{1}{2} \vol{\nu\tau}{\alpha\beta} \k{1}{\mu\nu\tau\sigma}{}
= k \vol{}{\mu\alpha\beta\sigma} ,\quad
k=\const
\label{trivcondition}
\endEQ
\end{lem}

\Proof{}
Any trivial current on $\Rsp{}(\divfrA,\divfrA')$ is characterized 
on $\Jsp{1}(A,A')$ by having the form 
\EQ
\curr{\mu} 
= \triv{\nu}{\mu\nu} 
+ \bb{}{\mu}{\alpha\beta} (\Fp{\alpha\beta}{} - \duF{\alpha\beta}{}) ,\quad
\Fp{\alpha\beta}{}:= \tAp{[\beta,\alpha]}{} ,\quad
\F{\alpha\beta}{}:= \tA{[\beta,\alpha]}{} , 
\label{currRHS}
\endEQ
where
\EQs
&& 
\curl{\mu\nu}{} = 
\T{1}{\mu\nu}{\alpha\beta}(x) \tA{\alpha}{} \tA{\beta}{}
+ \T{2}{\mu\nu}{\alpha\beta}(x) \tAp{\alpha}{} \tAp{\beta}{} 
+ \Tp{\mu\nu}{\alpha\beta}(x) \tAp{\alpha}{} \tA{\beta}{} , 
\\
&&
\bb{}{\mu}{\alpha\beta} = 
\bb{1}{\mu}{\alpha\beta\sigma}(x) \tA{\sigma}{}
+ \bb{2}{\mu}{\alpha\beta\sigma}(x) \tAp{\sigma}{} , 
\endEQs
with the coefficients subject to the index symmetries 
$\T{i}{\mu\nu}{\alpha\beta} = \T{i}{[\mu\nu]}{(\alpha\beta)}$, 
$\Tp{\mu\nu}{\alpha\beta} = \Tp{[\mu\nu]}{\alpha\beta}$, 
$\bb{i}{\mu}{\alpha\beta\sigma} = \bb{i}{\mu}{[\alpha\beta]\sigma}$. 
To proceed we expand \eqref{currRHS} 
and substitute the decompositions 
$\tA{\beta,\alpha}{} = 
\F{\alpha\beta}{} 
+ \trfr \tA{(\beta,\alpha)}{}$
and 
$\tAp{\beta,\alpha}{} = 
\Fp{\alpha\beta}{} 
+ \trfr \tAp{(\beta,\alpha)}{}$.
The coefficients of all terms other than 
$\F{\alpha\beta}{}\tA{\sigma}{}$ and $\F{\alpha\beta}{}\tAp{\sigma}{}$
cannot match \eqref{currLHS} and must therefore vanish, 
which leads to the conditions
\EQs
&&
\T{1}{\mu}{(\nu\alpha)\beta} = 
\T{2}{\mu}{(\nu\alpha)\beta} = 
\Tp{\mu}{(\nu|\beta|\alpha)} =
\Tp{\mu}{(\nu\alpha)\beta} = 
0 , 
\label{conds}\\
&& 
2\T{2}{\mu}{[\alpha\beta]\sigma} + \bb{2}{\mu}{\alpha\beta\sigma} = 0 , 
\label{rel1}\\
&&
\Tp{\mu}{[\alpha\beta]\sigma} + \bb{1}{\mu}{\alpha\beta\sigma} = 0 . 
\label{rel2}
\endEQs
The index symmetries imposed by \eqref{conds}
immediately show that 
$\T{1}{\mu}{\nu\alpha\beta}$ and $\T{2}{\mu}{\nu\alpha\beta}$ vanish
while $\Tp{\mu}{\nu\alpha\beta}$ must be totally antisymmetric
and hence 
\EQ
\Tp{\mu}{\nu\alpha\beta} = k(x) \invvol{\mu}{\nu\alpha\beta}
\label{kcond}
\endEQ
for some $k(x)$. 
It then follows from \eqref{rel1} that 
$\bb{2}{\mu}{\alpha\beta\sigma}$ also vanishes. 
Finally, 
by now equating \eqref{currLHS} to \eqref{currRHS}
and collecting like terms, we obtain 
\EQs
\k{1}{\mu}{\alpha\beta\sigma} 
= -\frac{1}{2} \vol{\alpha\beta}{\rho\tau} \bb{1}{\mu}{\rho\tau\sigma} ,\quad
\k{2}{\mu}{\rho\tau\sigma} 
= - \Tp{\mu}{\rho\tau\sigma} ,\quad
\coder{\rho} \k{2}{\mu}{\rho\tau\sigma} =0 . 
\label{kconds}
\endEQs
Then the relations \eqrefs{rel2}{kconds} establish 
the conditions \eqref{trivcondition} stated in the Lemma. 
\endProof

To apply this Lemma,
we observe the new \conscurr/s \eqsref{ME:dualitycurr}{ME:Cvecdualcurr} 
written in terms of $\tA{}{\nu},\tAp{}{\nu},\F{}{\mu\nu}$
have the form \eqref{currLHS} where 
\EQ
\k{1}{}{\mu\alpha\beta\sigma} = 
\flat{\mu[\alpha} \aa{\beta]\sigma} 
- \frac{1}{2} \vol{\mu\alpha\beta}{\nu} \aap{\nu\sigma}  ,\quad
\k{2}{}{\mu\alpha\beta\sigma} = 
\flat{\mu[\alpha} \aap{\beta]\sigma} 
+ \frac{1}{2} \vol{\mu\alpha\beta}{\nu} \aa{\nu\sigma} 
\endEQ
as given by \tableref{ME:currsform}.
The algebraic conditions \eqref{trivcondition} are readily 
found to fail for each of these currents. 
Therefore our main result is now proven.

 \mystretch{\STR}
 \begin{table}[h]
 \begin{center}
 \TAB{|c|c|c|} \hline
 Conserved current & 
 $\aa{\mu\nu}$ & $\aap{\mu\nu}$ 
 \\ \hline\hline
 Duality-rotation 
 & $0$ & $\flat{\mu\nu}$ 
 \\ \hline
 Internal \rb/ 
 & $\gam{\mu\nu}{}$ & $\dugam{\mu\nu}{}$
 \\ \hline
 Internal conformal 
 & $\z{\mu\nu}{} +\frac{1}{4} \Omega \flat{\mu\nu}$
 & $\duz{\mu\nu}{}$
 \\ \hline
 Internal dual conformal 
 & $-\duz{\mu\nu}{}$
 & $\z{\mu\nu}{} + \frac{1}{4} \Omega \flat{\mu\nu}$ 
 \\ \hline
 \endTAB
\caption{Form of nonlocal currents of \Meq/
\label{ME:currsform}}
 \end{center}
 \end{table}

\begin{thm}
\label{ME-nonlocalcurr}
\Meq/ admits the (nontrivial) nonlocal \conslaw/s
\eqsref{ME:dualitycurr}{ME:Cvecdualcurr}. 
These \conslaw/s span a 15-dimensional vector space
whose basis consists of one of duality-rotation type,
six of \rbvec/ type,
four of \cvec/ type and four of dual type. 
\end{thm}

We remark that these nonlocal \conslaw/s also arise directly from 
the nonlocal symmetries \eqsref{ME:rbvecsymm}{ME:Cvecdualsymm}
through use of the formula \eqref{MEcurrentformula}
in Theorem~\ref{thm:MEcurrentformula}
that generates \conscurr/s of \Meq/ from symmetries of \Meq/.
As a consequence, firstly, 
all the currents \eqsref{ME:dualitycurr}{ME:Cvecdualcurr} 
are invariant under the \dutr/ \eqref{JPS-duality} 
on the potentials 
(with the simultaneous induced transformation \eqref{ME-duality-F} 
on the electromagnetic field). 
More significantly,
the \cvec/ currents and dual currents
can be unified with the \rbvec/ currents
by the introduction of a \KY/ tensor,
corresponding to the analogous form of the nonlocal \syms/ 
\eqsref{ME:KYsymm}{ME:KYdualsymm}
stated in Theorem~\ref{ME-nonlocalsymm}. 
As we recall, the \KY/ tensor is identified with
the dual of the curl of a \Kvec/ of the form \eqref{unifyKV}
given by the sum of a \rb/ \Kvec/ and a \CKvec/. 

\begin{cor}
The nonlocal \conslaw/s 
\eqref{ME:rbveccurr}, \eqref{ME:Cveccurr}, \eqref{ME:Cvecdualcurr} 
admitted by \Meq/ 
have the unified form 
\EQs
\kycurr{\mu}
= &&
\ky{\nu}{\sigma}( 
\tAp{}{\sigma} \F{\mu\nu}{} - \tA{}{\sigma} \duF{\mu\nu}{} )
-\duky{\nu}{\sigma}( 
\tA{}{\sigma} \F{\mu\nu}{} + \tAp{}{\sigma} \duF{\mu\nu}{} )
\nonumber\\&&
+ \frac{1}{3} \x{\sigma}{} \coder{\tau}\duky{\sigma\tau}{}
( \tA{}{\nu} \F{\mu\nu}{} + \tAp{}{\nu} \duF{\mu\nu}{} )
\nonumber\\
= &&
\frac{1}{2} \ky{}{\nu\sigma} ( 
\A{\mu}{} \duF{}{\nu\sigma} - \Ap{\mu}{} \F{}{\nu\sigma} )
- 2 \ky{}{\nu[\sigma} ( 
\A{}{\sigma} \duF{\mu]}{\nu} - \Ap{}{\sigma} \F{\mu]}{\nu} )
+\frac{1}{6} \coder{\tau}\duky{\tau\nu}{} 
\vol{}{\mu\nu\alpha\beta} \A{}{\alpha} \Ap{}{\beta} 
\nonumber\\&&
-\wavecurr{\mu}(\chi(A),\ky{}{\nu\sigma}\duF{}{\nu\sigma})
+\wavecurr{\mu}(\chi'(A'),\ky{}{\nu\sigma}\F{}{\nu\sigma})
\label{ME:KYcurr}\\&&
\modcurl, 
\nonumber
\endEQs
and 
\EQs
\dukycurr{\mu}
= &&
-\ky{\nu}{\sigma}( 
\tA{}{\sigma} \F{\mu\nu}{} + \tAp{}{\sigma} \duF{\mu\nu}{} )
-\duky{\nu}{\sigma}( 
\tAp{}{\sigma} \F{\mu\nu}{} - \tA{}{\sigma} \duF{\mu\nu}{} )
\nonumber\\&&
+ \frac{1}{3} \x{\sigma}{} \coder{\tau}\duky{\sigma\tau}{}
( \tAp{}{\nu} \F{\mu\nu}{} - \tA{}{\nu} \duF{\mu\nu}{} )
\nonumber\\
= &&
\frac{1}{2} \ky{}{\nu\sigma} ( 
\A{\mu}{} \F{}{\nu\sigma} + \Ap{\mu}{} \duF{}{\nu\sigma} )
- 2 \ky{}{\nu[\sigma} ( 
\A{}{\sigma} \F{\mu]}{\nu} + \Ap{}{\sigma} \duF{\mu]}{\nu} )
\nonumber\\&&
- \frac{1}{3} \x{\sigma}{} \coder{\tau}\duky{\sigma\tau}{}
( \A{}{\nu} \duF{\mu\nu}{} - \Ap{}{\nu} \F{\mu\nu}{} )
\nonumber\\&&
-\wavecurr{\mu}(\chi(A),\ky{}{\nu\sigma}\F{}{\nu\sigma})
-\wavecurr{\mu}(\chi'(A'),\ky{}{\nu\sigma}\duF{}{\nu\sigma})
\label{ME:KYdualcurr}\\&&
\modcurl, 
\nonumber
\endEQs
depending on a \KY/ tensor \eqref{KYT}.
Under the \dutr/ \eqref{JPS-duality}, \eqref{ME-duality-F} 
on the potentials and electromagnetic field, 
the currents \eqrefs{ME:KYcurr}{ME:KYdualcurr} are invariant. 
In the case when the \KY/ tensor is constant 
they are interchanged under replacing 
$\ky{}{\mu\nu}$ with its dual $\duky{}{\mu\nu}$. 
Thus these currents span a 14-dimensional vector space
which is duality-invariant.
\end{cor}

\section{Concluding remarks}
\label{sec:concl}

In conclusion we mention a few applications of our main results. 

The new nonlocal infinitesimal \syms/ we have obtained for 
\Meq/ in Minkowski space are of point-type \cite{BluAnc:2002book}
to within a duality transformation, 
when expressed in terms of the joint electric and magnetic potentials
for the electromagnetic field. 
(Consequently, they can be realized as genuine point transformations
on the complexified jet space of the \potsys/.)
Thus these symmetries can be used to derive corresponding 
new group-invariant solutions of \Meq/
and to generalize physically interesting solutions 
(\eg/ plane waves and monopoles)
under the action of the finite symmetry group of transformations. 

The associated new nonlocal \conscurr/s we have derived from these \syms/ 
give rise to constants of motion for the electromagnetic field.
Explicit expressions for them 
given by gauge-invariant integrals of the field 
can be obtained by extending the methods used in \Ref{AncBlu:1997JMP}
to 3+1 dimensions.
In particular, we expect the resulting constants of motion to be
functionally independent of energy, momentum, angular/boost momentum,
and conformal quantities as well as chiral quantities
that arise as constants of motion from the local \conscurr/s of \Meq/.

For future work, 
it is planned to extend our results to classify
all nonlocal \syms/ and nonlocal \conscurr/s of \Meq/
produced via a complete classification 
of \syms/ and \conscurr/s of local form
in the potentials and their derivatives to any finite order,
admitted by the \jps/ in Lorentz gauge. 
A further extension of obvious interest would be to derive 
nonlocal \syms/ and nonlocal \conscurr/s of the electromagnetic field 
on curved background spacetimes, for instance
the Schwarzschild and Kerr black hole spacetimes. 
This analysis is tractable using the spinor techniques of 
\Ref{AncPoh:2003,AncPoh:2001}. 

Finally, our methods in this paper readily apply to other 
linear physical field equations, 
notably the linearized gravity wave equation in \Minksp/
and its spin $s$ generalization.
A classification of all local \syms/ and local \conscurr/s 
for the linear spin $s>0$ field equations in \Minksp/ 
has been carried out in \Ref{AncPoh:2004,AncPoh:2003,AncPoh:2002}. 
A systematic investigation of spin $s$ \potsys/s 
would be of significant interest. 

\acknowledgments
D.T. is supported by NSERC through a Canada Graduate Scholarship.

{
\baselineskip=14pt
\bibliography{ME-nonlocal}
}

\end{document}